# Idea and Theory of Particle Access

Bo LI*, Zhongjiang YAN, and Mao YANG

School of Electronics and Information, Northwestern Polytechnical University,

Xi'an, CHINA, 710072

{libo.npu*, zhjyan, yangmao}@nwpu.edu.cn

***Abstract*: Aiming at some problems existing in the current quality of service (QoS) mechanism of large-scale networks (i.e. poor scalability, coarse granularity for provided service levels, poor fairness between different service levels, and improving delay performance at the expense of sacrificing some resource utilization), the paper puts forward the idea and theory of particle access. In the proposed particle access mechanism, the network first granulates the information flow (that is, the information flow is subdivided into information particles, each of which is given its corresponding attributes), and allocates access resources to the information particle group which is composed of all the information particles to be transmitted, so as to ensure that the occupied bandwidth resources is minimized on the premise of meeting the delay requirements of each information particle. Moreover, in the paper, the concepts of both information particle and information particle group are defined; The relationships between the average access bandwidth requirement, the peak average access bandwidth requirement, and the minimum reachable access bandwidth of an information particle group are analyzed; The influences of time attribute and attribute of bearing capacity of an information particle group on the minimum reachable access bandwidth are analyzed. The equal scaling theorem for information particle groups is proved; Effective methods for the calculation of the minimum reachable access bandwidth of an information particle group are given, and a particle access algorithm based on dynamically adjusting the minimum reachable access bandwidth is proposed; Furthermore, several properties, such as the upper and the lower bounds of the minimum reachable access bandwidth, the property of linear** **superpositions of multiple information particle groups, the property of linear splitting of an information particle group, and a closed form of expression of the minimum reachable access bandwidth and so on, of the minimum reachable access bandwidth(s) of information particle group(s) are proved; Finally, the joint optimization of the mechanism of network wave and the mechanism of particle access is studied. For a given information particle group transmitted over a given primary path, a method to obtain the minimum reachable transmission bandwidth, which can guarantee the end-to-end delays of all the information particles contained in the information particle group, is proposed; It is proven that, given a primary path and an information particle group it bears, to guarantee the end-to-end transmission performance of the information particle group over the primary path, there exists a lower bound for the peak reachable transmission bandwidth of the primary path, and this lower bound is achievable. The research in this paper reveals that the combination of the mechanism of network wave and the mechanism of particle access is an effective way to solve the problem of end-to-end multi-hop transmission. Therefore, the corresponding information processing mechanism integrating the network wave mechanism and particle access mechanism in information processing networks is named as the "Wave-Particle Duality mechanism in Networks" (i.e. Mechanism of WPDN). Correspondingly, the related theory investigating this mechanism is referred to as the "Theory of Wave-Particle Duality in Networks" (i.e. Theory of WPDN).**

## I. Introduction

As we all know, compared with “circuit switching” technology, “packet switching” technology can significantly improve the utilization of network resources, so it has become one of the key technologies supporting today’s Internet. In packet switching systems, because the network resources are fully shared among various traffic flows, the problem of QoS guarantee becomes more challenging [1]. The basic starting point of QoS mechanism is to ensure various performance requirements of different traffic flows under the limited network resources. Therefore, two mechanisms of QoS guarantee, namely “integrated services” and “differentiated services” are proposed.

The core idea of integrated services [2] is to reserve a certain amount of resources in all the forwarding nodes belonging to the end-to-end path connecting the source node and the destination node of a traffic flow to meet the resource requirement of the traffic flow (this kind of QoS mechanism of providing resource reservation service for traffic flows is also called as “flow based” mechanism). In essence, integrated services can be regarded as an imitation of circuit switching mechanism over packet switching mechanism, so good QoS guarantee can be achieved for traffic flows by using the mechanism of integrated services. However, once this mechanism is applied in large-scale communication networks (such as the Internet), one will inevitably face the need to reserve resources for hundreds of traffic flows on each router at the same time, and maintain the corresponding statuses of all these traffic flows, which poses a great challenge to the design and the implementation of a network router! **In short, the main problem faced by the mechanism of integrated services is its poor scalability.**

The mechanism of differentiated services [3] is proposed to solve the problem of poor scalability faced by the mechanism of integrated services. The basic idea is to only divide packets of traffic flows in the network into several categories, and provide differentiated services for different types of packets at the intermediate forwarding nodes (this kind of QoS mechanism of providing differentiated services for different types of packets is also called as “class based” mechanism). Although this mechanism solves the problem of poor scalability faced by integrated services to a certain extent, there are still some noteworthy problems existing in the mechanism of differentiated services.

First, let us consider the possible problems of differentiated services in the case that packets of traffic flows are only divided into fewer types. For example, among the proposed mechanisms for differentiated services, a simple mechanism that can provide differentiated services is proposed, that is, the so-called “expedited forwarding” mechanism [4]. In the expedited forwarding mechanism, packets of traffic flows are only divided into two types, namely the “regular” type and the “expedited” type. Relative to the “regular” packets, the “expedited” packets have absolute priority for the utilization of network resources. The problem is that even for packets, belonging to the same “expedited” type or “regular” type, significantly different performance requirements still exist. In this way, it is somewhat “rough” to only divide the packets into two categories, so it is difficult to finely guarantee the QoS requirement of each traffic flow. **In short, by using the mechanism of differentiated services, if the packets of traffic flows are only divided into fewer types, it is difficult for the network to provide refined QoS guarantee.**

Secondly, let us think over the possible problems of the mechanism of differentiated services when packets of traffic flows are divided into many more types. For example, in differentiated services, a relatively complex mechanism, the so-called “assured forwarding” mechanism have been proposed [5]. In short, in the assured forwarding mechanism, packets are first divided into four different priorities, and then three levels of packet dropping are further defined. This is equivalent to finally dividing the packets of traffic flows into twelve service levels and providing differentiated services for different levels accordingly. For a mechanism just like the assured forwarding, since it supports more service levels, it can provide more refined differentiated services. However, facing with a large number of service types, how to ensure the fairness of resource allocation among all these service levels becomes more and more difficult to be dealt with. That is to say, with the increase of the number of service levels, the fairness relationships between different service levels that need to be properly handled with in the resource allocation strategy become more and more complicated (theoretically, assured forwarding needs to handle $\sum_{i=1}^{11} i = 66$ pairs of fairness relationships between all the service levels!), which makes it a very challenging problem to reconcile the fairness between various service levels. **In short, by adopting the mechanism of differentiated services, if packets of traffic flows are divided into many more types, it is more difficult to ensure fairness between different service levels.**

The fairness problems of the mechanism of differentiated services pointed out above will further bring about other problems. On the one hand, with the emergence of new types of network services, it is likely to be found out that the available service levels cannot provide satisfactory QoS guarantee for the new appeared services. In order to solve this problem, it is natural that one thinks of designing new service levels for new types of services and adds them to the existing class based QoS mechanism. Therefore, one needs to consider not only how to design the corresponding resource allocation strategy for the new service levels according to their specific performance requirements, but also how to make the newly defined service levels coexist "peacefully" with the existing service levels in the shared QoS guarantee mechanism. Obviously, with the increasing number of service levels defined, it will become more and more difficult to design an extra new service level which can coexist "peacefully" with the existing service levels. Once again, it can be seen that the problem of scalability arises. On the other hand, if the network allows an application to set its service level on its own will, the application vendor is very likely to intentionally label the packets generated by its developed application to some service level with high priority, which inevitably leads to the abuse of some service levels that are easier to acquire network resources, and further amplifies the harm caused by the existing unfairness between service levels, and even causes the final collapse of the network!

In recent years, people pay more and more attention to the research on how to solve the problem of delay guarantee in wireless access networks [6]. Among many solutions, some of them improve QoS for delay sensitive traffic by improving the total system throughput, scheduling strategies, or queuing strategies, and so on. However, these strategies cannot provide deterministic delay guarantee. Although some access strategies can provide deterministic delay guarantee, it is achieved at the expense of sacrificing some network access resources. For example, in some access strategies based on resource reservation, the network reserves a certain amount of resources for some expected packets of delay sensitive traffic flows in advance, so as to significantly improve the performance of these traffic flows. However, the arrival characteristics of some delay sensitive traffic have great uncertainty, which leads to the phenomenon that some pre reserved resources are wasted (that is, the expected packets do not arrive within the time period corresponding to the reserved resources). **In short, in order to guarantee the QoS for delay sensitive traffic, available solutions often sacrifice the utilization of network resources in exchange for the improvement of delay performance.**

Aiming at the problems existing in the mechanism of QoS guarantee in large-scale networks,

the paper proposes the idea and theory of particle access. The major contributions are listed as follows:

1) The basic idea of particle access is proposed, and the concepts of both information particle and information particle group are given;
2) The relationships between the average access bandwidth requirement, the peak average access bandwidth requirement, and the minimum reachable access bandwidth of an information particle group are analyzed;
3) The influences of time attribute and attribute of bearing capacity of an information particle group on the minimum reachable access bandwidth are analyzed. The equal scaling theorem for information particle groups is proved;
4) Effective methods for the calculation of the minimum reachable access bandwidth of an information particle group are proposed, and a particle access algorithm based on dynamically adjusting the minimum reachable access bandwidth is given;
5) The upper bound and the lower bound of the minimum reachable access bandwidth of an information particle group are given, respectively;
6) The property of linear superpositions of multiple information particle groups under specific constraints is proved. The property of linear splitting of an information particle group is also proved;
7) A closed form expression of representing the minimum reachable access bandwidth of an information particle subgroup as the product of its vector of switching functions and its vector of average access bandwidth requirements is given;
8) For a given information particle group transmitted over a given primary path, a method to obtain the minimum reachable transmission bandwidth, which can guarantee the end-to-end delays of all the information particles contained in the information particle group, is proposed;
9) The existence and achievability of the lower bound of the peak reachable transmission bandwidth of a primary path are proven.

The paper is organized into nine parts: in Section I, the problems existing in the current QoS mechanisms are pointed out; In Section II, the basic idea of the paper is proposed, that is, the idea of particle access is introduced into the access mechanism of networks; In Section III, the basic concept of information particle is defined; In Section IV, the concept of information particle group is further defined; In Section V, basic properties of the minimum reachable access bandwidth of an information particle group are proved; In Section VI, the intrinsic mechanism of the influences of time attribute and attribute of bearing capacity of an information particle group on the minimum reachable access bandwidth is analyzed; In Section VII, based on the basic framework of the theory built in the paper, effective methods for the calculation of the minimum reachable access bandwidth of an information particle group are given, and a particle access algorithm based on dynamically adjusting the minimum reachable access bandwidth is proposed. Furthermore, several properties of the minimum reachable access bandwidth(s) of information particle group(s) are proved; In Section VIII, the joint optimization method of the mechanism of network wave and the mechanism of particle access is proposed. And the existence and achievability of the lower bound of the peak reachable transmission bandwidth of a primary path are proven. Finally, conclusions are drawn in Section IX.

## II. Basic idea

This section describes the basic idea of particle access. We organize the descriptions of this section into answers to several key problems, so as to clearly outline the basic idea.

1) How to use the concept of information particle to realize a simple and universal modeling for network traffic flows?

The four main performance measurements of the QoS guarantee for end-to-end network traffic flows are [1]: bandwidth, delay, jitter, and packet loss rate. If the bandwidth and the performance of end-to-end delay in the network are acceptable, the delay jitter and the packet loss rate can be improved by using some caching technologies residing on the receiver side and some packet retransmission technologies, respectively. Thus, among the four performance measurements, bandwidth and delay are more important. Furthermore, if the number of information bits $\ell$ born by a packet and its tolerable end-to-end delay $L$ are given, one can easily estimate that the bandwidth demand of the packet is $B = \frac{\ell}{L}$. Therefore, considering that there is a simple conversion relationship between bandwidth and delay, and the delay is often the basis for estimating the required bandwidth, it is evident that among the four QoS measurements mentioned above, delay is the most critical one. This is why we firmly grasp the measurement of delay in the proposed traffic flow model which serves as the basis of the idea of particle access.

In order to build a mechanism which can effectively support the end-to-end QoS guarantee in large-scale networks, it is necessary to set up a simple and general model for the end-to-end traffic flows, based on which an efficient and flexible QoS mechanism can be built. In section I, we have analyzed the problems faced by flow based and class based QoS mechanisms. To solve these problems, it is believed that the new traffic model should not dwell on the level of each traffic flow or on the level of each traffic class, but should further focus on the more basic and fine-grained level, that is, the level of each packet, which makes the descriptions of the characteristics of end-to-end traffic flows are more fundamental. Therefore, the following assumptions are made:

a) An end-to-end traffic flow can be divided into several information blocks with a certain size (i.e. bearing capacity $\ell$ of information, unit: bit);

b) Each information block corresponds to a value of end-to-end delay tolerance (i.e. the

deadline $L$ of the information block, unit: second);

c) For an information block with a bearing capacity of $\ell$ and a deadline of $L$, if the network can successfully transmit all the information bits it bears to the receiving end before $L$, the transmission of the information block is regarded as valid.

In this paper, we define the information block with the attributes of bearing capacity and deadline as an information particle (to highlight that it is the cornerstone for constructing the QoS mechanism), and accordingly, an end-to-end traffic flow is modeled as an information particle flow composed of information particles (see sections III and IV for relevant details).

2) How to achieve efficient processing of information particles?

In the mechanism of particle access, network routers need to deal with the access and transmission of information particle groups composed of a large number of information particles. Specifically, it is necessary for routers to decide how much access bandwidth is allocated to the information particle groups and how to realize the orderly and efficient transmission of information particles within a given bandwidth. Generally speaking, there are three possible methods to be adopted:

a) Determining the required bandwidth and transmission time instant for each information particle according to the attributes of each of them;

b) According to the attributes of each information particle, some statistical characteristics of the corresponding information particle group are evaluated, and a rough estimation of the bandwidth required by the whole information particle group is obtained. Of course, in addition to evaluating the total bandwidth required by the information particle group, it is also necessary to further schedule the transmission of each information particle contained in the information particle group;

c) According to the attributes of each information particle, the bandwidth required by the corresponding information particle group is accurately evaluated, and based on which, the transmission of each information particle contained in the information particle group is further scheduled.

It is easy to be found that in method "a", because a single information particle is processed one by one, it is bound to bring about huge processing overhead. Therefore, method "a" is the least desirable one of the three methods. As for method “b”, it has less processing overhead and can bring about better scalability, but the statistical evaluation method adopted leads to large deviation in the estimation of required bandwidth. If the estimation of bandwidth is too large relative to the real one, it will cause the unnecessary waste of access resources. On the other hand, if the estimated access bandwidth is too small relative to the real one, it will affect the effective transmissions of some information particles, and eventually affect the QoS of the end-to-end traffic flow. For method “c”, if one can find a simple and accurate method to evaluate the required bandwidth and a simple and effective processing strategy for information particles, this method is undoubtedly the most ideal one. Because we have found out a simple and accurate bandwidth evaluation method and a simple strategy which can realize efficient transmission of information particles under a given access bandwidth, method “c” is naturally adopted to realize efficient particle access in the paper (see sections V, VI and VII for relevant details).

3) How to allocate access bandwidth for an information particle group? How to achieve efficient transmission of an information particle group?

For the former problem, our idea is to find out what is the minimum access bandwidth (defined as “minimum reachable access bandwidth of an information particle group” in the paper) that can make the transmission of all the information particles valid under the condition that the corresponding information particle group is given (that is, when the basic attributes of the information particles constituting the information particle group are known). For the latter problem, the basic idea of the paper is to find out a simple and effective processing strategy that can achieve the minimum reachable access bandwidth. It can be seen that only by solving these two key problems at the same time can one say that an efficient method to transmit information particle group has been found.

In sections V and VI, we analyze the relationships between the average access bandwidth requirement, the peak average access bandwidth requirement, and the minimum reachable access bandwidth of an information particle group, and point out that at least there exist a processing strategy of EDF (early deadline first), which can achieve efficient transmissions of all information particles under the constraint of the minimum reachable access bandwidth. Furthermore, in section VII, effective methods of obtaining the minimum reachable access bandwidth of an information particle group are proposed.

4) How to dynamically adjust the access bandwidth of an information particle group?

With the passage of time and the advancement of the transmission procedure of each information particle constituting an information particle group, the key attributes of the information particles contained in the information particle group (such as the attribute of bearing capacity and the deadline attribute) show some time variability, which leads to the time variability of the minimum reachable access bandwidth required by the information particle group. Therefore, we define the time point when the minimum reachable access bandwidth changes as the inflection point of the required access bandwidth of the information particle group. In Section VII, we systematically analyze the properties of the inflection point of the required access bandwidth, and put forward a method to determine the inflection point and the corresponding newly changed minimum reachable access bandwidth at that point. Furthermore, based on the theoretical results proved in the paper, in Section VII, a particle access algorithm based on dynamically adjusting the minimum reachable access bandwidth is proposed. The algorithm dynamically adjusts the minimum reachable access bandwidth of the information particle group at inflection point of the required access bandwidth, so as to ensures that the access efficiency is maintained to be one during the whole procedure of the access and transmission of the information particle group (that is, the optimal utilization of the allocated bandwidth can be achieved).

5) What are the advantages of particle access?

After comparing the problems pointed out in Section I with the basic idea of particle access described in this section, it is not difficult for us to summarize the advantages of particle access as follows:

a) Due to the efficient computation of the access bandwidth and the exerted processing strategy of an information particle group, particle access has good scalability;
b) Since the service quality of an information flow is guaranteed at the level of each information particle, particle access can achieve more fine-grained QoS guarantee;
c) Since the idea of providing differentiated services at the level of each traffic flow or at the level of each traffic class is abandoned, particle access can completely solve the intrinsic problems of fairness existing between different traffic flows or traffic classes;

d) Since it has the ability of dynamically adjusting the minimum reachable access bandwidth, the mechanism of particle access can dynamically maximize the utilization of bandwidth resources for an information particle group while still meeting the delay requirement of each information particle.

## III. Definition of an information particle

In this section, the concept of an information particle, the basic attributes of an information particle (including the bearing capacity, the initial time instant and the deadline, etc.), the effective survival time interval, the effective survival time span, an effective processing strategy performed by the system for an information particle, the universal set of effective processing strategies for an information particle, and the average access bandwidth requirement of an information particle are defined.

**Definition III.1: A relative time axis with the current time instant being set as the zero time point**

On the time axis considered in this paper, time instant $t = 0$ is defined as the current time instant of the system (i.e., the present time instant). Time instant $t > 0$ denotes the instant obtained by forward shifting $t$ seconds relative to the current time instant (i.e., a future instant). Time instant $t < 0$ corresponds to the instant obtained by a backward shift of $|t|$ seconds from the current time instant (i.e., a past instant).

**Note:**

1) Unless otherwise specified, only the case of $t \geq 0$ is considered in this paper;
2) Unless otherwise specified, all time points considered correspond to the current time instant of the system.

**Definition III.2: An information particle with a specific index number**

An information block that bears a certain amount of information (measured in bits), is assigned a uniquely identifiable index number $i$ $(\in \mathbb{Z}^+)$ and awaits system processing, is referred to as the information particle with an index number $i$.

**Note:**

1) In this paper, an information particle serves as the most fundamental unit to be processed by the system;
2) Information particles with distinct index numbers correspond to different information particles,

even if they possess identical bearing capacity, initial time instant and deadline. Conversely, different information particles are assigned to different index numbers;

3) The index number referred to herein bears no relation to outcomes derived from any sorting operation. Hence, it must be rigorously distinguished from the sequence number of information particles generated via the specific sorting operation defined subsequently in **Definition IV.2**;
4) Unless otherwise specified hereinafter, "information particle $i$" refers to the information particle with index number $i$.

**Definition III.3: Bearing capacity of an information particle at a specific time instant (unit: bit)**

The amount of information, pending to be processed by the system, borne by information particle $i$ at time instant $t$ is defined as the bearing capacity of the information particle $i$ at time instant $t$, which is denoted as $\ell_i(t)$ ($\ell_i(t) \in \mathbb{R}, \ell_i(t) \geq 0$). In particular, where no ambiguity arises, the bearing capacity of information particle $i$ at the current time instant is denoted as $\ell_i$.

**Note:**

1) It is assumed in this paper that the information borne by an information particle $i$ is continuously processed by the system, and the effectively processed portion of information is no longer included in subsequent measurements of the bearing capacity for this information particle. Accordingly, the bearing capacity $\ell_i(t)$ constitutes a time-varying process and may decrease continuously over time;
2) For the convenience of subsequent mathematical manipulation, bearing capacity $\ell_i(t)$ is treated as a real number;
3) In this paper, only the case where $\ell_i > 0$ is considered.

**Definition III.4: Initial time instant of an information particle (unit: second)**

The earliest time instant at which the system can perform effective processing on an information particle $i$ is defined as its initial time instant, which is denoted as $t_{b,i}$.

**Note:**

1) No effective processing can be performed on the information particle by the system when condition $t < t_{b,i}$ holds.

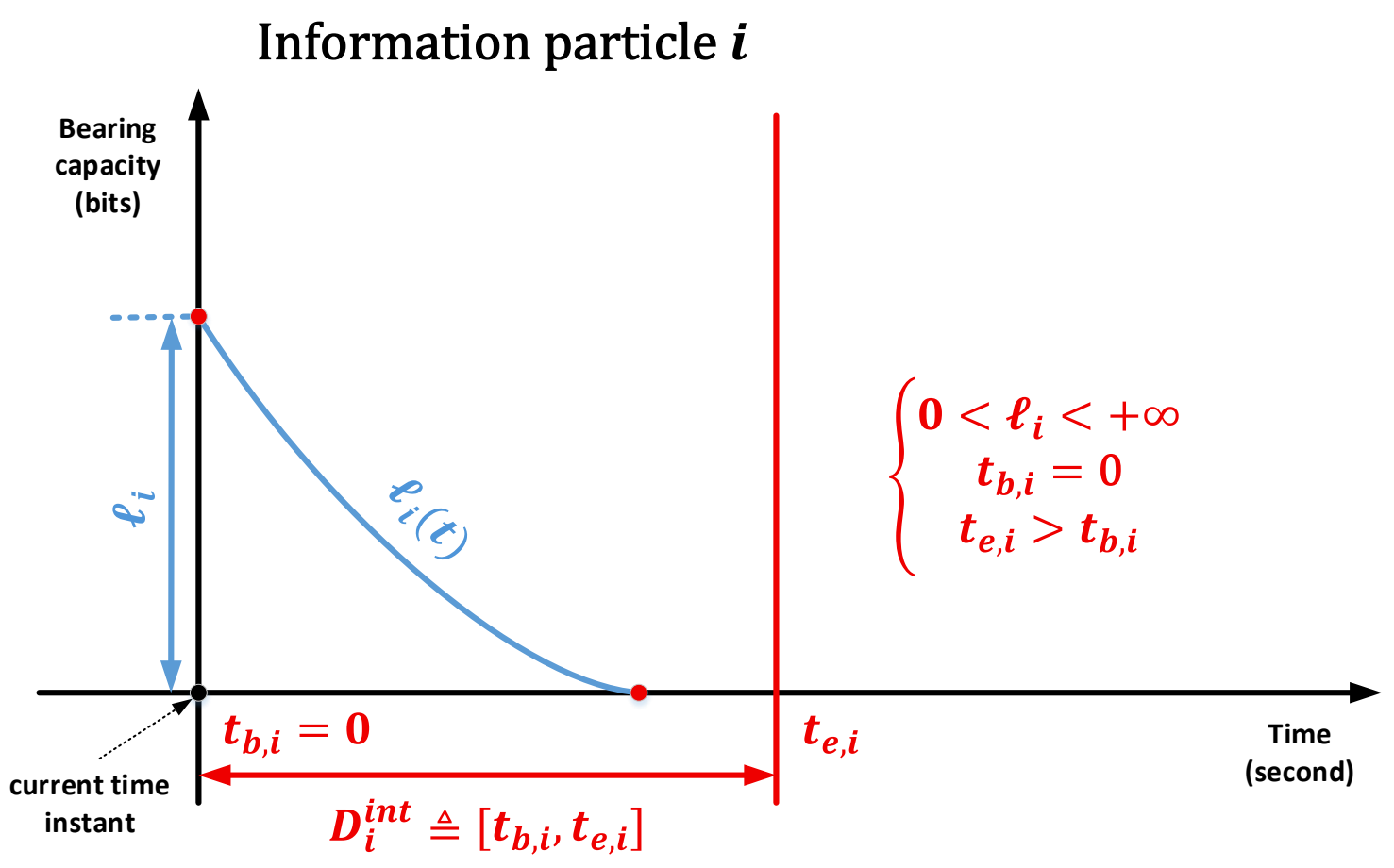

**Figure III.1** An valid information particle $i$ at the current time instant

**Definition III.5: Deadline of an information particle (unit: second)**

The latest time instant at which effective processing can be performed on an information particle $i$ by the system is defined as its deadline, which is denoted as $t_{e,i}$.

**Note:**

1) Effective processing cannot be performed on an information particle by the system when condition $t > t_{e,i}$ is satisfied;
2) As shown in **Figure III.1**, at the current time instant (i.e. $t = 0$), any valid information particle $i$ considered in this paper shall satisfy:
   a) $0 < \ell_i < +\infty$;
   b) $t_{b,i} = 0$;
   c) $t_{e,i} > t_{b,i}$;
3) It is assumed in this paper that the absolute time instant corresponding to the deadline of a valid information particle $i$ remains unchanged over time.

**Definition III.6: Effective survival time interval of an information particle**

The closed time interval $D_i^{int} \triangleq [t_{b,i}, t_{e,i}]$ is defined as the effective survival time interval of an information particle $i$.

**Note:**

1) Effective processing of an information particle can only be performed by the system within this time interval;
2) It is not excluded that effective processing cannot be performed on the information particle $i$ by the system even within its effective survival time interval (e.g., the system stays suspended during certain periods within this time interval, and so on).

**Definition III.7: Effective survival time span of an information particle (unit: second)**

The length of the time interval covered by the effective survival time interval $D_i^{int}$ of an information particle $i$ is defined as its effective survival time span, namely $L_{D_i^{int}} \triangleq t_{e,i} - t_{b,i}$.

**Definition III.8: An effective processing strategy performed by the system for an information particle**

Within the effective survival time interval $D_i^{int}$ of an information particle $i$, the time-varying process $v_i^{ES}(t)$ $((t \geq 0) \wedge (t \in D_i^{int}))$ is defined as one feasible effective processing strategy that may be performed by the system for the information particle $i$ if the following four conditions are satisfied simultaneously:

1) **Existence:** For any time instant $t$ $(t \in D_i^{int})$, there exists a real number $v_i^{ES}(t)$ corresponding to it;
2) **Non-negativity:** $v_i^{ES}(t) \geq 0$ $(t \in D_i^{int})$;
3) **Boundedness:** $v_i^{ES}(t) < +\infty$ $(t \in D_i^{int})$;
4) **Integrability:** $\int_{t \in D_i^{int}} v_i^{ES}(t)\, dt = \ell_i$.

**Note:**

1) The value $v_i^{ES}(t)$ at time instant $t$ $(t \in D_i^{int})$ denotes the instantaneous processing rate of the system for the information particle $i$ at that instant;
2) An effective processing strategy is a time-varying process composed of the instantaneous

processing rates corresponding to all time instants within time interval $D_i^{int}$;

3) For information particle $i$, the system will not perform effective processing on it outside its effective survival time interval $D_i^{int}$. Therefore, for the sake of completeness, we further define: $v_i^{ES}(t) \triangleq 0 \ ((t \geq 0) \wedge (t \notin D_i^{int}))$. In this way, the time-varying process $v_i^{ES}(t)$ is well-defined over the range of $t \geq 0$.

**Definition III.9: The universal set of effective processing strategies for an information particle**

The set $S_{v_i^{ES}}^U$ composed of all effective processing strategies of an information particle $i$ is defined as its universal set of effective processing strategies, i.e.,

$$S_{v_i^{ES}}^U \triangleq \left\{ v_i^{ES}(t) \left| \begin{cases} (v_i^{ES}(t) \in \mathbb{R}) \wedge (0 \leq v_i^{ES}(t) < +\infty) & ((t \geq 0) \wedge (t \in D_i^{int})) \\ \wedge \left(\int_{t \in D_i^{int}} v_i^{ES}(t)\, dt = \ell_i\right) & \\ v_i^{ES}(t) = 0 & ((t \geq 0) \wedge (t \notin D_i^{int})) \end{cases} \right. \right\}$$

(III.1)

**Definition III.10: Average access bandwidth requirement of an information particle (unit: bit/second)**

For information particle $i$, its average access bandwidth requirement in the system is defined as

$$B_{av,i} \triangleq \frac{\ell_i}{L_{D_i^{int}}}$$

(III.2)

**Definition III.11: Arrival status of an information particle**

For an information particle $i$, if its initial time instant is $t_{b,i} = 0$, its arrival status is regarded as “arrived”. Unless otherwise specified, this paper only considers the case that the information particle is “arrived”.

## IV. Definition of an information particle group

In this section, we define the concept of an information particle group, the incremental sorting operation performed on an information particle group, the bearing capacity of an information particle group, the effective survival time interval of an information particle group, the effective survival time span, the universal set of effective processing strategies performed on an information particle group, an effective processing strategy performed on an information particle group, the average access bandwidth requirement of an information particle group, the peak average access bandwidth requirement, the principal subgroup of an information particle group, the reachable access bandwidth of an information particle group, the minimum reachable access bandwidth of an information particle group, the arrival status of an information particle group, the access efficiency of an information particle group under a given access bandwidth, the inflection point of the required access bandwidth, the largest principal subgroup of an information particle group, a switching function of an information particle group, the vector of switching functions of an information particle subgroup, and the vector of average access bandwidth requirements of an information particle subgroup and so on.

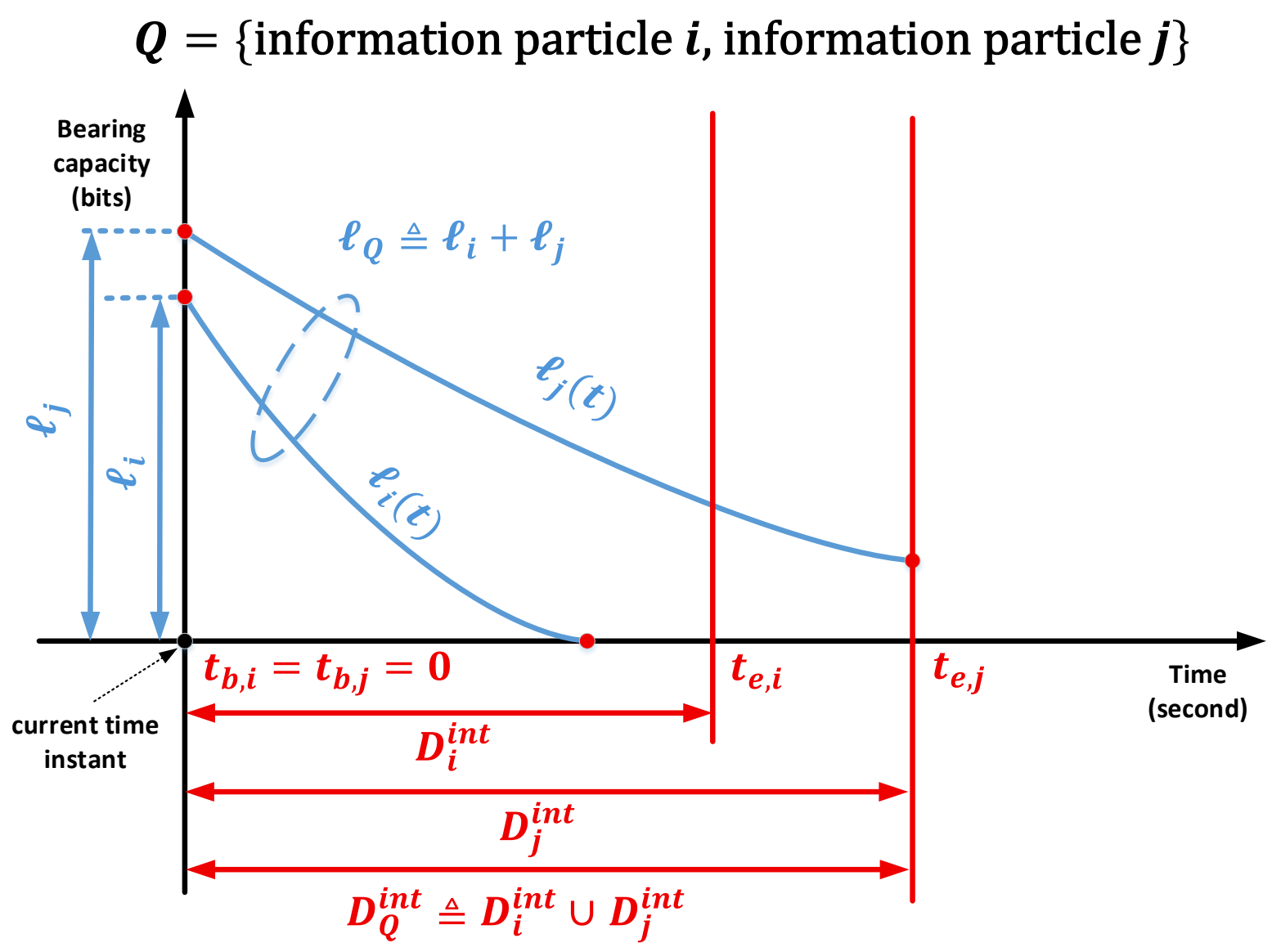

**Figure IV.1** An information particle group $Q$ composed of information particle $i$ and information particle $j$

**Definition IV.1: An information particle group**

As shown in **Figure IV.1**, a nonempty set containing $N\ (\geq 1)$ valid information particles (see the second note of **Definition III.5**) is defined as an information particle group, which is denoted as $Q$.

**Note:**

1) If an information particle $i$ belongs to an information particle group $Q$, this relationship is denoted as $i \in Q$.

**Definition IV.2: Incremental sorting operation performed on an information particle group**

Given an information particle group $Q$, and let it be composed of $N\ (\geq 1)$ valid information particles. If the information particles are sorted from small to large according to their deadlines, and the information particle ranked first are given the sequence number "1" (its deadline is $t_{e,1}$), the information particle ranked second are given the sequence number "2" (its deadline is $t_{e,2}\ (\geq t_{e,1})$),..., and the information particle ranked $N$ are given the sequence number "$N$" (its deadline is $t_{e,N}\ (\geq t_{e,N-1})$). This sorting operation for the information particles contained in $Q$ is called incremental sorting operation, which is denoted as $Op_{Inc}(Q)$.

**Note:**

1) The following convention is established hereinafter: after the incremental sorting operation $Op_{Inc}(Q)$ is performed on an information particle group $Q$, any reference to "information particle $i$" within $Q$ shall actually refer to the information particle with sequence number $i$ in $Q$ following the incremental sorting operation.

**Definition IV.3: Bearing capacity of an information particle group (unit: bit)**

As shown in **Figure IV.1**, for an information particle group $Q$, the sum of the bearing capacities of all the information particles contained therein, i.e. $\ell_Q \triangleq \sum_{i \in Q} \ell_i$, is defined as its bearing capacity.

**Note:**

1) With reference to the definition of the bearing capacity $\ell_i$ of an information particle $i$ at the

current time instant (**Definition III.3**), it can be concluded that $\ell_Q$ is essentially the bearing capacity of information particle group $Q$ at the current time instant.

**Definition IV.4: Effective survival time interval of an information particle group**

As shown in **Figure IV.1**, for an information particle group $Q$, the union of the effective survival time intervals of all the information particles contained therein, i.e. $D_Q^{int} \triangleq \bigcup_{i \in Q} D_i^{int}$, is defined as its effective survival time interval.

**Definition IV.5: Effective survival time span of an information particle group (unit: second)**

For an information particle group $Q$, the total time length $L_{D_Q^{int}}$ covered by its effective survival time interval $D_Q^{int}$ is defined as its effective survival time span, i.e. $L_{D_Q^{int}} \triangleq len(D_Q^{int})$.

**Definition IV.6: The universal set of effective processing strategies performed on an information particle group**

Given an information particle group $Q$ containing $N\ (\geq 1)$ valid information particles, and the incremental sorting operation $Op_{Inc}(Q)$ is performed thereon. Let the universal set of effective processing strategies of information particle $i\ (i \in Q)$ with sequence number $i$, which is assigned after sorting operation, be $S_{v_i^{ES}}^U$ (**Definition III.9**). The universal set $S_{Q^{ES}}^U$ of effective processing strategies performed on the information particle group $Q$ is defined as:

$$\begin{aligned} S_{Q^{ES}}^U &\triangleq S_{v_1^{ES}}^U \times \ldots \times S_{v_N^{ES}}^U \\ &= \left\{ (v_1^{ES}(t), \ldots, v_N^{ES}(t)) \middle| v_i^{ES}(t) \in S_{v_i^{ES}}^U, t \geq 0, 1 \leq i \leq N \right\} \end{aligned} \tag{IV.1}$$

**Note:**

1) It can be seen from the above definition that the set $S_{Q^{ES}}^U$ is the Cartesian product of sets $S_{v_i^{ES}}^U\ (1 \leq i \leq N)$.

**Definition IV.7: An effective processing strategy performed on an information particle group**

Given an information particle group $Q$ consisting of $N\ (\geq 1)$ valid information particles, whose universal set of effective processing strategies is $S_{Q^{ES}}^{U}$. Any element contained in $S_{Q^{ES}}^{U}$ is referred to as one effective processing strategy performed on the information particle group $Q$.

**Definition IV.8: The average access bandwidth requirement of an information particle group (unit: bits/second)**

For an information particle group $Q$, define $B_{av,Q} \triangleq \ell_Q / L_{D_Q^{int}}$ as its average access bandwidth requirement.

**Definition IV.9: The peak average access bandwidth requirement of an information particle group (unit: bits/second)**

The maximum value of the average access bandwidth requirements of all the nonempty subsets (including $Q$ itself) contained in an information particle group $Q$ is defined as the peak average access bandwidth requirement of $Q$, which is denoted as $B_{av,Q}^{pk}$.

**Note:**

1) There may be more than one information particle subgroup achieving the peak average access bandwidth requirement.

**Definition IV.10: A principal subgroup of an information particle group**

If the average access bandwidth requirement of an information particle subgroup $Q'\ (Q' \subseteq Q)$ is $B_{av,Q'} = B_{av,Q}^{pk}$, then $Q'$ is a principal information particle subgroup of $Q$.

**Note:**

1) There may be more than one information particle subgroup in $Q$.

**Definition IV.11: A reachable access bandwidth of an information particle group (unit: bits/second)**

Given an information particle group $Q$ containing $N\ (\geq 1)$ valid information particles, and the incremental sorting operation $Op_{Inc}(Q)$ is performed thereon. Within the effective survival time interval $D_Q^{int}$ of $Q$, if there exist at least one valid processing strategy $s_Q = (v_1^{ES}(t), \dots, v_N^{ES}(t))\ (s_Q \in S_{Q^{ES}}^U)$ for $Q$, so that the relation $\sum_{i \in Q} v_i^{ES}(t) \leq B_{re,Q}\ (B_{re,Q} \in \mathbb{R}, 0 < B_{re,Q} < +\infty, t \in D_Q^{int})$ holds, it is said that the access bandwidth $B_{re,Q}$ is reachable for $Q$ (i.e., $B_{re,Q}$ is a reachable access bandwidth of $Q$).

**Note:**

1) In this paper, only valid information particles (the note section of **Definition III.5**) are considered. Therefore, it is not difficult to prove that the reachable access bandwidth $B_{re,Q} > 0$.

**Definition IV.12: The universal set of reachable access bandwidths and the minimum reachable access bandwidth for an information particle group**

The set consisting of all reachable access bandwidths of the information particle group $Q$ over its effective survival time interval $D_Q^{int}$ is defined as the universal set of reachable access bandwidths for $Q$, which is denoted as $S_{B_{re,Q}}^U$. Furthermore, define the minimum among the elements contained in the set $S_{B_{re,Q}}^U$ as the minimum reachable access bandwidth of the information particle group $Q$, denoted by $B_{re,Q}^{min}$.

**Note:**

1) In **Theorem V.1**, we shall prove that set $S_{B_{re,Q}}^U$ indeed possesses a minimum among its elements. Hence, the above definition of the minimum access bandwidth is rational.

**Definition IV.13: The arrival status of an information particle group**

For an information particle group $Q$, if the arrival statuses of all the information particles contained in it are “arrived”, the arrival status of $Q$ is defined to be “arrived”. Unless otherwise specified, this paper only considers the case that the information particle group is “arrived”.

**Definition IV.14: The access efficiency of an information particle group under a given reachable access bandwidth and within its effective survival time interval**

For an information particle group $Q$, given its effective survival time interval $D_Q^{int}$ and a reachable access bandwidth $B_{re,Q}$ ($\in S_{B_{re,Q}}^U$), the corresponding access efficiency is defined as:

$$\eta_{Q|B_{re,Q}} \triangleq \frac{\ell_Q}{\left(B_{re,Q} \cdot L_{D_Q^{int}}\right)}$$

(IV.2)

Combined with **Definition IV.8**, it can be further obtained that:

$$\eta_{Q|B_{re,Q}} = \frac{B_{av,Q}}{B_{re,Q}}$$

(IV.3)

**Definition IV.15: The inflection point of the required access bandwidth of an information particle group**

Given an information particle group $Q$, an incremental sorting operation $Op_{Inc}(Q)$ is performed on it. If the information particle with the sorting sequence number of $N_{cr,Q}$ $(\geq 1)$ is the one with the largest sequence number among all the information particles contained in all the principal information particle subgroups in $Q$, then $N_{cr,Q}$ is defined as the inflection point of the required access bandwidth of $Q$.

**Note:**

1) The physical meaning of $N_{cr,Q}$ will be clarified in Section VII.

**Definition IV.16: The largest principal subgroup of an information particle group**

Given an information particle group $Q$, an incremental sorting operation $Op_{Inc}(Q)$ is performed on it. Define the principal information particle subgroup containing the information particle with the sequence number of $N_{cr,Q}$ as the largest principal subgroup of $Q$, which is denoted as $Q_{cr}^{r*}$.

**Definition IV.17: An information particle subgroup $Q_i$ of an information particle group $Q$**

Given an information particle group $Q$ composed of $N\ (\geq 1)$ information particles, and an incremental sorting operation $Op_{Inc}(Q)$ is performed on it. $Q_i\ (Q_i \subseteq Q)$ is defined as the information particle subgroup composed of all the information particles whose sorting sequence numbers are not greater than $i\ (1 \leq i \leq N)$.

**Definition IV.18: An information particle group with a specific index number**

$Q^i$ denotes an information particle group with an index number $i\ (\in \mathbb{Z}^+)$.

**Definition IV.19: A switching function of an information particle group $\boldsymbol{Q}$ with a sequence number $\boldsymbol{i}$**

Given an information particle group $Q$ composed of $N\ (N \geq 2)$ information particles, and an incremental sorting operation $Op_{Inc}(Q)$ is performed on it. A switching function $\delta_{i,Q}$ of an information particle group $Q$ with a sequence number $i\ (1 \leq i \leq N-1)$ is defined as:

$$\delta_{i,Q} \triangleq \begin{cases} 1 & B_{re,Q_i}^{min} \geq B_{av,Q_{i+1}} \\ 0 & B_{re,Q_i}^{min} < B_{av,Q_{i+1}} \end{cases} \tag{IV.4}$$

**Note:**

1) According to **Definition IV.17** and **Definition IV.19**, we know that $\delta_{i,Q_{i+1}} = \delta_{i,Q_{i+2}} = \ldots = \delta_{i,Q_{N-1}} = \delta_{i,Q_N} = \delta_{i,Q}$, where $Q_{i+1} \subseteq Q_{i+2} \subseteq \ldots \subseteq Q_{N-1} \subseteq Q_N = Q$.

**Definition IV.20: The vector of switching functions of an information particle subgroup $\boldsymbol{Q_n}\ (\boldsymbol{Q_n} \subseteq \boldsymbol{Q})$**

Given an information particle group $Q$ composed of $N\ (N \geq 2)$ information particles, and an incremental sorting operation $Op_{Inc}(Q)$ is performed on it. Let $Q_n\ (Q_n \subseteq Q)$ denote the information particle subgroup composed of all the information particles whose sorting sequence numbers are not greater than $n\ (2 \leq n \leq N)$. The vector $\boldsymbol{\delta}_{Q_n}$ of switching functions of $Q_n$ is defined as a $n \times 1$ column vector. That is,

$$\boldsymbol{\delta}_{Q_n}{}^T \triangleq \begin{cases} \left(\delta_{n-1,Q_n}, 1-\delta_{n-1,Q_n}\right)_{1\times n} & (n=2) \\ \begin{pmatrix} \prod_{j=1}^{n-1}\delta_{j,Q_n}, \left(1-\delta_{1,Q_n}\right)\prod_{j=2}^{n-1}\delta_{j,Q_n}, \\ \dots, \left(1-\delta_{i-1,Q_n}\right)\prod_{j=i}^{n-1}\delta_{j,Q_n}, \dots, \\ \left(1-\delta_{n-2,Q_n}\right)\prod_{j=n-1}^{n-1}\delta_{j,Q_n}, 1-\delta_{n-1,Q_n} \end{pmatrix}_{1\times n} & (2 < n \le N) \end{cases}$$

(IV.5)

**Definition IV.21： The vector of average access bandwidth requirements of an information particle subgroup $\boldsymbol{Q_n}$ $(\boldsymbol{Q_n} \subseteq \boldsymbol{Q})$**

Given an information particle group $Q$ composed of $N$ $(N \ge 2)$ information particles, and an incremental sorting operation $Op_{Inc}(Q)$ is performed on it. Let $Q_n$ $(Q_n \subseteq Q)$ denote the information particle subgroup composed of all the information particles whose sorting sequence numbers are not greater than $n$ $(2 \le n \le N)$. The vector $\boldsymbol{B}_{av,Q_n}$ of average access bandwidth requirements of $Q_n$ is defined as a $n \times 1$ column vector. That is,

$$\boldsymbol{B}_{av,Q_n}{}^T \triangleq \left(B_{av,Q_1}, B_{av,Q_2}, \dots, B_{av,Q_{n-1}}, B_{av,Q_n}\right)_{1\times n}$$

(IV.6)

## V. The relationships between the average access bandwidth requirement, the peak average access bandwidth requirement, and the minimum reachable access bandwidth of an information particle group

In this section, the relationship between the reachable access bandwidth of an information particle group and the average access bandwidth requirements of its subgroups is proved (**Lemma V.1.1**); It is proved that the access efficiency of an information particle group is not greater than one under a given reachable access bandwidth and within its effective survival time interval (**Corollary V.1.1.1**); The relationship between the reachable access bandwidth of an information particle group and its peak average access bandwidth requirement is proved (**Corollary V.1.1.2**); The necessary and sufficient conditions for an access bandwidth to be reachable are given (***Lemma V.1.2***); The reachability of the peak average access bandwidth requirement is proved (**Corollary V.1.2.1**); The theorem on the existence and reachability of the lower bound for reachable access bandwidth of an information particle group (***Theorem V.1***) and the range of reachable access bandwidth for an information particle group (***Theorem V.2***) are proved.

**Lemma V.1.1: Given an information particle group $Q$, let any of its reachable access bandwidth be $B_{re,Q}$ ($\in S^U_{B_{re,Q}}$). If the average access bandwidth requirement of any non-empty subset $Q'$ (including $Q$ itself) contained in $Q$ is $B_{av,Q'}$, there is $B_{re,Q} \geq B_{av,Q'}$.**

**Proof:**

First, the incremental sorting operation $Op_{Inc}(Q)$ (**Definition IV.2**) is performed on the information particle group $Q$, which contains $N\ (\geq 1)$ valid information particles, and each information particle contained in $Q$ is assigned a corresponding sorting sequence number accordingly.

Let the universal set of effective processing strategies performed on $Q$ be $S^U_{Q^{ES}}$ (**Definition IV.6**). And, let $B_{re,Q}$ be any reachable access bandwidth of $Q$. According to the definition of a reachable access bandwidth (**Definition IV.11**), there exists at least one effective processing strategy $s_Q = (v_1^{ES}(t), \ldots, v_N^{ES}(t))$ ($s_Q \in S^U_{Q^{ES}}$) for $Q$ within $S^U_{Q^{ES}}$ such that the following inequality relation holds

$$B_{re,Q} \geq \sum_{i \in Q} v_i^{ES}(t) \ \ (t \in D_Q^{int})$$

(V.1)

Furthermore, let $Q' \subseteq Q$ ($Q' \neq \emptyset$), then we have

$$B_{re,Q} \geq \sum_{i \in Q} v_i^{ES}(t) \geq \sum_{i \in Q'} v_i^{ES}(t) \ \ (t \in D_{Q'}^{int}, D_{Q'}^{int} \subseteq D_Q^{int})$$

(V.2)

Integrate over the above inequality within the effective survival time interval $D_{Q'}^{int}$, it can be obtained that

$$\int_{t \in D_{Q'}^{int}} B_{re,Q} dt \geq \int_{t \in D_{Q'}^{int}} \left( \sum_{i \in Q'} v_i^{ES}(t) \right) dt$$

(V.3)

According to the definition of the bearing capacity $\ell_{Q'}$ of an information particle subgroup $Q'$ (**Definition IV.3**), there is

$$\int_{t \in D_{Q'}^{int}} \left( \sum_{i \in Q'} v_i^{ES}(t) \right) dt = \ell_{Q'}$$

(V.4)

Substitute the above relationship into **Equation V.3**, and combined with **Definition IV.5** and **Definition IV.8**, we get

$$B_{re,Q} \geq \frac{\ell_{Q'}}{L_{D_{Q'}^{int}}} = B_{av,Q'}$$

(V.5)

■

**Corollary V.1.1.1: The access efficiency $\eta_{Q|B_{re,Q}}$ of an information particle group $Q$ under a given reachable access bandwidth $B_{re,Q}$ and within its effective survival time interval $D_Q^{int}$ is not greater than one, i.e. $\eta_{Q|B_{re,Q}} \leq 1$.**

**Proof:**

Let the information particle subgroup $Q' = Q$, we can have $B_{re,Q} \geq B_{av,Q}$ from **Lemma V.1.1**. Furthermore, in combination with **Definition IV.14**, there is $\eta_{Q|B_{re,Q}} = \frac{B_{av,Q}}{B_{re,Q}} \leq 1$. ■

**Corollary V.1.1.2: Given an information particle group $\boldsymbol{Q}$, $\boldsymbol{B_{re,Q}}$ ($\in \boldsymbol{S^U_{B_{re,Q}}}$) is any of its reachable access bandwidth. Let its peak average access bandwidth requirement is $\boldsymbol{B^{pk}_{av,Q}}$, there is $\boldsymbol{B_{re,Q} \geq B^{pk}_{av,Q}}$.**

**Proof:**

This can be directly proved from **Lemma V.1.1** and the definition of the peak average access bandwidth requirement for an information particle group (**Definition IV.9**). ■

**Lemma V.1.2: Necessary and sufficient conditions for reachable access bandwidths.**

**Given an information particle group $\boldsymbol{Q}$ (composed of $\boldsymbol{N\,(\geq 1)}$ information particles). An incremental sorting operation $\boldsymbol{Op_{Inc}(Q)}$ is performed on $\boldsymbol{Q}$ (Definition IV.2). Let $\boldsymbol{Q_i\ (Q_i \subseteq Q)}$ be the information particle subgroup composed of all the information particles in $\boldsymbol{Q}$ whose sorting sequence numbers are not greater than $\boldsymbol{i\ (1 \leq i \leq N)}$. The bearing capacity and the effective survival time span of $\boldsymbol{Q_i}$ are $\boldsymbol{\ell_{Q_i}}$ and $\boldsymbol{L_{D^{int}_{Q_i}}}$, respectively. The sufficient and necessary conditions for an access bandwidth $\boldsymbol{B}$ to be reachable for $\boldsymbol{Q}$ is that**

$$\frac{\boldsymbol{\ell_{Q_i}}}{\boldsymbol{B}} \leq \boldsymbol{L_{D^{int}_{Q_i}}} \tag{V.6}$$

**holds for all $\boldsymbol{Q_i\ (Q_i \subseteq Q, 1 \leq i \leq N)}$.**

**Proof:**

First, we prove that if the access bandwidth $B$ is reachable, then $\frac{\ell_{Q_i}}{B} \leq L_{D^{int}_{Q_i}}$ is true for all $Q_i\ (Q_i \subseteq Q, 1 \leq i \leq N)$. Let the universal set of effective processing strategies performed on $Q$ be $S^U_{Q^{ES}}$ (**Definition IV.6**). According to the definition of the reachable access bandwidth of an information particle group $Q$ (**Definition IV.11**), there exists at least one effective processing strategy $s_Q = (v^{ES}_1(t), \dots, v^{ES}_N(t))\ (s_Q \in S^U_{Q^{ES}})$ for $Q$ within $S^U_{Q^{ES}}$ such that the following inequality relation holds:

$$\begin{aligned}\int_{t\in D_{Q_i}^{int}} B dt &\geq \int_{t\in D_{Q_i}^{int}} \left(\sum_{j=1}^{N} v_j^{ES}(t)\right) dt \\ &\geq \int_{t\in D_{Q_i}^{int}} \left(\sum_{j=1}^{i} v_j^{ES}(t)\right) dt \\ &= \ell_{Q_i}\ (1 \leq i \leq N)\end{aligned}$$

(V.7)

From the above equation, it can be obtained that

$$B \cdot L_{D_{Q_i}^{int}} \geq \ell_{Q_i}\ (1 \leq i \leq N)$$

(V.8)

Therefore, the following equation holds for all the information particle subgroups $Q_i$ $(1 \leq i \leq N)$

$$\frac{\ell_{Q_i}}{B} \leq L_{D_{Q_i}^{int}}$$

(V.9)

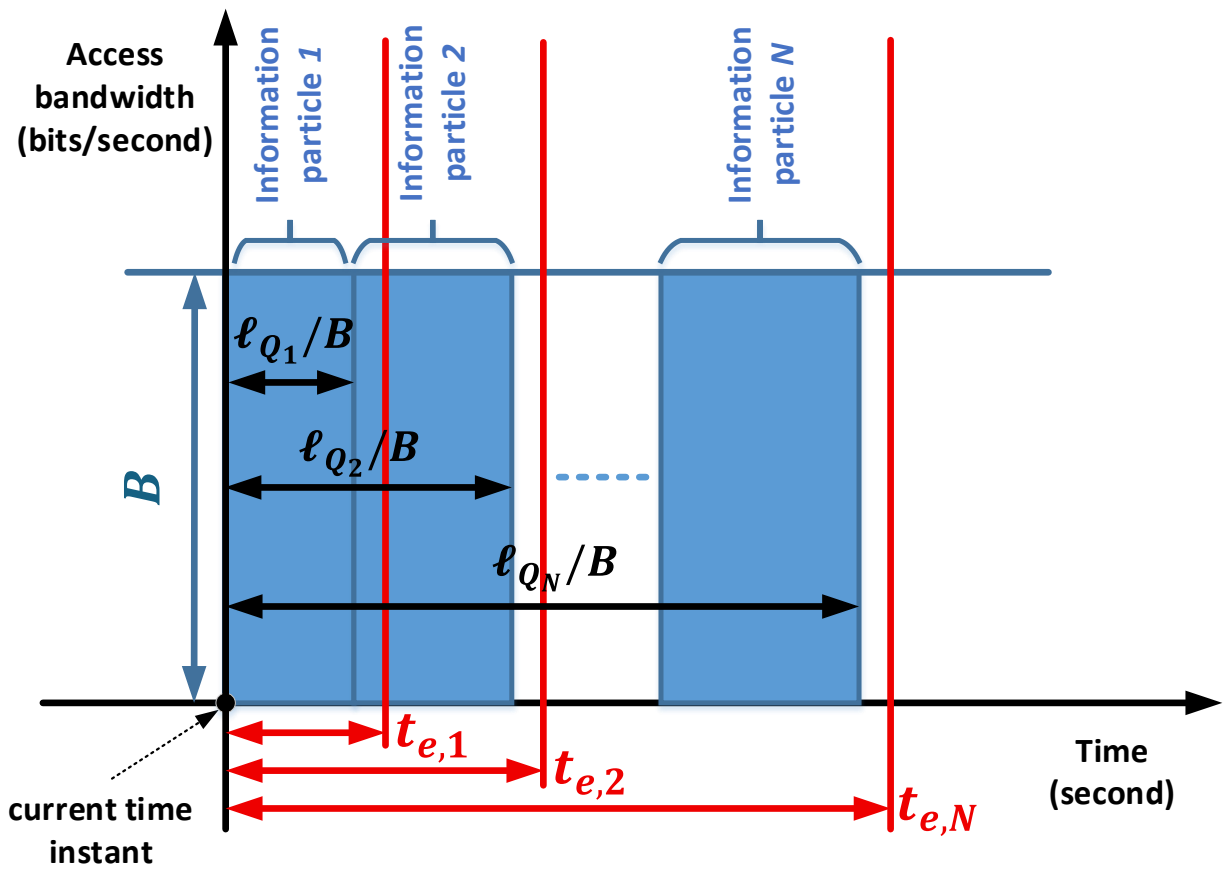

**Figure V.1** The reachable access bandwidth $B$ corresponding to processing the information particle group $Q$ with the EDF strategy

Secondly, we prove that if $\frac{\ell_{Q_i}}{B} \leq L_{D_{Q_i}^{int}}$ is true for all $Q_i$ $(Q_i \subseteq Q, 1 \leq i \leq N)$, then the access bandwidth $B$ is reachable. As shown in **Figure V.1**, if $\frac{\ell_{Q_i}}{B} \leq L_{D_{Q_i}^{int}}$ is true for all $Q_i$ $(1 \leq i \leq N)$, it means that for all the information particles $i$ $(1 \leq i \leq N)$ in $Q$, if EDF (Early Deadline First) processing strategy is adopted and each particle in $Q$ is processed one by one at the instantaneous processing rate $B$, the process of each information particle can be completed before its deadline, i.e. $t_{e,i}$ $(1 \le i \le N)$. In other words, with the access bandwidth $B$, it is found that at least EDF processing strategy is effective for the process of the information particle group $Q$. Therefore, $B$ is a reachable access bandwidth for $Q$. ■

**Note:**

1) By slightly deforming the conclusion of **Lemma V.1.2**, it can be seen that the sufficient and necessary conditions for an access bandwidth $B$ to be reachable for $Q$ is that $\frac{\ell_{Q_i}}{L_{D_{Q_i}^{int}}} \le B$ holds for all information particle subgroups $Q_i$ $(Q_i \subseteq Q, 1 \le i \le N)$. According to the definition of the average access bandwidth requirement of an information particle group (**Definition IV.8**), it can be known that **Lemma V.1.2** actually points out that the sufficient and necessary conditions for an access bandwidth $B$ to be reachable is that $B_{av,Q_i} \triangleq \frac{\ell_{Q_i}}{L_{D_{Q_i}^{int}}} \le B$ holds for all the information particle subgroups $Q_i$ $(Q_i \subseteq Q, 1 \le i \le N)$.

**Corollary V.1.2.1: Given an information particle group $Q$, the access bandwidth $B_{av,Q}^{pk}$ is reachable, i.e. $B_{av,Q}^{pk} \in S_{B_{re,Q}}^{U}$.**

**Proof:**

Given an information particle group $Q$, consider that it is composed of $N$ $(\ge 1)$ information particles. Let $Q_i$ $(Q_i \subseteq Q)$ indicate the information particle subgroup composed of all the information particles in the information particle group $Q$ whose sorting sequence numbers are not greater than $i$ $(1 \le i \le N)$ after the operation $Op_{Inc}(Q)$. According to the definition of the peak average access bandwidth requirement $B_{av,Q}^{pk}$ of an information particle group $Q$ (**Definition IV.9**), there is

$$B_{av,Q}^{pk} \ge B_{av,Q_i} = \frac{\ell_{Q_i}}{L_{D_{Q_i}^{int}}} \ (1 \le i \le N) \tag{V.10}$$

Therefore, the following formula holds for all $Q_i$ $(1 \le i \le N)$

$$\frac{\ell_{Q_i}}{B_{av,Q}^{pk}} \le L_{D_{Q_i}^{int}} \tag{V.11}$$

According to **Lemma V.1.2**, $B_{av,Q}^{pk}$ is reachable. ■

**Theorem V.1: Existence and reachability theorem for the lower bound of reachable access bandwidth of an information particle group.**

**Given an information particle group $\boldsymbol{Q}$, its peak average access bandwidth requirement is $\boldsymbol{B_{av,Q}^{pk}}$. Then $\boldsymbol{B_{av,Q}^{pk}}$ is the minimum among the elements contained in the universal set $\boldsymbol{S_{B_{re,Q}}^{U}}$ of reachable access bandwidths for the information particle group $\boldsymbol{Q}$, i.e. $\boldsymbol{B_{re,Q}^{min} = B_{av,Q}^{pk}}$.**

**Proof:**

This follows directly from **Corollary V.1.1.2**, **Corollary V.1.2.1** and the definition of the minimum reachable access bandwidth (**Definition IV.12**). ■

**Note：**

1) **Theorem V.1** states that $B_{av,Q}^{pk}$ is a lower bound for all reachable access bandwidths of the information particle group $Q$. Moreover, since $B_{av,Q}^{pk}$ itself belongs to set $S_{B_{re,Q}}^{U}$, this lower bound is necessarily reachable;
2) The above theorem provides the basic method to determine the minimum reachable access bandwidth of an information particle group. That is, the minimum reachable access bandwidth of an information particle group can be obtained by determining the peak average access bandwidth requirement of the information particle group;
3) According to the definition of the peak average access bandwidth requirement of an information particle group (**Definition IV.9**), it is known that it will not vary with the reassignment of the index numbers of information particles contained in the corresponding information particle group (such as performing incremental sorting operation $Op_{Inc}(Q)$ etc.). Since the minimum reachable access bandwidth of an information particle group is equal to its peak average access bandwidth requirement, the minimum reachable access bandwidth will not vary with the reassignment of the index numbers of information particles contained in the information particle group.

**Corollary V.1.1: Given an information particle group $\boldsymbol{Q}$, and its minimum reachable access bandwidth is $\boldsymbol{B_{re,Q}^{min}}$, then any access bandwidth $\boldsymbol{B}$ $(\boldsymbol{B \geq B_{re,Q}^{min}})$ is reachable.**

**Proof:**

It follows from **Theorem V.1** that the access bandwidth $B_{re,Q}^{min}\,(= B_{av,Q}^{pk})$ is reachable. According to **Lemma V.1.2**, since access bandwidth $B_{re,Q}^{min}$ is reachable, all the information particle subgroups $Q_i$ $(Q_i \subseteq Q, 1 \leq i \leq N)$ of $Q$ should satisfy the relationship

$$B_{av,Q_i} = \frac{\ell_{Q_i}}{L_{D_{Q_i}^{int}}} \leq B_{re,Q}^{min}$$

(V.12)

Furthermore, for any access bandwidth $B$ $(B > B_{re,Q}^{min})$, all the information particle subgroups $Q_i$ $(Q_i \subseteq Q, 1 \leq i \leq N)$ of $Q$ should satisfy

$$B_{av,Q_i} = \frac{\ell_{Q_i}}{L_{D_{Q_i}^{int}}} \leq B_{re,Q}^{min} < B$$

(V.13)

Therefore, once again, according to **Lemma V.1.2**, the above considered access bandwidth $B$ $(B > B_{re,Q}^{min})$ is reachable. ■

**Theorem V.2: Range of reachable access bandwidths for an information particle group.**

**Given an information particle group $\boldsymbol{Q}$ with minimum reachable access bandwidth $\boldsymbol{B_{re,Q}^{min}}$, the universal set of reachable access bandwidths for the information particle group $\boldsymbol{Q}$ is $\boldsymbol{S_{B_{re,Q}}^{U} = [B_{re,Q}^{min}, +\infty)}$.**

**Proof:**

For the access bandwidth of $B \geq B_{re,Q}^{min}$, it is reachable according to **Corollary V.1.1**, i.e., $B \in S_{B_{re,Q}}^{U}$.

For the access bandwidth of $B < B_{re,Q}^{min}$, we adopt proof by contradiction. Assume that $B$ is reachable. Then from **Corollary V.1.1.2** and **Theorem V.1**, we have $B \geq B_{av,Q}^{pk} = B_{re,Q}^{min}$. This contradicts the assumption.

From the above arguments, we obtain $S^{U}_{B_{re,Q}} = [B^{min}_{re,Q}, +\infty)$. ■

## VI. The influences of the time attribute and the attribute of bearing capacity of an information particle group on the minimum reachable access bandwidth

In this section, it is proved that with the increase of the deadline of one information particle in a given information particle group, the average access bandwidth requirement (**Theorem VI.1**) and the minimum reachable access bandwidth (**Corollary VI.1.1**) of the information particle group will not increase (or even decrease); It is proved that (**Theorem VI.2**) after adding a new information particle into a given information particle group, the minimum reachable access bandwidth of the information particle group will not decrease (or even increase); It is proved that with the decrease of the bearing capacity of an information particle in a given information particle group, the average access bandwidth requirement (**Theorem VI.4**) and the minimum reachable access bandwidth (**Corollary VI.4.1**) of the information particle group will not increase (or even decrease); It is proved that for a given information particle group that meets a certain bandwidth constraint (i.e., the given access bandwidth is greater than the minimum reachable access bandwidth of the information particle group), it can still meet the bandwidth constraint by adjusting the deadline attribute (***Theorem VI.3***, i.e. regulation theorem for deadline of an information particle under bandwidth constraint) or the bearing capacity attribute (***Theorem VI.5***, i.e. regulation theorem for bearing capacity of an information particle under bandwidth constraint) of a newly arrived information particle; The equal scaling theorem for information particle groups (***Theorem VI.6***) is proved.

**Theorem VI.1: Given an information particle group $\boldsymbol{Q}$ composed of $\boldsymbol{N}\ (\geq \mathbf{1})$ information particles. Select any of an information particle $\boldsymbol{i}\ (\boldsymbol{i} \in \boldsymbol{Q})$ in $\boldsymbol{Q}$ and increase its deadline from $\boldsymbol{t_{e,i}}$ to $\boldsymbol{t'_{e,i}}$ (i.e., $\boldsymbol{t'_{e,i} > t_{e,i}}$). If the average access bandwidth requirement of $\boldsymbol{Q}$ varies accordingly from $\boldsymbol{B_{av,Q}}$ to $\boldsymbol{B'_{av,Q}}$ with the variation of the deadline of $\boldsymbol{i}$, there is $\boldsymbol{B_{av,Q} \geq B'_{av,Q}}$.**

**Proof:**

Suppose that as the deadline of the information particle $i$ increases from $t_{e,i}$ to $t'_{e,i}$, the effective survival time span of the information particle group $Q$ varies from $L_{D_Q^{int}}$ to $L'_{D_Q^{int}}$. It can be obtained from the definition of the average access bandwidth requirement (**Definition IV.8**) that

$$B_{av,Q} = \frac{\ell_Q}{L_{D_Q^{int}}}$$

(VI.1)

$$B'_{av,Q} = \frac{\ell_Q}{L'_{D_Q^{int}}}$$

(VI.2)

With the deadline of the information particle $i$ ($i \in Q$) increasing from $t_{e,i}$ to $t'_{e,i}$, there is

$$L'_{D_Q^{int}} \geq L_{D_Q^{int}}$$

(VI.3)

Based on the above three equations, we have $B_{av,Q} \geq B'_{av,Q}$. ■

**Corollary VI.1.1: Given an information particle group $Q$ composed of $N\,(\geq 1)$ information particles. Select any of an information particle $i$ ($i \in Q$) in $Q$ and increase its deadline from $t_{e,i}$ to $t'_{e,i}$ (i.e., $t'_{e,i} > t_{e,i}$). If the minimum reachable access bandwidth of $Q$ varies accordingly from $B_{re,Q}^{min}$ to $B_{re,Q}^{min}{}'$ with the variation of the deadline of $i$, there is $B_{re,Q}^{min} \geq B_{re,Q}^{min}{}'$.**

**Proof:**

As the deadline of the information particle $i$ increases from $t_{e,i}$ to $t'_{e,i}$, the peak average access bandwidth requirement of $Q$ changes from $B_{av,Q}^{pk}$ to $B_{av,Q}^{pk}{}'$. The conclusion in **Theorem VI.1** is also applicable to any of a nonempty subset $Q'$ (including $Q$ itself) of $Q$. That is, as the deadline of the information particle $i$ increases from $t_{e,i}$ to $t'_{e,i}$, the average access bandwidth requirement of $Q'$ will not increase. Therefore, combined with the definition of the peak average access bandwidth requirement (**Definition IV.9**), there is

$$B_{av,Q}^{pk}{}' \leq B_{av,Q}^{pk}$$

(VI.4)

According to **Theorem V.1**, we have

$$B_{re,Q}^{min} = B_{av,Q}^{pk}$$

(VI.5)

$$B_{re,Q}^{min}{}' = B_{av,Q}^{pk}{}'$$

(VI.6)

Based on the above three equations, we have $B_{re,Q}^{min} \geq B_{re,Q}^{min}{}'$. ■

**Theorem VI.2: Given an information particle group $\boldsymbol{Q}$ composed of $\boldsymbol{N}\ (\geq \mathbf{1})$ information particles. Let its minimum reachable access bandwidth be $\boldsymbol{B_{re,Q}^{min}}$. A new information particle $\boldsymbol{i}\ (\boldsymbol{i} \notin \boldsymbol{Q})$ is added into $\boldsymbol{Q}$. Therefore, a new information particle group $\boldsymbol{Q'}\ (\boldsymbol{Q'} = \{\boldsymbol{i}\} \cup \boldsymbol{Q})$ is formed. Assume that the minimum reachable access bandwidth of $\boldsymbol{Q'}$ is $\boldsymbol{B_{re,Q'}^{min}}$, there is $\boldsymbol{B_{re,Q'}^{min} \geq B_{re,Q}^{min}}$.**

**Proof:**

Based on the definition of the peak average access bandwidth requirement (**Definition IV.9**), there is

$$B_{av,Q'}^{pk} \geq B_{av,Q}^{pk}$$

(VI.7)

According to **Theorem V.1**, we have

$$B_{re,Q}^{min} = B_{av,Q}^{pk}$$

(VI.8)

$$B_{re,Q'}^{min} = B_{av,Q'}^{pk}$$

(VI.9)

Based on the above three equations, we have $B_{re,Q'}^{min} \geq B_{re,Q}^{min}$. ■

**Lemma VI.3.1: Given an information particle group $\boldsymbol{Q}$ composed of $\boldsymbol{N}\ (\geq \mathbf{1})$ information particles. A new information particle $\boldsymbol{i}\ (\boldsymbol{i} \notin \boldsymbol{Q})$ is added into $\boldsymbol{Q}$, and a new information particle group $\boldsymbol{Q'}\ (\boldsymbol{Q'} = \{\boldsymbol{i}\} \cup \boldsymbol{Q})$ is formed. It is assumed that the peak average access bandwidth requirements of $\boldsymbol{Q}$ and $\boldsymbol{Q}'$ are $\boldsymbol{B_{av,Q}^{pk}}$ and $\boldsymbol{B_{av,Q'}^{pk}}$, respectively. The deadline and the bearing capacity of the newly added information particle $\boldsymbol{i}$ are $\boldsymbol{t_{e,i}}$ and $\boldsymbol{\ell_i}$, respectively. With the exception of $\boldsymbol{t_{e,i}}$, all attribute parameters for member information particles in $\boldsymbol{Q}'$ remain unchanged. Then, the following relation holds**

$$\lim_{t_{e,i} \to +\infty} \boldsymbol{B_{av,Q'}^{pk}} = \boldsymbol{B_{av,Q}^{pk}}$$

**Proof:**

Let there be $M\ (\geq 1)$ nonempty information particle subgroups contained in the information particle group $Q$, which are denoted by $Q^1, \dots, Q^M$, respectively. Accordingly, the information particle subgroup $Q^{j+M} \triangleq \{i\} \cup Q^j\ (1 \leq j \leq M)$ denotes the information particle subgroup formed by adding the information particle $i$ to the information particle subgroup $Q^j$. In particular, let the information particle subgroup $Q^{2M+1} \triangleq \{i\}$ denote the information particle subgroup containing only the information particle $i$. As a result, there exist $2M+1$ nonempty information particle subgroups contained in the information particle group $Q'$, which are denoted by $Q^1, \dots, Q^M, Q^{1+M}, \dots, Q^{2M}, Q^{2M+1}$, respectively. According to the definition of the peak average access bandwidth requirement for an information particle group (**Definition IV.9**), we can proceed with the following derivation:

$$
\begin{aligned}
\lim_{t_{e,i}\to+\infty} B_{av,Q'}^{pk} &= \lim_{t_{e,i}\to+\infty}\left(max\left\{\underset{j=1,\dots,M}{max}\left\{B_{av,Q^j}\right\}, \underset{j=1,\dots,M}{max}\left\{B_{av,Q^{j+M}}\right\}, B_{av,Q^{2M+1}}\right\}\right) \\
&= \lim_{t_{e,i}\to+\infty}\left(max\left\{\underset{j=1,\dots,M}{max}\left\{\left(B_{av,Q^j}, B_{av,Q^{j+M}}\right)\right\}, B_{av,Q^{2M+1}}\right\}\right) \\
&= max\left\{\underset{j=1,\dots,M}{max}\left\{\left(B_{av,Q^j}, \lim_{t_{e,i}\to+\infty}\left(B_{av,Q^{j+M}}\right)\right)\right\}, \lim_{t_{e,i}\to+\infty}\left(B_{av,Q^{2M+1}}\right)\right\} \\
&= max\left\{\underset{j=1,\dots,M}{max}\left\{\left(B_{av,Q^j}, \lim_{t_{e,i}\to+\infty}\left(\frac{\ell_{Q^{j+M}}}{L_{D_{Q^{j+M}}^{int}}}\right)\right)\right\}, \lim_{t_{e,i}\to+\infty}\left(\frac{\ell_{Q^{2M+1}}}{L_{D_{Q^{2M+1}}^{int}}}\right)\right\} \\
&= max\left\{\underset{j=1,\dots,M}{max}\left\{\left(B_{av,Q^j}, \lim_{t_{e,i}\to+\infty}\left(\frac{\ell_{Q^j}+\ell_i}{max\left(L_{D_{Q^j}^{int}}, t_{e,i}\right)}\right)\right)\right\}, \lim_{t_{e,i}\to+\infty}\left(\frac{\ell_i}{t_{e,i}}\right)\right\} \\
&= max\left\{\underset{j=1,\dots,M}{max}\left\{\left(B_{av,Q^j}, 0\right)\right\}, 0\right\} \\
&= \underset{j=1,\dots,M}{max}\left\{B_{av,Q^j}\right\} = B_{av,Q}^{pk}
\end{aligned}
$$

(VI.10.1)

in which the following notation is used

$$
\underset{j=1,\dots,M}{max}\left\{B_{av,Q^j}\right\} \triangleq max\left\{B_{av,Q^1}, \dots, B_{av,Q^M}\right\}
$$

(VI.10.2)

and

$$
\underset{j=1,\dots,M}{max}\left\{\left(B_{av,Q^j}, B_{av,Q^{j+M}}\right)\right\} \triangleq max\left\{max\left\{B_{av,Q^1}, B_{av,Q^{1+M}}\right\}, \dots, max\left\{B_{av,Q^M}, B_{av,Q^{2M}}\right\}\right\}
$$

(VI.10.3)

■

**Theorem VI.3: Regulation theorem for deadline of an information particle under bandwidth constraint.**

**Given an information particle group $Q$ composed of $N\ (\geq 1)$ information particles. Let its minimum reachable access bandwidth be $B_{re,Q}^{min}$. A new information particle $i\ (i \notin Q)$ is added into $Q$, and a new information particle group $Q'\ (Q' = \{i\} \cup Q)$ is formed. The minimum reachable access bandwidth of $Q'$ is $B_{re,Q'}^{min}$, and its bearing capacity is $\ell_{Q'} < +\infty$. It is assumed that the adjustment of the minimum reachable access bandwidth $B_{re,Q'}^{min}$ of $Q'$ can be achieved by only changing the deadline $t_{e,i}$ of the newly added information particle $i$. And an access bandwidth $B\ (\in \mathbb{R}^+)$ which satisfies the relationship of $B > B_{re,Q}^{min}$ is given. It can be concluded that there always exists a value of $t_{e,i}\ (\in \mathbb{R}^+)$ which makes the corresponding $B_{re,Q'}^{min}$ satisfy the relationship of $B_{re,Q'}^{min} < B$.**

**Proof:**

Set the deadline $t_{e,i}$ to a positive real number $t^*\ (\in \mathbb{R}^+)$ with $t^* > t_{b,i}$. Accordingly, the corresponding minimum reachable access bandwidth of $Q'$ is denoted as $B_{re,Q'|t_{e,i}=t^*}^{min}$. Therefore, there are two cases to be considered, namely

**Case 1:** $B_{re,Q'|t_{e,i}=t^*}^{min} < B$

In this case, by setting $t_{e,i}$ to be equal to $t^*$, the proposition can be satisfied.

**Case 2:** $B_{re,Q'|t_{e,i}=t^*}^{min} \geq B$

According to **Lemma VI.3.1**, we have

$$\lim_{t_{e,i}\to+\infty} B_{av,Q'}^{pk} = B_{av,Q}^{pk}$$

Furthermore, based on **Theorem V.1**, we have

$$\begin{aligned} B_{re,Q'|t_{e,i}=+\infty}^{min} &\triangleq \lim_{t_{e,i}\to+\infty} B_{re,Q'}^{min} \\ &= \lim_{t_{e,i}\to+\infty} B_{av,Q'}^{pk} = B_{av,Q}^{pk} \\ &= B_{re,Q}^{min} \end{aligned} \qquad \text{(VI.11)}$$

Namely

$$B^{min}_{re,Q'|t_{e,i}=+\infty} = B^{min}_{re,Q} < B$$

(VI.12)

According to the definition of the limit, there must be a positive real number $\delta\ (> t^*)$ which makes $B^{min}_{re,Q'|t_{e,i}>\delta} < B$ hold when $t_{e,i} > \delta$. Therefore, the proposition in this case holds.

Since the above two cases must be one of them, the proposition is proved. ■

**Theorem VI.4: Given an information particle group $Q$ composed of $N\ (\geq 1)$ information particles. Select any of an information particle $i\ (i \in Q)$ in $Q$ and decrease its bearing capacity from $\ell_i$ to $\ell'_i$ (i.e., $0 < \ell'_i < \ell_i$). As the bearing capacity of $i$ varies, the average access bandwidth requirement of $Q$ changes from $B_{av,Q}$ to $B'_{av,Q}$. It can be concluded that $B_{av,Q} > B'_{av,Q}$.**

**Proof:**

As the bearing capacity of the information particle $i$ decreases from $\ell_i$ to $\ell'_i$, the bearing capacity of the information particle group $Q$ varies from $\ell_Q$ to $\ell'_Q$. It can be seen from the definition of the average access bandwidth requirement (**Definition IV.8**) that

$$B_{av,Q} = \frac{\ell_Q}{L_{D_Q^{int}}}$$

(VI.13)

$$B'_{av,Q} = \frac{\ell'_Q}{L_{D_Q^{int}}}$$

(VI.14)

Since the bearing capacity of the information particle $i\ (i \in Q)$ decreases from $\ell_i$ to $\ell'_i$, we have

$$\ell'_Q < \ell_Q$$

(VI.15)

Based on the above three equations, it can be concluded that $B_{av,Q} > B'_{av,Q}$. ■

**Corollary VI.4.1: Given an information particle group $Q$ composed of $N\ (\geq 1)$ information particles. Select any of an information particle $i\ (i \in Q)$ in $Q$ and decrease its bearing capacity from $\ell_i$ to $\ell'_i$ (i.e., $0 < \ell'_i < \ell_i$). As the bearing capacity of $i$ varies, the minimum reachable access bandwidth of $Q$ changes from $B^{min}_{re,Q}$ to $B^{min}_{re,Q}{}'$. It**

**can be concluded that $B_{re,Q}^{min} \geq B_{re,Q}^{min}{'}$.**

**Proof:**

As the bearing capacity of information particle $i$ decreases from $\ell_i$ to $\ell'_i$, the peak average access bandwidth requirement of the information particle group $Q$ changes from $B_{av,Q}^{pk}$ to $B_{av,Q}^{pk}{'}$. Based on **Theorem VI.4** and the definition of the peak average access bandwidth requirement (**Definition IV.9**), there is

$$B_{av,Q}^{pk}{'} \leq B_{av,Q}^{pk} \tag{VI.16}$$

According to **Theorem V.1**, we have

$$B_{re,Q}^{min} = B_{av,Q}^{pk} \tag{VI.17}$$

$$B_{re,Q}^{min}{'} = B_{av,Q}^{pk}{'} \tag{VI.18}$$

Based on the above three equations, it can be concluded that $B_{re,Q}^{min} \geq B_{re,Q}^{min}{'}$. ■

**Lemma VI.5.1 : Given an information particle group $Q$ composed of $N\,(\geq 1)$ information particles. A new information particle $i\,(i \notin Q)$ is added into $Q$, and a new information particle group $Q'\,(Q' = \{i\} \cup Q)$ is formed. It is assumed that the peak average access bandwidth requirements of $Q$ and $Q'$ are $B_{av,Q}^{pk}$ and $B_{av,Q'}^{pk}$, respectively. The deadline and the bearing capacity of the newly added information particle $i$ are $t_{e,i}$ and $\ell_i$, respectively. With the exception of $\ell_i$, all attribute parameters for member information particles in $Q'$ remain unchanged. Then, the following relation holds**

$$\lim_{\ell_i \to 0^+} B_{av,Q'}^{pk} = B_{av,Q}^{pk}$$

**Proof:**

We retain the notation employed in **Lemma VI.3.1**. Let there be $M\,(\geq 1)$ nonempty information particle subgroups contained in the information particle group $Q$, which are denoted by $Q^1, \dots, Q^M$, respectively. Accordingly, there exist $2M + 1$ nonempty information particle subgroups contained in the information particle group $Q'$, which are denoted by $Q^1, \dots, Q^M, Q^{1+M}, \dots, Q^{2M}, Q^{2M+1}$, respectively. According to the definition of the peak average access bandwidth requirement for an information particle group (**Definition IV.9**), we can proceed with the following derivation:

$$
\begin{aligned}
\lim_{\ell_i \to 0^+} B_{av,Q'}^{pk} &= \lim_{\ell_i \to 0^+} \left( max \left\{ \max_{j=1,\dots,M} \{B_{av,Q^j}\}, \max_{j=1,\dots,M} \{B_{av,Q^{j+M}}\}, B_{av,Q^{2M+1}} \right\} \right) \\
&= \lim_{\ell_i \to 0^+} \left( max \left\{ \max_{j=1,\dots,M} \{(B_{av,Q^j}, B_{av,Q^{j+M}})\}, B_{av,Q^{2M+1}} \right\} \right) \\
&= max \left\{ \max_{j=1,\dots,M} \left\{ \left( B_{av,Q^j}, \lim_{\ell_i \to 0^+} (B_{av,Q^{j+M}}) \right) \right\}, \lim_{\ell_i \to 0^+} (B_{av,Q^{2M+1}}) \right\} \\
&= max \left\{ \max_{j=1,\dots,M} \left\{ \left( B_{av,Q^j}, \lim_{\ell_i \to 0^+} \left( \frac{\ell_{Q^{j+M}}}{L_{D_{Q^{j+M}}^{int}}} \right) \right) \right\}, \lim_{\ell_i \to 0^+} \left( \frac{\ell_{Q^{2M+1}}}{L_{D_{Q^{2M+1}}^{int}}} \right) \right\} \\
&= max \left\{ \max_{j=1,\dots,M} \left\{ \left( \left( \frac{\ell_{Q^j}}{L_{D_{Q^j}^{int}}} \right), \lim_{\ell_i \to 0^+} \left( \frac{\ell_{Q^j} + \ell_i}{max \left( L_{D_{Q^j}^{int}}, t_{e,i} \right)} \right) \right) \right\}, \lim_{\ell_i \to 0^+} \left( \frac{\ell_i}{t_{e,i}} \right) \right\} \\
&= max \left\{ \max_{j=1,\dots,M} \left\{ \left( \left( \frac{\ell_{Q^j}}{L_{D_{Q^j}^{int}}} \right), \left( \frac{\ell_{Q^j}}{max \left( L_{D_{Q^j}^{int}}, t_{e,i} \right)} \right) \right) \right\}, 0 \right\} \\
&= max \left\{ \max_{j=1,\dots,M} \{B_{av,Q^j}\}, 0 \right\} \\
&= \max_{j=1,\dots,M} \{B_{av,Q^j}\} = B_{av,Q}^{pk}
\end{aligned}
\tag{VI.19}
$$

■

**Theorem VI.5: Regulation theorem for bearing capacity of an information particle under bandwidth constraint.**

**Given an information particle group $Q$ composed of $N\ (\geq 1)$ information particles. Let its minimum reachable access bandwidth be $B_{re,Q}^{min}$. A new information particle $i\ (i \notin Q)$ is added into $Q$, and a new information particle group $Q'\ (Q' = \{i\} \cup Q)$ is formed. The minimum reachable access bandwidth of $Q'$ is $B_{re,Q'}^{min}$. It is assumed that the adjustment of the minimum reachable access bandwidth $B_{re,Q'}^{min}$ of $Q'$ can be achieved by only changing the bearing capacity $\ell_i$ of a newly added information particle $i$. And an access bandwidth $B\ (\in \mathbb{R}^+)$ which satisfies the relationship of $B > B_{re,Q}^{min}$ is given. It can**

**be concluded that there always exists a value of $\ell_i$ $(\in \mathbb{R}^+)$ which makes the corresponding $B_{re,Q'}^{min}$ satisfy the relationship of $B_{re,Q'}^{min} < B$.**

**Proof:**

Set the bearing capacity $\ell_i$ to a positive real number $\ell^*$ $(\in \mathbb{R}^+)$. Accordingly, the corresponding minimum reachable access bandwidth of $Q'$ is denoted as $B_{re,Q'|\ell_i=\ell^*}^{min}$. Therefore, there are two cases to be considered, namely

**Case 1:** $B_{re,Q'|\ell_i=\ell^*}^{min} < B$

In this case, by setting $\ell_i$ to $\ell^*$, the proposition can be satisfied.

**Case 2:** $B_{re,Q'|\ell_i=\ell^*}^{min} \geq B$

According to **Lemma VI.5.1**, we have

$$\lim_{\ell_i \to 0^+} B_{av,Q'}^{pk} = B_{av,Q}^{pk}$$

Furthermore, based on **Theorem V.1**, we have

$$\begin{aligned} B_{re,Q'|\ell_i=0^+}^{min} &\triangleq \lim_{\ell_i \to 0^+} B_{re,Q'}^{min} \\ &= \lim_{\ell_i \to 0^+} B_{av,Q'}^{pk} = B_{av,Q}^{pk} \\ &= B_{re,Q}^{min} \end{aligned}$$

(VI.20)

Namely

$$B_{re,Q'|\ell_i=0^+}^{min} = B_{re,Q}^{min} < B$$

(VI.21)

According to the definition of the limit, there must be a positive real number $\delta$ $(< \ell^*)$ which makes $B_{re,Q'|\ell_i<\delta}^{min} < B$ hold when $0 < \ell_i < \delta$. Therefore, the proposition in this case holds.

Since the above two cases must be one of them, the proposition is proved. ■

**Lemma VI.6.1: Given two information particle groups $Q^A$ and $Q^B$ each consisting of $N$ $(\geq 1)$ information particles. Perform the incremental sorting operation $Op_{Inc}(Q)$ on the information particle groups $Q^A$ and $Q^B$ respectively, and assign corresponding sequence numbers to all the information particles contained in $Q^A$ and $Q^B$ accordingly.**

**Let $t_{e,i}^{A}$ and $t_{e,i}^{B}$ denote the deadlines of the information particles with sequence number $i\,(1 \le i \le N)$ in $Q^{A}$ and $Q^{B}$, respectively. Let $\ell_{i}^{A}$ and $\ell_{i}^{B}$ denote the bearing capacities of the information particles with sequence number $i\,(1 \le i \le N)$ in $Q^{A}$ and $Q^{B}$, respectively. Suppose the following relations hold between $Q^{A}$ and $Q^{B}$**

$$t_{e,i}^{B} = \alpha \cdot t_{e,i}^{A}\ (\alpha \in \mathbb{R}^{+}, 1 \le i \le N)$$

**(VI.22)**

**and**

$$\ell_{i}^{B} = \beta \cdot \ell_{i}^{A}\ (\beta \in \mathbb{R}^{+}, 1 \le i \le N)$$

**(VI.23)**

**where $\alpha$ and $\beta$ are two given positive real constants. Furthermore, let the subgroup $Q^{A(k)}\ (Q^{A(k)} \subseteq Q^{A})$ of information particles with index number $k$ in $Q^{A}$ consist of the information particles with sequence numbers $i_1, \dots, i_j, \dots, i_m\ (i_1 < \cdots < i_j < \cdots < i_m, 1 \le i_j \le N, 1 \le j \le m)$ in $Q^{A}$, which is denoted as:**

$$Q^{A(k)} \triangleq \{i_1, \dots, i_j, \dots, i_m\}_{Q^{A}}^{k}$$

**(VI.24)**

**Correspondingly, let the subgroup $Q^{B(k')}\ (Q^{B(k')} \subseteq Q^{B})$ of information particles with index number $k'$ in $Q^{B}$ ($k'$ is not necessarily equal to $k$) also consist of the information particles with sequence numbers $i_1, \dots, i_j, \dots, i_m\ (i_1 < \cdots < i_j < \cdots < i_m, 1 \le i_j \le N, 1 \le j \le m)$ in $Q^{B}$, which is denoted as:**

$$Q^{B(k')} \triangleq \{i_1, \dots, i_j, \dots, i_m\}_{Q^{B}}^{k'}$$

**(VI.25)**

**If the subgroup $Q^{A(k)}$ of information particles is a principal subgroup of the information particle group $Q^{A}$ (Definition IV.10), then the subgroup $Q^{B(k')}$ of information particles is also a principal subgroup of the information particle group $Q^{B}$.**

**Proof:**

From the definition of the average access bandwidth requirement of an information particle group (**Definition IV.8**), the average access bandwidth requirements of $Q^{A(k)}$ and $Q^{B(k')}$ are respectively given by

$$B_{av,Q^{A(k)}} = \frac{\ell_{Q^{A(k)}}}{L_{D^{int}_{Q^{A(k)}}}} = \frac{\sum_{j=1}^{m} \ell^{A}_{i_j}}{L_{D^{int}_{Q^{A(k)}}}}$$

(VI.26)

and

$$B_{av,Q^{B(k')}} = \frac{\ell_{Q^{B(k')}}}{L_{D^{int}_{Q^{B(k')}}}} = \frac{\sum_{j=1}^{m} \ell^{B}_{i_j}}{L_{D^{int}_{Q^{B(k')}}}}$$

(VI.27)

Combining equations **VI.22**, **VI.23**, **VI.26** and **VI.27**, we can obtain that

$$\begin{aligned} B_{av,Q^{B(k')}} &= \frac{\sum_{j=1}^{m} \ell^{B}_{i_j}}{L_{D^{int}_{Q^{B(k')}}}} \\ &= \frac{\beta \cdot \sum_{j=1}^{m} \ell^{A}_{i_j}}{\alpha \cdot L_{D^{int}_{Q^{A(k)}}}} \\ &= \frac{\beta}{\alpha} \cdot B_{av,Q^{A(k)}} \end{aligned}$$

(VI.28)

Let any subgroup $Q^{B(k'_-)}$ $(Q^{B(k'_-)} \subseteq Q^{B}, Q^{B(k'_-)} \neq Q^{B(k')})$ of information particles with index number $k'_-$ $(k'_- \neq k')$ in $Q^{B}$ that is different from $Q^{B(k')}$ consist of the information particles with sequence numbers $l_1, \dots, l_j, \dots, l_{m_-}$ $(l_1 < \cdots < l_j < \cdots < l_{m_-}, 1 \leq l_j \leq N, 1 \leq j \leq m_-)$ in $Q^{B}$, which is denoted as

$$Q^{B(k'_-)} \triangleq \{l_1, \dots, l_j, \dots, l_{m_-}\}^{k'_-}_{Q^{B}}$$

(VI.29)

From the definition of the average access bandwidth requirement of an information particle group (**Definition IV.8**), the average access bandwidth requirement of subgroup $Q^{B(k'_-)}$ of information particles is

$$B_{av,Q^{B(k'_-)}} = \frac{\ell_{Q^{B(k'_-)}}}{L_{D^{int}_{Q^{B(k'_-)}}}} = \frac{\sum_{j=1}^{m_-} \ell^{B}_{l_j}}{L_{D^{int}_{Q^{B(k'_-)}}}}$$

(VI.30)

Corresponding to the construction structure of subgroup $Q^{B(k'_-)}$ of information particles, let subgroup $Q^{A(k_-)}$ $(Q^{A(k_-)} \subseteq Q^{A}, Q^{A(k_-)} \neq Q^{A(k)})$ of information particles with index number $k_-$ $(k_- \neq k)$ in $Q^{A}$, which is also composed of information particles with sequence numbers $l_1, \dots, l_j, \dots, l_{m_-}$ $(l_1 < \cdots < l_j < \cdots < l_{m_-}, 1 \leq l_j \leq N, 1 \leq j \leq m_-)$, be denoted as

$$Q^{A(k_-)} \triangleq \{l_1, \dots, l_j, \dots, l_{m_-}\}_{Q^A}^{k_-}$$

(VI.31)

The average access bandwidth requirement of subgroup $Q^{A(k_-)}$ of information particles is

$$B_{av,Q^{A(k_-)}} = \frac{\ell_{Q^{A(k_-)}}}{L_{D^{int}_{Q^{A(k_-)}}}} = \frac{\sum_{j=1}^{m_-} \ell^A_{l_j}}{L_{D^{int}_{Q^{A(k_-)}}}}$$

(VI.32)

Combining equations **VI.22**, **VI.23**, **VI.30** and **VI.32**, we obtain

$$\begin{aligned} B_{av,Q^{B(k'_-)}} &= \frac{\sum_{j=1}^{m_-} \ell^B_{l_j}}{L_{D^{int}_{Q^{B(k'_-)}}}} \\ &= \frac{\beta \cdot \sum_{j=1}^{m_-} \ell^A_{l_j}}{\alpha \cdot L_{D^{int}_{Q^{A(k_-)}}}} \\ &= \frac{\beta}{\alpha} \cdot B_{av,Q^{A(k_-)}} \end{aligned}$$

(VI.33)

Considering that subgroup $Q^{A(k)}$ of information particles is a principal subgroup of the information particle group $Q^A$ (**Definition IV.10**), we have

$$B_{av,Q^{A(k_-)}} \le B_{av,Q^{A(k)}}$$

(VI.34)

Combining equations **VI.28**, **VI.33** and **VI.34**, we obtain

$$B_{av,Q^{B(k'_-)}} \le B_{av,Q^{B(k')}}$$

(VI.35)

Therefore, subgroup $Q^{B(k')}$ of information particles is also a principal subgroup of the information particle group $Q^B$. ■

**Theorem VI.6: Equal scaling theorem for information particle groups.**

**Given two information particle groups $Q^A$ and $Q^B$ each consisting of $N\ (\ge 1)$ information particles. Perform the incremental sorting operation $Op_{Inc}(Q)$ on the information particle groups $Q^A$ and $Q^B$ respectively, and assign corresponding sequence numbers to all the information particles contained in $Q^A$ and $Q^B$ accordingly. Let $t^A_{e,i}$ and $t^B_{e,i}$ denote the deadlines of the information particles with sequence number $i\ (1 \le$**

**$i \leq N$) in $Q^A$ and $Q^B$, respectively. Let $\ell_i^A$ and $\ell_i^B$ denote the bearing capacities of the information particles with sequence number $i$ ($1 \leq i \leq N$) in $Q^A$ and $Q^B$, respectively. Suppose the following relations hold between $Q^A$ and $Q^B$**

$$\boldsymbol{t_{e,i}^B = \alpha \cdot t_{e,i}^A \ (\alpha \in \mathbb{R}^+, 1 \leq i \leq N)}$$

**and**

$$\boldsymbol{\ell_i^B = \beta \cdot \ell_i^A \ (\beta \in \mathbb{R}^+, 1 \leq i \leq N)}$$

**where $\alpha$ and $\beta$ are two given positive real constants. Then the minimum reachable access bandwidths $B_{re,Q^A}^{min}$ and $B_{re,Q^B}^{min}$ of information particle groups $Q^A$ and $Q^B$ satisfy the following relation**

$$\boldsymbol{B_{re,Q^B}^{min} = \frac{\beta}{\alpha} \cdot B_{re,Q^A}^{min}}$$

**(VI.36)**

**Proof:**

Suppose subgroup $Q^{A(k)}$ of information particles (see **Equation VI.24**) is a principal subgroup of the information particle group $Q^A$. According to **Lemma VI.6.1**, subgroup $Q^{B(k')}$ of information particles (see **Equation VI.25**) is a principal subgroup of the information particle group $Q^B$. Combining with **Theorem V.1**, we can conclude that

$$B_{re,Q^B}^{min} = B_{av,Q^B}^{pk} = B_{av,Q^{B(k')}}$$

(VI.37)

Furthermore, from **Equation VI.28**, we obtain

$$\begin{aligned} B_{re,Q^B}^{min} &= B_{av,Q^{B(k')}} = \frac{\beta}{\alpha} \cdot B_{av,Q^{A(k)}} \\ &= \frac{\beta}{\alpha} \cdot B_{av,Q^A}^{pk} = \frac{\beta}{\alpha} \cdot B_{re,Q^A}^{min} \end{aligned}$$

(VI.38)

■

# VII. A particle access algorithm based on dynamically adjusting the minimum reachable access bandwidth

In this section, calculation methods for the minimum reachable access bandwidth of an information particle group are proposed (***Theorem VII.1*** and ***Corollary VII.1.1***); In **Corollary VII.1.2**, it is proved that when the access bandwidth of a given information particle group is set to be equal to its minimum reachable access bandwidth, the access efficiency of any of its principal information particle subgroup reaches one (that is, for any of its principal information particle subgroup, the allocated access bandwidth can be fully utilized); The problem of each principal information particle subgroup being composed of which information particles is studied (**Corollary VII.1.3**), and the inclusion and included relationships among several principal information particle subgroups are proved (**Corollary VII.1.4**); The necessary and sufficient conditions for several information particles forming or not forming a principal information particle subgroup are proved (**Corollary VII.1.5**); The fact that the minimum reachable access bandwidth will decrease at the inflection point of the access bandwidth requirement of an information particle group is proved (***Corollary VII.1.6***); Based on the theoretical basis built in this section, a particle access algorithm based on dynamically adjusting the minimum reachable access bandwidth is proposed (***Algorithm VII.1***); The upper bound (***Corollary VII.1.7***) and the lower bound (***Corollary VII.1.8***) of the minimum reachable access bandwidth of an information particle group are given, respectively; The property of linear superpositions of multiple information particle groups under specific constraints is proved (***Corollary VII.1.9***); The property of linear splitting of an information particle group is also proved (***Corollary VII.1.10***); A closed form expression of representing the minimum reachable access bandwidth of an information particle subgroup as the product of its vector of switching functions and its vector of average access bandwidth requirements is given (**Corollary VII.1.11**); And it is proved that the vector of switching functions has and only has one component with its value being equal to one (**Corollary VII.1.12**).

## *A. Calculation methods for the minimum reachable access bandwidth of an information particle group*

**Theorem VII.1: Iterative calculation method for the minimum reachable access bandwidth**

**of an information particle group.**

**Given an information particle group $\boldsymbol{Q}$ composed of $\boldsymbol{N}\,(>\mathbf{1})$ information particles, and an incremental sorting operation $\boldsymbol{Op_{Inc}(Q)}$ is performed on it. Let $\boldsymbol{B_{av,Q_i}}$ indicate the average access bandwidth requirement of the information particle subgroup $\boldsymbol{Q_i}\ (\boldsymbol{Q_i} \subseteq \boldsymbol{Q})$ composed of all the information particles whose sorting sequence numbers are not greater than $\boldsymbol{i}\ (\mathbf{1} \le \boldsymbol{i} \le \boldsymbol{N})$. Then there is**

$$\boldsymbol{B_{re,Q_l}^{min} = max\{B_{re,Q_{l-1}}^{min}, B_{av,Q_l}\}\ (1 < l \le N)} \tag{VII.1}$$

**Proof:**

We retain the notation employed in **Lemma VI.3.1**. Let there be $M\,(\ge 1)$ nonempty information particle subgroups contained in the information particle group $Q_{l-1}\ (1 < l \le N)$, which are denoted by $Q_{l-1}^1, \dots, Q_{l-1}^M$, respectively. Accordingly, there exist $2M+1$ nonempty information particle subgroups contained in the information particle group $Q_l\ (1 < l \le N)$ ($Q_l$ is formed by appending the information particle $l$ to $Q_{l-1}$), which are denoted by $Q_{l-1}^1, \dots, Q_{l-1}^M, Q_{l-1}^{1+M}, \dots, Q_{l-1}^{2M}, Q_{l-1}^{2M+1}$, respectively. In these symbolic expressions, the following definitions are employed

$$Q_{l-1}^{j+M} \triangleq \{l\} \cup Q_{l-1}^j\ (1 \le j \le M) \tag{VII.2.1}$$

and

$$Q_{l-1}^{2M+1} \triangleq \{l\} \tag{VII.2.2}$$

According to the definition of the peak average access bandwidth requirement for an information particle group (**Definition IV.9**), we can proceed with the following derivation:

$$\begin{aligned} B_{av,Q_l}^{pk} &= max\left\{\max_{j=1,\dots,M}\left\{B_{av,Q_{l-1}^j}\right\}, \max_{j=1,\dots,M}\left\{B_{av,Q_{l-1}^{j+M}}\right\}, B_{av,Q_{l-1}^{2M+1}}\right\} \\ &= max\left\{B_{av,Q_{l-1}}^{pk}, \max_{j=1,\dots,M}\left\{\left(\frac{\ell_{Q_{l-1}^{j+M}}}{L_{D_{Q_{l-1}^{j+M}}^{int}}}\right)\right\}, \left(\frac{\ell_{Q_{l-1}^{2M+1}}}{L_{D_{Q_{l-1}^{2M+1}}^{int}}}\right)\right\} \\ &= max\left\{B_{av,Q_{l-1}}^{pk}, \max_{j=1,\dots,M}\left\{\left(\frac{\ell_{Q_{l-1}^j} + \ell_l}{L_{D_{Q_l}^{int}}}\right)\right\}, \left(\frac{\ell_l}{L_{D_{Q_l}^{int}}}\right)\right\} \end{aligned} \tag{VII.3.1}$$

As the following inequality holds, that is,

$$\frac{\ell_{Q_{l-1}^j} + \ell_l}{L_{D_{Q_l}^{int}}} > \frac{\ell_l}{L_{D_{Q_l}^{int}}} \ (1 \le j \le M)$$

(VII.3.2)

From this, we proceed with the derivation of **Equation VII.3.1** and obtain

$$\begin{aligned} B_{av,Q_l}^{pk} &= max\left\{B_{av,Q_{l-1}}^{pk}, \max_{j=1,\dots,M}\left\{\left(\frac{\ell_{Q_{l-1}^j} + \ell_l}{L_{D_{Q_l}^{int}}}\right)\right\}\right\} \\ &= max\left\{B_{av,Q_{l-1}}^{pk}, \frac{\ell_l + \max_{j=1,\dots,M}\left\{\ell_{Q_{l-1}^j}\right\}}{L_{D_{Q_l}^{int}}}\right\} \end{aligned}$$

(VII.4.1)

It follows from the definition of the bearing capacity of an information particle group (**Definition IV.3**) that among all information particle subgroups contained in an information particle group $Q_{l-1}$, the one with the maximum bearing capacity is itself (i.e., $\ell_{Q_{l-1}}$). Hence, we have

$$\ell_{Q_{l-1}} = \max_{j=1,\dots,M}\left\{\ell_{Q_{l-1}^j}\right\}$$

(VII.4.2)

From this, we continue the derivation given in **Equation VII.4.1** to obtain

$$\begin{aligned} B_{av,Q_l}^{pk} &= max\left\{B_{av,Q_{l-1}}^{pk}, \frac{\ell_l + \ell_{Q_{l-1}}}{L_{D_{Q_l}^{int}}}\right\} \\ &= max\left\{B_{av,Q_{l-1}}^{pk}, \frac{\ell_{Q_l}}{L_{D_{Q_l}^{int}}}\right\} \\ &= max\left\{B_{av,Q_{l-1}}^{pk}, B_{av,Q_l}\right\} \end{aligned}$$

(VII.5)

According to **Theorem V.1**, we have

$$B_{re,Q_l}^{min} = max\{B_{re,Q_{l-1}}^{min}, B_{av,Q_l}\}$$

■

Note:

1) For an information particle group $Q$ containing $N$ information particles, we have $Q_N = Q$.

**Corollary VII.1.1: Direct calculation method for the minimum reachable access bandwidth of an information particle group.**

**Given an information particle group $Q$ composed of $N\ (\ge 1)$ information particles,**

**and an incremental sorting operation $Op_{Inc}(Q)$ is performed on it. Let $B_{av,Q_i}$ indicate the average access bandwidth requirement of the information particle subgroup $Q_i$ ($Q_i \subseteq Q$) composed of all the information particles whose sorting sequence numbers are not greater than $i$ ($1 \le i \le N$). Then there is**

$$B_{re,Q}^{min} = \max_{l=1,\dots,N}\{B_{av,Q_l}\}$$

**(VII.6)**

**Proof:**

For the case of $N = 1$, since the proposition conclusion obviously holds, its proof is omitted. Next, let us consider the case of $N \ge 2$. The proof is performed by mathematical induction. For case $N = 2$, we obtain from **Equation VII.5** that

$$B_{av,Q_2}^{pk} = max\{B_{av,Q_1}^{pk}, B_{av,Q_2}\}$$

(VII.7)

According to the definition of the peak average access bandwidth requirement of information particle group $Q$ (**Definition IV.9**), it can be readily obtained that

$$B_{av,Q_1}^{pk} = B_{av,Q_1}$$

(VII.8)

Substitute the above formula into **Equation VII.7**, and we obtain that

$$B_{av,Q_2}^{pk} = max\{B_{av,Q_1}, B_{av,Q_2}\}$$

(VII.9)

From **Theorem V.1**, we obtain that

$$B_{re,Q_2}^{min} = B_{av,Q_2}^{pk} = max\{B_{av,Q_1}, B_{av,Q_2}\}$$

The above completes the proof that the proposition holds when $N = 2$.

Now assume that the proposition holds under the case of $N = k$ ($k \ge 2$), i.e.,

$$B_{re,Q_k}^{min} = \max_{l=1,\dots,k}\{B_{av,Q_l}\}$$

(VII.10)

For the case of $N = k + 1$, we derive the result from **Equation VII.5**, **Theorem V.1** and **Equation VII.10** that

$$
\begin{aligned}
B^{pk}_{av,Q_{k+1}} &= max\{B^{pk}_{av,Q_k}, B_{av,Q_{k+1}}\} \\
&= max\{B^{min}_{re,Q_k}, B_{av,Q_{k+1}}\} \\
&= max\{\max_{l=1,\ldots,k}\{B_{av,Q_l}\}, B_{av,Q_{k+1}}\} \\
&= \max_{l=1,\ldots,k+1}\{B_{av,Q_l}\}
\end{aligned}
$$

(VII.11)

Furthermore, we obtain from **Theorem V.1** that

$$
B^{min}_{re,Q_{k+1}} = B^{pk}_{av,Q_{k+1}} = \max_{l=1,\ldots,k+1}\{B_{av,Q_l}\}
$$

Therefore, the above proves that the proposition also holds for the case of $N = k + 1$. ■

*B. Some properties of information particle subgroups*

**Corollary VII.1.2: Given an information particle group $Q$ composed of $N\,(\geq 1)$ information particles. Let its access bandwidth be equal to its minimum reachable access bandwidth $B_{re,Q}^{min}$. The access efficiency of any of its principal information particle subgroup $Q'_{cr}$ $(Q'_{cr} \subseteq Q)$ reaches one, i.e. $\eta_{Q'_{cr}|B_{re,Q}^{min}} = 1$.**

**Proof:**

According to the definition of the access efficiency (**Definition IV.14**), the definition of a principal information particle subgroup (**Definition IV.10**) and **Theorem V.1**, we have

$$\begin{aligned} \eta_{Q'_{cr}|B_{re,Q}^{min}} &= \frac{B_{av,Q'_{cr}}}{B_{re,Q}^{min}} \\ &= \frac{B_{av,Q}^{pk}}{B_{re,Q}^{min}} \\ &= 1 \end{aligned} \tag{VII.12}$$

■

**Note:**

1) Access efficiency $\eta_{Q'_{cr}|B_{re,Q}^{min}} = 1$, which means that when the access bandwidth of an information particle group $Q$ is limited to its minimum reachable access bandwidth $B_{re,Q}^{min}(= B_{av,Q}^{pk})$, within the effective survival time interval $D_{Q'_{cr}}^{int}$ covered by the principal information particle subgroup $Q'_{cr}$, the access resources with bandwidth $B_{re,Q}^{min}$ are completely occupied by the processing of $Q'_{cr}$, that is, the bandwidth resources are fully utilized.

**Corollary VII.1.3: Continuity theorem for a principal information particle subgroups (I).**

**Given an information particle group $Q$ composed of $N\,(\geq 1)$ information particles, and an incremental sorting operation $Op_{Inc}(Q)$ is performed on it. If the maximum sequence number of all the information particles contained in a principal information particle subgroup $Q'_{cr}$ $(Q'_{cr} \subseteq Q)$ is $k\,(k \geq 1)$, $Q'_{cr}$ must contain all the information particles in $Q$ whose sequence numbers are greater than or equal to 1 and less than or**

**equal to $k$ ($k \geq 1$).**

**Proof:**

For the case of $k$=1, the proposition is obviously true. Proof omitted.

Consider the case of $k > 1$. Reduction to absurdity is adopted. Assuming that there exist an information particle $j\,(1 \leq j < k, \ell_j > 0)$ in $Q$ which does not belong to the principal information particle subgroup $Q'_{cr}$. $j$ and $Q'_{cr}$ can then be combined to form another information particle subgroup $Q''_{cr} \triangleq Q'_{cr} \cup \{j\}$. According to the definition of the average access bandwidth requirement (**Definition IV.8**), we have

$$\begin{aligned} B_{av,Q''_{cr}} &= \frac{\sum_{i \in Q''_{cr}} \ell_i}{L_{D^{int}_{Q''_{cr}}}} \\ &= \frac{\sum_{i \in Q''_{cr}} \ell_i}{L_{D^{int}_{Q'_{cr}}}} \\ &= \frac{\ell_j + \sum_{i \in Q'_{cr}} \ell_i}{L_{D^{int}_{Q'_{cr}}}} \\ &= \frac{\ell_j}{L_{D^{int}_{Q'_{cr}}}} + B_{av,Q'_{cr}} \\ &> B_{av,Q'_{cr}} \end{aligned}$$

(VII.13)

The above equation is in contradiction with the premise (i.e., the information particle subgroup $Q'_{cr}$ is a principal information particle subgroup of $Q$). ■

**Note:**

1) The relationship $L_{D^{int}_{Q''_{cr}}} = L_{D^{int}_{Q'_{cr}}}$ is used in the above equation. This is because that the sequence number of the information particle $j$ is smaller than that of the information particle $k$, so the effective survival time interval $D^{int}_{Q'_{cr}}$ of the information particle subgroup $Q'_{cr}$ and the effective survival time interval $D^{int}_{Q''_{cr}}$ of $Q''_{cr}$ are completely overlapped;
2) The above proposition reveals that there exists continuity among the sequence numbers of information particles constituting a principal information particle subgroup.

**Corollary VII.1.4: Continuity theorem for a principal information particle subgroups (II).**

**Given an information particle group $Q$ composed of $N$ ($> 1$) information particles,**

**and an incremental sorting operation $Op_{Inc}(Q)$ is performed on it. Let the corresponding maximum sequence numbers of information particles contained in principal information particle subgroups $Q'_{cr1}$ ($Q'_{cr1} \subseteq Q$) and $Q'_{cr2}$ ($Q'_{cr2} \subseteq Q$) are $k_1$ ($k_1 \geq 1$) and $k_2$ ($k_2 \geq 1$) respectively, and $k_1 < k_2$. It can be concluded that $Q'_{cr1}$ is a true subset of $Q'_{cr2}$, i.e. $Q'_{cr1} \subset Q'_{cr2}$.**

**Proof:**

According to **Corollary VII.1.3**, $Q'_{cr1}$ contains all the information particles with sequence numbers greater than or equal to 1 and less than or equal to $k_1$ in $Q$, and $Q'_{cr2}$ contains all the information particles with sequence numbers greater than or equal to 1 and less than or equal to $k_2$ ($k_2 > k_1$) in $Q$, so it is evident that $Q'_{cr2}$ contains $Q'_{cr1}$. ■

**Note:**

1) The above proposition indicates that for an information particle group containing finite elements, the relationships among its principal information particle subgroups resemble Russian dolls, i.e., one principal information particle subgroup is nested within another. As the information particle group contains only a finite number of principal information particle subgroups, there must exist a principal information particle subgroup (i.e., the largest principal information particle subgroup, which is defined in **Definition IV.16**) containing all the other principal information particle subgroup.

**Corollary VII.1.5: Sufficient and necessary condition for an information particle subgroup to be a principal information particle subgroup.**

**Given an information particle group $Q$ composed of $N$ ($\geq 1$) information particles, and an incremental sorting operation $Op_{Inc}(Q)$ is performed on it. Let $Q_k$ ($Q_k \triangleq \{i | 1 \leq i \leq k, i \in Q\}$) be the information particle subgroup composed of all the information particles in $Q$ whose sequence number is not greater than $k$ ($k \geq 1$), and the effective survival time span of $Q_k$ is $L_{D_{Q_k}^{int}}$. It can be concluded that**

**1) The sufficient and necessary condition for $Q_k$ being a principal information particle subgroup of $Q$ is**

$$\boldsymbol{L_{D_{Q_k}^{int}} = \frac{\left(\sum_{i=1}^{k} \ell_i\right)}{B_{av,Q}^{pk}}}$$

**(VII.14)**

2) **The sufficient and necessary condition for $\boldsymbol{Q_k}$ being not a principal information particle subgroup of $\boldsymbol{Q}$ is**

$$\boldsymbol{L_{D_{Q_k}^{int}} > \frac{\left(\sum_{i=1}^{k} \ell_i\right)}{B_{av,Q}^{pk}}}$$

**(VII.15)**

**Proof:**

First, let us prove the conclusion (1). If $Q_k$ is a principal information particle subgroup of $Q$, according to the definition of a principal information particle subgroup (**Definition IV.10**), the average access bandwidth requirement of $Q_k$ is $B_{av,Q_k} = B_{av,Q}^{pk}$, i.e. $B_{av,Q_k} = \frac{(\sum_{i=1}^{k} \ell_i)}{L_{D_{Q_k}^{int}}} = B_{av,Q}^{pk}$. Therefore, there is $L_{D_{Q_k}^{int}} = \frac{(\sum_{i=1}^{k} \ell_i)}{B_{av,Q}^{pk}}$. Furthermore, if $L_{D_{Q_k}^{int}} = \frac{(\sum_{i=1}^{k} \ell_i)}{B_{av,Q}^{pk}}$ holds, there is $B_{av,Q_k} = \frac{(\sum_{i=1}^{k} \ell_i)}{L_{D_{Q_k}^{int}}} = B_{av,Q}^{pk}$. According to the definition of a principal information particle subgroup (**Definition IV.10**), $Q_k$ is a principal information particle subgroup of $Q$. Therefore, conclusion (1) is proved.

Next, let us prove the conclusion (2). If $Q_k$ is not a principal information particle subgroup of $Q$, according to the definition of a principal information particle subgroup (**Definition IV.10**), the average access bandwidth requirement of $Q_k$ is $B_{av,Q_k} < B_{av,Q}^{pk}$, i.e. $B_{av,Q_k} = \frac{(\sum_{i=1}^{k} \ell_i)}{L_{D_{Q_k}^{int}}} < B_{av,Q}^{pk}$. Therefore, there is $L_{D_{Q_k}^{int}} > \frac{(\sum_{i=1}^{k} \ell_i)}{B_{av,Q}^{pk}}$. Furthermore, if $L_{D_{Q_k}^{int}} > \frac{(\sum_{i=1}^{k} \ell_i)}{B_{av,Q}^{pk}}$ holds, there is $B_{av,Q_k} = \frac{(\sum_{i=1}^{k} \ell_i)}{L_{D_{Q_k}^{int}}} < B_{av,Q}^{pk}$. According to the definition of a principal information particle subgroup (**Definition IV.10**), $Q_k$ is not a principal information particle subgroup of $Q$. Therefore, conclusion (2) is proved. ■

**Note:**

1) On one hand, based on the above proposition, if the information particles contained in an information particle group $Q$ are processed according to the EDF processing strategy with the access bandwidth being set to be $B_{av,Q}^{pk}$, and moreover, if the completion time of the processing of the information particle with sequence number of $k\ (k \geq 1)$ is $\frac{(\sum_{i=1}^{k} \ell_i)}{B_{av,Q}^{pk}} < L_{D_{Q_k}^{int}}$, which indicates that the completion time of the processing of the information particle $k$ is ahead of its deadline $t_{e,k}$, it can be concluded that the information particle subgroup $Q_k\ (Q_k = \{i|1 \leq i \leq k, i \in Q\})$ must not be a principal information particle subgroup of $Q$;

2) On the other hand, based on the above proposition, if the completion time of the processing of the information particle with sequence number of $k\ (k \geq 1)$ is $\frac{(\sum_{i=1}^{k} \ell_i)}{B_{av,Q}^{pk}} = L_{D_{Q_k}^{int}}$, which indicates that the completion time of the processing of the information particle $k$ is just the same as its deadline $t_{e,k}$, it can be concluded that the information particle subgroup $Q_k\ (Q_k = \{i|1 \leq i \leq k, i \in Q\})$ must be a principal information particle subgroup of $Q$.

**Corollary VII.1.6: A property of the inflection point of the access bandwidth requirement of an information particle group.**

**Given an information particle group $Q$ composed of $N\ (\geq 1)$ information particles, and an incremental sorting operation $Op_{Inc}(Q)$ is performed on it. Let $Q_k\ (Q_k \triangleq \{i|1 \leq i \leq k, i \in Q\})$ be the information particle subgroup composed of all the information particles in $Q$ whose sequence number is not greater than $k\ (k \geq 1)$, and the effective survival time span of $Q_k$ is $L_{D_{Q_k}^{int}}$. If the inflection point of the access bandwidth requirement of $Q$ (Definition IV.15) is $N_{cr,Q}\ (\geq 1)$, and $N_{cr,Q} < N$. Let**

$$\alpha \triangleq \max_{k=N_{cr,Q}+1,\dots,N} \left\{ \frac{\sum_{i=N_{cr,Q}+1}^{k} \ell_i}{L_{D_{Q_k}^{int}} - L_{D_{Q_{N_{cr,Q}}}^{int}}} \right\} \qquad \text{(VII.16)}$$

**then $\alpha < B_{av,Q}^{pk}\ (B_{av,Q}^{pk} = B_{re,Q}^{min})$ holds.**

**Proof:**

Assuming that when $k = j\ (N_{cr,Q} + 1 \leq j \leq N)$, the value of $\frac{\sum_{i=N_{cr,Q}+1}^{k} \ell_i}{\left(L_{D_{Q_k}^{int}} - L_{D_{Q_{N_{cr,Q}}}^{int}}\right)}$ reaches the maximum, i.e.,

$$\alpha = \frac{\sum_{i=N_{cr,Q}+1}^{j} \ell_i}{L_{D_{Q_j}^{int}} - L_{D_{Q_{N_{cr,Q}}}^{int}}}$$

(VII.17)

According to the definition of the largest principal information particle subgroup of $Q$ (**Definition IV.16**) and **Corollary VII.1.3**, it is known that the information particle subgroup $Q_{N_{cr,Q}}$ ($Q_{N_{cr,Q}} \triangleq \{i|1 \leq i \leq N_{cr,Q}, i \in Q\}$) is the largest principal information particle subgroup of $Q$, so the following equation holds

$$\begin{aligned} B_{av,Q}^{pk} &= B_{av,Q_{N_{cr,Q}}} \\ &= \frac{\sum_{i=1}^{N_{cr,Q}} \ell_i}{L_{D_{Q_{N_{cr,Q}}}^{int}}} \end{aligned}$$

(VII.18)

Moreover, let us define an auxiliary variable $\sigma$ as

$$\sigma \triangleq (\alpha - B_{av,Q}^{pk})(L_{D_{Q_j}^{int}} - L_{D_{Q_{N_{cr,Q}}}^{int}})$$

(VII.19)

By combing **Equation VII.17** and **Equation VII.19**, it can be obtained that

$$\begin{aligned} B_{av,Q}^{pk} &= \alpha - \frac{\sigma}{L_{D_{Q_j}^{int}} - L_{D_{Q_{N_{cr,Q}}}^{int}}} \\ &= \frac{\sum_{i=N_{cr,Q}+1}^{j} \ell_i - \sigma}{L_{D_{Q_j}^{int}} - L_{D_{Q_{N_{cr,Q}}}^{int}}} \end{aligned}$$

(VII.20)

According to **Equation VII.18** and **Equation VII.20**, there is

$$\begin{aligned} B_{av,Q}^{pk} &= \frac{\sum_{i=1}^{N_{cr,Q}} \ell_i}{L_{D_{Q_{N_{cr,Q}}}^{int}}} \\ &= \frac{\sum_{i=N_{cr,Q}+1}^{j} \ell_i - \sigma}{L_{D_{Q_j}^{int}} - L_{D_{Q_{N_{cr,Q}}}^{int}}} \\ &= \frac{\sum_{i=1}^{N_{cr,Q}} \ell_i + \sum_{i=N_{cr,Q}+1}^{j} \ell_i - \sigma}{L_{D_{Q_{N_{cr,Q}}}^{int}} + (L_{D_{Q_j}^{int}} - L_{D_{Q_{N_{cr,Q}}}^{int}})} \\ &= \frac{\sum_{i=1}^{j} \ell_i - \sigma}{L_{D_{Q_j}^{int}}} \\ &= B_{av,Q_j} - \frac{\sigma}{L_{D_{Q_j}^{int}}} \end{aligned}$$

(VII.21)

Based on the definition of $N_{cr,Q}$ (**Definition IV.15**), we have $B_{av,Q_j} < B_{av,Q}^{pk}$, which is further substituted into **Equation VII.21**, it can be obtained that

$$\sigma = \left(B_{av,Q_j} - B_{av,Q}^{pk}\right) \cdot L_{D_{Q_j}^{int}} < 0$$

(VII.22)

From **Equation VII.19**, **Equation VII.22** and **Theorem V.1**, we have the conclusion that

$$\alpha < B_{av,Q}^{pk} \ (B_{av,Q}^{pk} = B_{re,Q}^{min})$$

■

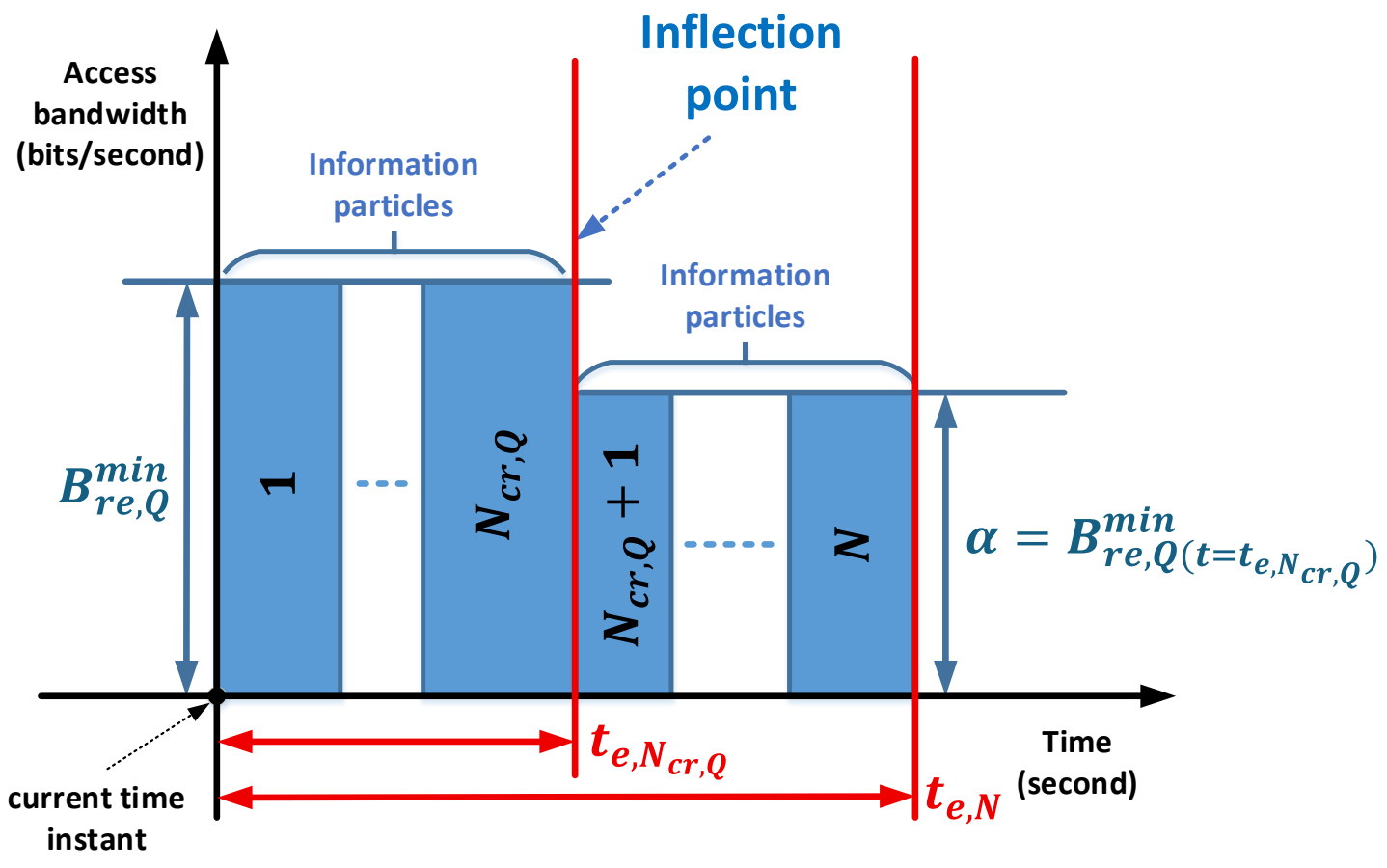

**Figure VII.1** Variations of the minimum reachable access bandwidth of an information particle group $Q$ with time

**Note:**

1) As shown in **Figure VII.1**, based on **Corollary VII.1.2, Corollary VII.1.3 and Corollary VII.1.4**, and **Definitions IV.15** and **IV.16**, it can be seen that when the access bandwidth of an information particle group $Q$ is limited to its minimum reachable access bandwidth $B_{re,Q}^{min}$ $(= B_{av,Q}^{pk})$ and EDF processing strategy is adopted, it is within the effective survival time interval $D_{Q^*_{cr}}^{int}$ covered by its largest principal information particle subgroup $Q^*_{cr}$ (its corresponding effective survival time span is denoted as $L_{D_{Q_{N_{cr,Q}}}^{int}}$) that the access resources with bandwidth of $B_{re,Q}^{min}$ $(= B_{av,Q}^{pk})$ are completely occupied by the processing of information carried by $N_{cr,Q}$ information particles belonging to $Q^*_{cr}$. The completion time of the processing of the information particle with sequence number $N_{cr,Q}$ is $\frac{\left(\sum_{i=1}^{N_{cr,Q}} \ell_i\right)}{B_{av,Q}^{pk}} = L_{D_{Q_{N_{cr,Q}}}^{int}}$, which indicates that the completion time of the processing of the information particle $N_{cr,Q}$ is exactly the same as its deadline $t_{e,N_{cr,Q}}$. Therefore, at this time instant, it can be observed that under a given access bandwidth $B_{re,Q}^{min}$ $(= B_{av,Q}^{pk})$ and within its effective survival time interval $D_{Q^*_{cr}}^{int}$, the access efficiency of the largest principal information particle subgroup $Q^*_{cr}$ is $\eta_{Q^*_{cr}|B_{re,Q}^{min}} = 1$;

2) Here, the physical meaning of $\alpha$ (**Equation VII.16**) proposed in the above proposition is further explained and discussed. When considering the information particle group $Q$ in the preceding article, we have not clearly pointed out that the determination of the values of its attributes is closely related to the specific time point. In fact, we do refer to specific time points to determine the exact attribute values of an information particle group and those of each information particles contained in it.

Without losing generality, first let us consider the time instant at $t = 0$. At that time, we observed that the information particle group $Q(t = 0)$ (Note: here we explicitly marked the corresponding time for the information particle group) is composed of $N$ valid information particles. By performing the incremental sorting operation $Op_{Inc}(Q(t = 0))$, sorting the $N$ information particles contained therein and giving each of these information particles a corresponding sequence number, the information particle group $Q(t = 0)$ can be specifically determined as a set composed of $N$ information particles, i.e. information particle $i \in Q(t = 0)$ $(1 \le i \le N)$. Relative to the current time instant $t = 0$, the bearing capacity of the information particle $i$ in $Q(t = 0)$ is $\ell_i$ (>0), and its deadline is $t_{e,i}$ (>0).

How about the situation within the time interval $[0, t_{e,N_{cr,Q(t=0)}}]$? According to **Theorem VII.1** and **Corollary VII.1.1**, the minimum reachable access bandwidth corresponding to $Q(t = 0)$ can be obtained, i.e. $B_{re,Q(t=0)}^{min} = B_{av,Q(t=0)}^{pk} = \max_{i=1,\ldots,N}\{B_{av,Q_i(t=0)}\}$. Based on **Definitions IV.15 and IV.16**, the inflection point $N_{cr,Q(t=0)}$ of the access bandwidth requirement and the corresponding largest principal information particle subgroup $Q^*_{cr}(t = 0) \triangleq \{i|1 \le i \le N_{cr,Q(t=0)}, i \in Q(t = 0)\}$ of $Q(t = 0)$ can also be obtained. Relative to the current time instant $t = 0$, let the deadline of the information particle with sequence number of $N_{cr,Q(t=0)}$ be $t_{e,N_{cr,Q(t=0)}}$. Based on **Corollary 2, 3, and 4 of Theorem VII.1**, it can be seen that within the time interval $[0, t_{e,N_{cr,Q(t=0)}}]$, if the access bandwidth of the whole information particle group is limited to $B^{min}_{re,Q(t=0)}$ $(= B^{pk}_{av,Q(t=0)})$ and the EDF processing strategy is adopted, the access resources with a bandwidth of $B^{min}_{re,Q(t=0)}$ $(= B^{pk}_{av,Q(t=0)})$ will be completely occupied by the processing of the information carried by the information particles in $Q(t = 0)$ whose sequence numbers are not greater than $N_{cr,Q(t=0)}$, and all the information carried by these information particles will be processed before time $t_{e,N_{cr,Q(t=0)}}$. In other words, within the time interval $[0, t_{e,N_{cr,Q(t=0)}}]$, given the access bandwidth and the adopted EDF processing strategy, we can fully know the processing procedure of the information particle group.

From time $t = t_{e,N_{cr,Q(t=0)}}$, the initial information particle group $Q(t = 0)$ contains only the remaining $N - N_{cr,Q(t=0)}$ information particles (assume that no other new information particles are generated within the time interval $[0, t_{e,N_{cr,Q(t=0)}}]$), and at this time instant it becomes the information particle group $Q(t = t_{e,N_{cr,Q(t=0)}})$. Then, what are the minimum reachable access bandwidth and the peak average access bandwidth requirement of the information particle group $Q(t = t_{e,N_{cr,Q(t=0)}})$ relative to the time instant $t = t_{e,N_{cr,Q(t=0)}}$? Obviously, the conclusion in **Theorem VII.1** can still be applied to the information particle group $Q(t = t_{e,N_{cr,Q(t=0)}})$. **Based on Theorem VII.1, it is not difficult to be found that $\alpha$ is the minimum reachable access bandwidth of the information particle group $Q(t = t_{e,N_{cr,Q(t=0)}})$. Therefore, the conclusion of Corollary VII.1.6 is that the minimum reachable access bandwidth $B^{min}_{re,Q(t=0)}$ $(= B^{pk}_{av,Q(t=0)})$ corresponding to the information particle group $Q(t = 0)$ must be greater than the minimum reachable access bandwidth $B^{min}_{re,Q(t=t_{e,N_{cr,Q(t=0)}})}$ $(= B^{pk}_{av,Q(t=t_{e,N_{cr,Q(t=0)}})})$ corresponding to the information particle group $Q(t = t_{e,N_{cr,Q(t=0)}})$. Therefore, time instant $t = t_{e,N_{cr,Q(t=0)}}$ is precisely the inflection point where the minimum reachable access bandwidth of the information particle group changes, which is why we define the sequence number of the information particle with sequence number $N_{cr,Q(t=0)}$ in $Q(t = 0)$ as the inflection point of its access bandwidth requirement.** Of course, in the above discussion, in order to explain that the properties of an information particle group vary with time, we explicitly indicate the corresponding time instants. For the sake of brevity, in the other part of this paper, we still omit the annotation of the corresponding time without causing ambiguity;

3) The conclusion of **Corollary VII.1.6** shows that when EDF processing strategy is adopted, the required minimum reachable access bandwidth may decrease as information particles are processed one by one. In order to ensure the delay requirements of each information particles and fully utilize the access bandwidth of the system, it is necessary to design an information particle group access algorithm which can dynamically adjust the minimum reachable access bandwidth (see **Algorithm VII.1**).

### *C. Dynamic access algorithm of an information particle group*

**Algorithm VII.1: Particle access algorithm based on dynamically adjusting the minimum reachable access bandwidth.**

1: Update time $t$ to the current time instant;

2: Initialize the set $Q$ ($Q$ is the set of information particles to be processed (i.e. information particle group)) to include only the information particles to be processed at the current time instant;

3: WHILE ($Q \neq \emptyset$) DO

4: Refresh the attributes (i.e. bearing capacity and deadline) of all the information particles contained in $Q$ relative to the current time instant $t$;

5: Perform incremental sorting operation $Op_{Inc}(Q)$ on $Q$;

6: Based on **Theorem VII.1**, the minimum reachable access bandwidth of the current information particle group $Q$ is calculated and updated as $B_{re,Q}^{min}$;

7: Based on **Definition IV.10** and **Definition IV.15**, calculate and update the inflection point of the access bandwidth requirement of the current information particle group $Q$ as $N_{cr,Q}$;

8: Take $B_{re,Q}^{min}$ as the access bandwidth, the first $N_{cr,Q}$ information particles in $Q$ are processed successively according to the EDF processing strategy (once the information particles are completely processed, they will be deleted from the set $Q$);

9: Update time instant as $t \Leftarrow t + t_{e,N_{cr,Q}}$;

10: END WHILE

**Note:**

1) If new information particles arrive during the procedure of the processing, the current processing operation should be interrupted immediately, and the algorithm is executed again from **step 1**.

*D. Some properties of the minimum reachable access bandwidths of one or more information particle groups*

**Corollary VII.1.7: The upper bound of the minimum reachable access bandwidth $B_{re,Q}^{min}$ of an information particle group.**

**Given an information particle group $Q$ containing $N\ (\geq 1)$ information particles, let its minimum reachable access bandwidth be $B_{re,Q}^{min}$. The average access bandwidth requirement of the information particle $i\ (1 \leq i \leq N)$ contained in $Q$ is $B_{av,i} \triangleq \ell_i / L_{D_i^{int}}$ (Definition III.10), then there is**

$$B_{re,Q}^{min} \leq \sum_{i=1}^{N} B_{av,i}$$

**(VII.23)**

**When $L_{D_1^{int}} = L_{D_2^{int}} = \ldots = L_{D_N^{int}}$, the equation holds.**

**Proof:**

It is proved by mathematical induction.

As for the case of $N = 1$, the information particle group $Q$ contains only one information particle, that is, the information particle $i\ (i = 1)$. According to the definition of the peak average access bandwidth requirement of an information particle group (**Definition IV.9**), we have $B_{av,Q}^{pk} = B_{av,i}\ (i = 1)$. Furthermore, according to **Theorem V.1**, we have $B_{re,Q}^{min} = B_{av,Q}^{pk}$. Based on the above, it can be obtained that $B_{re,Q}^{min} = B_{av,i}\ (i = 1)$. Therefore, when $N = 1$, the proposition holds (i.e. the equation holds).

As for the case of $N = n\ (n \geq 1)$, it is assumed that the proposition holds (i.e. $B_{re,Q_n}^{min} \leq \sum_{i=1}^{n} B_{av,i}$ holds). In the following, let us consider the case of $N = n + 1$. Without loss of generality, firstly an incremental sorting operation $Op_{Inc}(Q)$ is performed on $Q$ (Note: performing an incremental sorting operation will not change the minimum reachable access bandwidth of the corresponding information particle group). The sorted information particle group $Q$ still contains exactly the same $n + 1$ information particles, just the sequence numbers of each information particle in it may differ before and after being sorted. Let $B_{re,Q_{n+1}}^{min}$, $B_{av,Q_{n+1}}^{pk}$, and $B_{av,Q_{n+1}}$ represent the minimum reachable access bandwidth, the peak average access bandwidth requirement, and the average access bandwidth requirement of an information particle subgroup $Q_{n+1}$ (**Definition IV.17**) of $Q$, respectively. According to **Equation VII.5** and **Theorem V.1**, we have

$$\begin{aligned} B^{pk}_{av,Q_{n+1}} &= max\{B^{pk}_{av,Q_n}, B_{av,Q_{n+1}}\} \\ \Rightarrow \quad B^{min}_{re,Q_{n+1}} &= max\{B^{min}_{re,Q_n}, B_{av,Q_{n+1}}\} \end{aligned}$$

(VII.24.1)

which can be further expressed as

$$B^{min}_{re,Q_{n+1}} = \delta_{n,Q} B^{min}_{re,Q_n} + (1 - \delta_{n,Q}) B_{av,Q_{n+1}}$$

(VII.24.2)

where $\delta_{n,Q}$ is the switching function (**Definition IV.19**) of $Q$ with a sequence number $n$.

Below, we will discuss the cases of $\delta_{n,Q} = 1$ and $\delta_{n,Q} = 0$, respectively.

When $\delta_{n,Q} = 1$ (i.e. $B^{min}_{re,Q_n} \geq B_{av,Q_{n+1}}$), due to the assumption that $B^{min}_{re,Q_n} \leq \sum_{i=1}^{n} B_{av,i}$ holds, there is

$$B^{min}_{re,Q_{n+1}} = B^{min}_{re,Q_n} \leq \sum_{i=1}^{n} B_{av,i} < \sum_{i=1}^{n+1} B_{av,i}$$

(VII.25)

It shows that when $\delta_{n,Q} = 1$, the proposition holds.

As for the case of $\delta_{n,Q} = 0$ (i.e. $B^{min}_{re,Q_n} < B_{av,Q_{n+1}}$), we have

$$\begin{aligned} B^{min}_{re,Q_{n+1}} &= B_{av,Q_{n+1}} = \frac{\ell_{Q_{n+1}}}{L_{D^{int}_{Q_{n+1}}}} \\ &= \frac{\sum_{i=1}^{n} \ell_i + \ell_{n+1}}{L_{D^{int}_{n+1}}} \\ &= \frac{\sum_{i=1}^{n} \ell_i}{L_{D^{int}_{n+1}}} + B_{av,n+1} \\ &\leq \sum_{i=1}^{n} \frac{\ell_i}{L_{D^{int}_i}} + B_{av,n+1} \\ &= \sum_{i=1}^{n+1} B_{av,i} \end{aligned}$$

(VII.26)

When $L_{D^{int}_1} = L_{D^{int}_2} = \ldots = L_{D^{int}_n} = L_{D^{int}_{n+1}}$, the equality in **Equation VII.26** holds. This indicates that for the case of $\delta_{n,Q} = 0$, the proposition also holds.

Based on the above, it can be obtained that the proposition holds for the case of $N = n + 1$. ■

**Note:**

1) According to the above **Corollary VII.1.7**, when the deadlines of all the $N\ (\geq 1)$ information particles contained in an information particle group $Q$ are equal, the minimum reachable access bandwidth can be quickly calculated out by using equation $B_{re,Q}^{min} = \sum_{i=1}^{N} B_{av,i}$.

**Corollary VII.1.8: The lower bound of the minimum reachable access bandwidth $B_{re,Q}^{min}$ of an information particle group.**

**Given an information particle group $Q$ containing $N\ (\geq 2)$ information particles, let its minimum reachable access bandwidth be $B_{re,Q}^{min}$. The average access bandwidth requirement of the information particle $i\ (1 \leq i \leq N)$ contained in $Q$ is $B_{av,i} \triangleq \ell_i / L_{D_i^{int}}$ (Definition III.10), then there is**

$$B_{re,Q}^{min} > \frac{\sum_{i=1}^{N} B_{av,i}}{N}$$

**(VII.27)**

**Proof:**

It is proved by mathematical induction.

As for the case of $N = 2$, the information particle group $Q$ contains only two information particles. Without loss of generality, firstly an incremental sorting operation $Op_{Inc}(Q)$ is performed on $Q$ (Note: performing an incremental sorting operation will not change the minimum reachable access bandwidth of the corresponding information particle group). According to **Equation VII.24.2**, we have

$$B_{re,Q_2}^{min} = \delta_{1,Q} B_{re,Q_1}^{min} + (1 - \delta_{1,Q}) B_{av,Q_2}$$

(VII.28)

Where $\delta_{1,Q}$ is the switching function (**Definition IV.19**) of $Q$ with a sequence number 1. Below, we will discuss the cases of $\delta_{1,Q} = 1$ and $\delta_{1,Q} = 0$, respectively.

When $\delta_{n,Q} = 1$ (i.e. $B_{re,Q_1}^{min} \geq B_{av,Q_2}$), there is

$$B_{re,Q_1}^{min} = B_{av,1} \geq B_{av,Q_2} > B_{av,2}$$

(VII.29)

And then, there is

$$
\begin{aligned}
2B_{re,Q_2}^{min} &= 2B_{re,Q_1}^{min} = 2B_{av,1} \\
&= B_{av,1} + B_{av,1} \\
&> B_{av,1} + B_{av,2}
\end{aligned}
$$

(VII.30)

Therefore, it can be obtained that $B_{re,Q_2}^{min} > \frac{B_{av,1}+B_{av,2}}{2}$ holds.

When $\delta_{1,Q} = 0$ (i.e. $B_{re,Q_1}^{min} < B_{av,Q_2}$), there is

$$
B_{re,Q_1}^{min} = B_{av,1} < B_{av,Q_2}
$$

(VII.31)

And then, there is

$$
\begin{aligned}
2B_{re,Q_2}^{min} &= 2B_{av,Q_2} = 2\frac{\ell_1 + \ell_2}{L_{D_{Q_2}^{int}}} \\
&= \frac{\ell_1 + \ell_2}{L_{D_{Q_2}^{int}}} + \frac{\ell_1 + \ell_2}{L_{D_2^{int}}} \\
&> \frac{\ell_1 + \ell_2}{L_{D_{Q_2}^{int}}} + B_{av,2} \\
&> B_{av,1} + B_{av,2}
\end{aligned}
$$

(VII.32)

Therefore, it can be obtained that $B_{re,Q_2}^{min} > \frac{B_{av,1}+B_{av,2}}{2}$ also holds in this case. Based on the above, it can be obtained that $B_{re,Q}^{min} > \frac{B_{av,1}+B_{av,2}}{2}$ holds for the case of $N = 2$.

As for the case of $N = n\ (n \geq 2)$, it is assumed that the proposition holds (i.e. $B_{re,Q}^{min} > \frac{\sum_{i=1}^{n} B_{av,i}}{n}$ holds). In the following, let us consider the case of $N = n + 1$. Without loss of generality, firstly an incremental sorting operation $Op_{Inc}(Q)$ is performed on $Q$ (Note: performing an incremental sorting operation will not change the minimum reachable access bandwidth of the corresponding information particle group). The sorted information particle group $Q$ still contains exactly the same $n + 1$ information particles, just the sequence numbers of each information particle in it may differ before and after being sorted. According to **Equation VII.24.2**, we have $B_{re,Q_{n+1}}^{min} = \delta_{n,Q} B_{re,Q_n}^{min} + (1 - \delta_{n,Q}) B_{av,Q_{n+1}}$, where $\delta_{n,Q}$ is the switching function (**Definition IV.19**) of $Q$ with a sequence number $n$.

Below, we will discuss the cases of $\delta_{n,Q} = 1$ and $\delta_{n,Q} = 0$, respectively.

When $\delta_{n,Q} = 1$ (i.e. $B_{re,Q_n}^{min} \geq B_{av,Q_{n+1}}$), there is

$$
\begin{aligned}
(n+1)B_{re,Q_{n+1}}^{min} &= (n+1)B_{re,Q_n}^{min} \\
&= nB_{re,Q_n}^{min} + B_{re,Q_n}^{min} \\
&> \sum_{i=1}^{n} B_{av,i} + B_{re,Q_n}^{min} \\
&\geq \sum_{i=1}^{n} B_{av,i} + B_{av,Q_{n+1}} \\
&= \sum_{i=1}^{n} B_{av,i} + \frac{\ell_{Q_{n+1}}}{L_{D_{Q_{n+1}}^{int}}} \\
&= \sum_{i=1}^{n} B_{av,i} + \frac{\sum_{i=1}^{n+1} \ell_i}{L_{D_{n+1}^{int}}} \\
&> \sum_{i=1}^{n+1} B_{av,i}
\end{aligned}
$$

(VII.33)

It shows that when $\delta_{n,Q} = 1$, $B_{re,Q_{n+1}}^{min} > \frac{\sum_{i=1}^{n+1} B_{av,i}}{n+1}$ holds.

As for the case of $\delta_{n,Q} = 0$ (i.e. $B_{re,Q_n}^{min} < B_{av,Q_{n+1}}$), we have

$$
\begin{aligned}
(n+1)B_{re,Q_{n+1}}^{min} &= (n+1)B_{av,Q_{n+1}} \\
&= nB_{av,Q_{n+1}} + B_{av,Q_{n+1}} \\
&> nB_{re,Q_n}^{min} + B_{av,Q_{n+1}} \\
&> nB_{re,Q_n}^{min} + B_{av,n+1} \\
&> \sum_{i=1}^{n} B_{av,i} + B_{av,n+1} \\
&= \sum_{i=1}^{n+1} B_{av,i}
\end{aligned}
$$

(VII.34)

It shows that when $\delta_{n,Q} = 0$, $B_{re,Q_{n+1}}^{min} > \frac{\sum_{i=1}^{n+1} B_{av,i}}{n+1}$ also holds.

Based on the above, it can be obtained that the proposition holds for the case of $N = n + 1$. ■

**Note:**

1) According to the above **Corollary VII.1.8**, for an information particle group $Q$ containing $N\ (\geq 2)$ information particles, an access bandwidth constraint of less than or equal to $\frac{\sum_{i=1}^{N} B_{av,i}}{N}$ must not be reachable. Therefore, this conclusion can be used to quickly determine the non-reachability of a given access bandwidth constraint.

**Corollary VII.1.9: Linear superposition characteristics of multiple information particle groups under specific constraints.**

**Given $M+1\,(M \geq 1)$ information particle groups $Q^i\,(1 \leq i \leq M+1)$, each of which contains $N\,(\geq 1)$ information particles and has been sorted by performing an incremental sorting operation $Op_{Inc}(Q^i)\,(1 \leq i \leq M+1)$. The deadlines of information particles in $Q^i\,(1 \leq i \leq M+1)$ with their sequence numbers being $j\,(1 \leq j \leq N)$ are all equal to $t_{e,j}$. And the bearing capacities of information particles in $Q^i\,(1 \leq i \leq M+1)$ with their sequence numbers being $j\,(1 \leq j \leq N)$ are $\ell_{j,Q^i}\,(1 \leq i \leq M+1)$, respectively, and satisfy the relationship**

$$\ell_{j,Q^{M+1}} = \sum_{i=1}^{M} \alpha_i \ell_{j,Q^i} \;\; (\alpha_i > 0)$$

**(VII.35)**

**where $\alpha_i\,(1 \leq i \leq M+1)$ are given positive real constants. $B^{min}_{re,Q^i_j}\,(1 \leq i \leq M+1; 1 \leq j \leq N)$ is the minimum reachable access bandwidth of the information particle subgroup $Q^i_j$ of $Q^i$ (according to Definition IV.17, $Q^i_j$ is the information particle subgroup composed of all the information particles in $Q^i$ whose sorting sequence numbers are not greater than $j$). The sequence of switching functions of the information particle group $Q^i\,(1 \leq i \leq M+1)$ is denoted as $\delta_{n,Q^i}\,(1 \leq i \leq M+1; 1 \leq n \leq N-1)$ (Definition IV.19), and it is assumed that the following relationship holds.**

$$\delta_{n,Q^1} = \delta_{n,Q^2} = \ldots = \delta_{n,Q^M} \;\; (1 \leq n \leq N-1)$$

**(VII.36)**

**Then, the following relationships must hold, which are**

$$B^{min}_{re,Q^{M+1}_j} = \sum_{i=1}^{M} \alpha_i B^{min}_{re,Q^i_j} \;\; (1 \leq j \leq N)$$

**(VII.37)**

**and**

$$\delta_{n,Q^1} = \delta_{n,Q^2} = \ldots = \delta_{n,Q^M} = \delta_{n,Q^{M+1}} \;\; (1 \leq n \leq N-1)$$

**(VII.38)**

**Proof:**

It is proved by mathematical induction.

Firstly, let us consider the case of $j = 1$. According to **Theorem V.1**, we have

$$\begin{aligned}
B^{min}_{re,Q_1^{M+1}} &= B^{pk}_{av,Q_1^{M+1}} = B_{av,Q_1^{M+1}} \\
&= \frac{\ell_{Q_1^{M+1}}}{L_{D^{int}_{Q_1^{M+1}}}} = \frac{\ell_{1,Q^{M+1}}}{t_{e,1}} \\
&= \frac{\sum_{i=1}^{M} \alpha_i \ell_{1,Q^i}}{t_{e,1}} \\
&= \sum_{i=1}^{M} \alpha_i \left(\frac{\ell_{1,Q^i}}{t_{e,1}}\right) \\
&= \sum_{i=1}^{M} \alpha_i B_{av,Q_1^i} \\
&= \sum_{i=1}^{M} \alpha_i B^{pk}_{av,Q_1^i} \\
&= \sum_{i=1}^{M} \alpha_i B^{min}_{re,Q_1^i}
\end{aligned}$$

(VII.39)

Therefore, as for the case of $j = 1$, it can be obtained that $B^{min}_{re,Q_1^{M+1}} = \sum_{i=1}^{M} \alpha_i B^{min}_{re,Q_1^i}$ holds.

As for the case of $j = n$ ($1 \le n < N$), it is assumed that the proposition holds (i.e. $B^{min}_{re,Q_n^{M+1}} = \sum_{i=1}^{M} \alpha_i B^{min}_{re,Q_n^i}$ holds). In the following, let us consider the case of $j = n + 1$. According to **Equation VII.24.2**, we have

$$B^{min}_{re,Q_{n+1}^{M+1}} = \delta_{n,Q^{M+1}} B^{min}_{re,Q_n^{M+1}} + (1 - \delta_{n,Q^{M+1}}) B_{av,Q_{n+1}^{M+1}}$$

(VII.40)

According to the question setting of $\delta_{n,Q^1} = \delta_{n,Q^2} = \ldots = \delta_{n,Q^M}$, we can discuss the cases of $\delta_{n,Q^1} = \delta_{n,Q^2} = \ldots = \delta_{n,Q^M} = 1$ and $\delta_{n,Q^1} = \delta_{n,Q^2} = \ldots = \delta_{n,Q^M} = 0$, respectively.

When $\delta_{n,Q^1} = \delta_{n,Q^2} = \ldots = \delta_{n,Q^M} = 1$ (i.e. $B^{min}_{re,Q_n^i} \ge B_{av,Q_{n+1}^i}$ ($1 \le i \le M$)), there is

$$\sum_{i=1}^{M} \alpha_i B^{min}_{re,Q_n^i} \ge \sum_{i=1}^{M} \alpha_i B_{av,Q_{n+1}^i}$$

(VII.41)

where, according to the above assumption, there is

$$B^{min}_{re,Q^{M+1}_n} = \sum_{i=1}^{M} \alpha_i B^{min}_{re,Q^i_n}$$

(VII.42)

And then, there is

$$\begin{aligned}\sum_{i=1}^{M} \alpha_i B_{av,Q^i_{n+1}} &= \sum_{i=1}^{M} \left( \alpha_i \frac{\ell_{Q^i_{n+1}}}{L_{D^{int}_{Q^i_{n+1}}}} \right) \\ &= \frac{\sum_{i=1}^{M} \left( \alpha_i \sum_{j=1}^{n+1} \ell_{j,Q^i} \right)}{t_{e,n+1}} \\ &= \frac{\sum_{j=1}^{n+1} \left( \sum_{i=1}^{M} \alpha_i \ell_{j,Q^i} \right)}{t_{e,n+1}} \\ &= \frac{\sum_{j=1}^{n+1} \ell_{j,Q^{M+1}}}{t_{e,n+1}} \\ &= B_{av,Q^{M+1}_{n+1}}\end{aligned}$$

(VII.43)

Substituting **Equation VII.42** and **Equation VII.43** into **Equation VII.41** yields

$$B^{min}_{re,Q^{M+1}_n} \geq B_{av,Q^{M+1}_{n+1}}$$

(VII.44)

Considering the definition of a switching function $\delta_{n,Q^{M+1}}$ (**Definition IV.19**), that is

$$\delta_{n,Q^{M+1}} \triangleq \begin{cases} 1 & B^{min}_{re,Q^{M+1}_n} \geq B_{av,Q^{M+1}_{n+1}} \\ 0 & B^{min}_{re,Q^{M+1}_n} < B_{av,Q^{M+1}_{n+1}} \end{cases}$$

(VII.45)

Therefore, according to the above two equations, it can be obtained that $\delta_{n,Q^{M+1}} = 1$, i.e. $\delta_{n,Q^1}$= $\delta_{n,Q^2}$=...=$\delta_{n,Q^M} = \delta_{n,Q^{M+1}}$ holds. By combining **Equation VII.40** and **Equation VII.42**, it can be obtained that

$$B^{min}_{re,Q^{M+1}_{n+1}} = B^{min}_{re,Q^{M+1}_n} = \sum_{i=1}^{M} \alpha_i B^{min}_{re,Q^i_n}$$

(VII.46)

As for the case of $\delta_{n,Q^1}$= $\delta_{n,Q^2}$=...=$\delta_{n,Q^M} = 0$ (i.e. $B^{min}_{re,Q^i_n} < B_{av,Q^i_{n+1}}$ $(1 \leq i \leq M)$), there is

$$\sum_{i=1}^{M} \alpha_i B^{min}_{re,Q^i_n} < \sum_{i=1}^{M} \alpha_i B_{av,Q^i_{n+1}}$$

(VII.47)

Substituting **Equation VII.42** and **Equation VII.43** into the above equation, there is

$$B_{re,Q_n^{M+1}}^{min} < B_{av,Q_{n+1}^{M+1}}$$

(VII.48)

Considering the definition of a switching function $\delta_{n,Q^{M+1}}$, it can be concluded that $\delta_{n,Q^{M+1}} = 0$, i.e. $\delta_{n,Q^1}$= $\delta_{n,Q^2}$=...=$\delta_{n,Q^M} = \delta_{n,Q^{M+1}}$ holds. By combining **Equation VII.40** and **Equation VII.43**, there is

$$B_{re,Q_{n+1}^{M+1}}^{min} = B_{av,Q_{n+1}^{M+1}} = \sum_{i=1}^{M} \alpha_i B_{av,Q_{n+1}^i}$$

(VII.49)

Based on the above analyses of the case of $\delta_{n,Q^1}$= $\delta_{n,Q^2}$=...=$\delta_{n,Q^M} = 1$ (**Equation VII.46**) and that of $\delta_{n,Q^1}$= $\delta_{n,Q^2}$=...=$\delta_{n,Q^M} = 0$ (**Equation VII.49**), it can be concluded that the following relationship holds

$$B_{re,Q_{n+1}^{M+1}}^{min} = \begin{cases} \sum_{i=1}^{M} \alpha_i B_{re,Q_n^i}^{min} & (\delta_{n,Q^1}=\delta_{n,Q^2}=...=\delta_{n,Q^M} = \delta_{n,Q^{M+1}} = 1) \\ \sum_{i=1}^{M} \alpha_i B_{av,Q_{n+1}^i} & (\delta_{n,Q^1}=\delta_{n,Q^2}=...=\delta_{n,Q^M} = \delta_{n,Q^{M+1}} = 0) \end{cases}$$

(VII.50)

The above equation can be further written as follows

$$B_{re,Q_{n+1}^{M+1}}^{min} = \begin{cases} \sum_{i=1}^{M} \alpha_i B_{re,Q_{n+1}^i}^{min} & (\delta_{n,Q^1}=\delta_{n,Q^2}=...=\delta_{n,Q^M} = \delta_{n,Q^{M+1}} = 1) \\ \sum_{i=1}^{M} \alpha_i B_{re,Q_{n+1}^i}^{min} & (\delta_{n,Q^1}=\delta_{n,Q^2}=...=\delta_{n,Q^M} = \delta_{n,Q^{M+1}} = 0) \end{cases}$$

(VII.51)

Based on the above, it can be obtained that $B_{re,Q_{n+1}^{M+1}}^{min} = \sum_{i=1}^{M} \alpha_i B_{re,Q_{n+1}^i}^{min}$ holds for the case of $j = n + 1$. ■

**Note:**

1) In the above **Corollary VII.1.9**, the information particle group $Q^{M+1}$ can be regarded as a kind of linear superposition of $M$ $(M \geq 1)$ information particle groups $Q^i$ $(1 \leq i \leq M)$ with similar structures (i.e. containing the same number of information particles, information particles contained having the same distribution of deadlines, and each information particle group having the same sequence of switching functions, etc.). **Corollary VII.1.9** further reveals that there exisit a simple linear relationship between the minimum reachable access bandwidth of information particle group $Q^{M+1}$ and the minimum reachable access bandwidths of information particle groups $Q^i$ $(1 \le i \le M)$;

2) Furthermore, it can be observed that the linear superposition of information particle groups with similar structures does not result in further compression of the demand for bandwidth resources.

**Corollary VII.1.10: Linear splitting characteristics of an information particle group.**

**Given $M+1$ $(M \ge 1)$ information particle groups $Q^i$ $(1 \le i \le M+1)$, each of which contains $N\,(\ge 1)$ information particles and has been sorted by performing an incremental sorting operation $Op_{Inc}(Q^i)$ $(1 \le i \le M+1)$. The deadlines of information particles in $Q^i$ $(1 \le i \le M+1)$ with their sequence numbers being $j$ $(1 \le j \le N)$ are all equal to $t_{e,j}$. And the bearing capacities of information particles in $Q^i$ $(1 \le i \le M+1)$ with their sequence numbers being $j$ $(1 \le j \le N)$ are $\ell_{j,Q^i}$ $(1 \le i \le M+1)$, respectively, and satisfy the relationship**

$$\ell_{j,Q^i} = \alpha_i \ell_{j,Q^{M+1}} \; (\alpha_i > 0, 1 \le i \le M)$$

**(VII.52)**

**where $\alpha_i$ $(1 \le i \le M)$ are given positive real constants. $B^{min}_{re,Q^i_j}$ $(1 \le i \le M+1; 1 \le j \le N)$ is the minimum reachable access bandwidth of the information particle subgroup $Q^i_j$ of $Q^i$. The sequence of switching functions of the information particle group $Q^i$ $(1 \le i \le M+1)$ is denoted as $\delta_{n,Q^i}$ $(1 \le i \le M+1; 1 \le n \le N-1)$ (Definition IV.19). Then, it can be concluded that the following relationships must hold**

$$B^{min}_{re,Q^i_j} = \alpha_i B^{min}_{re,Q^{M+1}_j} \; (1 \le i \le M, 1 \le j \le N)$$

**(VII.53)**

**and**

$$\delta_{n,Q^1} = \delta_{n,Q^2} = \ldots = \delta_{n,Q^{M+1}} \; (1 \le n \le N-1)$$

**(VII.54)**

**Proof:**

It is proved by mathematical induction.

Firstly, let us consider the case of $j = 1$. According to **Theorem V.1**, we have

$$\begin{aligned} B^{min}_{re,Q^i_1} &= \frac{\ell_{1,Q^i}}{t_{e,1}} \\ &= \frac{\alpha_i \ell_{1,Q^{M+1}}}{t_{e,1}} \\ &= \alpha_i B_{av,Q^{M+1}_1} \\ &= \alpha_i B^{pk}_{av,Q^{M+1}_1} \\ &= \alpha_i B^{min}_{re,Q^{M+1}_1} \quad (1 \le i \le M) \end{aligned}$$

(VII.55)

Therefore, as for the case of $j = 1$, it can be obtained that $B^{min}_{re,Q^i_1} = \alpha_i B^{min}_{re,Q^{M+1}_1}$ $(1 \le i \le M)$ holds.

As for the case of $j = n$ $(1 \le n < N)$, it is assumed that the proposition holds (i.e. $B^{min}_{re,Q^i_n} = \alpha_i B^{min}_{re,Q^{M+1}_n}$ $(1 \le i \le M)$ holds). In the following, let us consider the case of $j = n + 1$. We can discuss the cases of $\delta_{n,Q^{M+1}} = 1$ and $\delta_{n,Q^{M+1}} = 0$, respectively.

When $\delta_{n,Q^{M+1}} = 1$, there is

$$B^{min}_{re,Q^{M+1}_n} \ge B_{av,Q^{M+1}_{n+1}}$$

(VII.56)

where, according to the above assumption, there is

$$B^{min}_{re,Q^{M+1}_n} = \frac{B^{min}_{re,Q^i_n}}{\alpha_i} \ (1 \le i \le M)$$

(VII.57)

and

$$\begin{aligned} B_{av,Q^{M+1}_{n+1}} &= \frac{\sum_{j=1}^{n+1} \ell_{j,Q^{M+1}}}{t_{e,n+1}} \\ &= \frac{\sum_{j=1}^{n+1} \frac{\ell_{j,Q^i}}{\alpha_i}}{t_{e,n+1}} \\ &= \frac{B_{av,Q^i_{n+1}}}{\alpha_i} \qquad (1 \le i \le M) \end{aligned}$$

(VII.58)

Substituting the above two equations into **Equation VII.56** yields

$$B^{min}_{re,Q^i_n} \geq B_{av,Q^i_{n+1}} \; (1 \leq i \leq M)$$

(VII.59)

Combining with the definition of a switching function $\delta_{n,Q^i}$ (**Definition IV.19**), it can be obtained that $\delta_{n,Q^i} = 1 \; (1 \leq i \leq M)$, i.e. $\delta_{n,Q^1} = \delta_{n,Q^2} = \ldots = \delta_{n,Q^M} = \delta_{n,Q^{M+1}}$ holds. According to **Equation VII.24.2**, we have

$$B^{min}_{re,Q^i_{n+1}} = \delta_{n,Q^i} B^{min}_{re,Q^i_n} + (1 - \delta_{n,Q^i}) B_{av,Q^i_{n+1}}$$

(VII.60)

Since $\delta_{n,Q^i} = 1 \; (1 \leq i \leq M)$, it can be obtained that

$$\begin{aligned} B^{min}_{re,Q^i_{n+1}} &= B^{min}_{re,Q^i_n} \\ &= \alpha_i B^{min}_{re,Q^{M+1}_n} \\ &= \alpha_i B^{min}_{re,Q^{M+1}_{n+1}} \quad (1 \leq i \leq M) \end{aligned}$$

(VII.61)

For the case of $\delta_{n,Q^{M+1}} = 0$, there is

$$B^{min}_{re,Q^{M+1}_n} < B_{av,Q^{M+1}_{n+1}}$$

(VII.62)

Similarly, it can be concluded that

$$B^{min}_{re,Q^i_n} < B_{av,Q^i_{n+1}} \; (1 \leq i \leq M)$$

(VII.63)

Combining with the definition of a switching function $\delta_{n,Q^i}$, it can be obtained that $\delta_{n,Q^i} = 0 \; (1 \leq i \leq M)$, i.e. $\delta_{n,Q^1} = \delta_{n,Q^2} = \ldots = \delta_{n,Q^M} = \delta_{n,Q^{M+1}}$ also holds in this case. Similarly, from **Equation VII.24.2** and **Equation VII.58**, it can be obtained that

$$\begin{aligned} B^{min}_{re,Q^i_{n+1}} &= B_{av,Q^i_{n+1}} \\ &= \alpha_i B_{av,Q^{M+1}_{n+1}} \\ &= \alpha_i B^{min}_{re,Q^{M+1}_{n+1}} \quad (1 \leq i \leq M) \end{aligned}$$

(VII.64)

Based on the above analyses (**Equation VII.61** and **Equation VII.64**), it can be concluded that both $B^{min}_{re,Q^i_{n+1}} = \alpha_i B^{min}_{re,Q^{M+1}_{n+1}} \; (1 \leq i \leq M)$ and $\delta_{n,Q^1} = \delta_{n,Q^2} = \ldots = \delta_{n,Q^M} = \delta_{n,Q^{M+1}}$ hold for the case of $j = n + 1$. ■

**Note:**

1) In the above **Corollary VII.1.10**, if $\sum_{i=1}^{M}\alpha_i = 1$, it can be regarded as linearly splitting the information particle group $Q^{M+1}$ into $M$ $(M \geq 1)$ information particle groups $Q^i$ $(1 \leq i \leq M)$ with similar structures (i.e. containing the same number of information particles, information particles contained having the same distribution of deadlines, and each information particle group having the same sequence of switching functions, etc.), and at the same time maintaining the total bandwidth requirement unchanged, i.e. $B_{re,Q^{M+1}}^{min} = \sum_{i=1}^{M} B_{re,Q^i}^{min}$. This provides a feasible approach for dividing an information particle group with a high demand of bandwidth resources into several information particle groups each with lower demands for bandwidth resources, and then by simultaneously using multiple information particle group processing units each with smaller bandwidths to bear the splitted information particle group.

**Corollary VII.1.11: A closed form expression of the minimum reachable access bandwidth of an information particle subgroup.**

**Given an information particle group $\boldsymbol{Q}$, which contains $\boldsymbol{N}\,(\geq 2)$ information particles and has been sorted by performing an incremental sorting operation $\boldsymbol{Op_{Inc}(Q)}$. The vector of switching functions (Definition IV.20) and the vector of average access bandwidth requirements (Definition IV.21) of an information particle subgroup $\boldsymbol{Q_n}$ $(\boldsymbol{Q_n} \subseteq \boldsymbol{Q}, 2 \leq \boldsymbol{n} \leq \boldsymbol{N})$ are $\boldsymbol{\delta_{Q_n}}$ and $\boldsymbol{B_{av,Q_n}}$, respectively. Then, it can be concluded that the following relationship must hold**

$$\boldsymbol{B_{re,Q_n}^{min} = \delta_{Q_n}{}^{T} \cdot B_{av,Q_n}} \tag{VII.65}$$

**Proof:**

As for the case of $n = 2$, according to **Equation VII.24.2**, **Definition IV.20**, and **Definition IV.21**, we have

$$
\begin{aligned}
B_{re,Q_2}^{min} &= \delta_{1,Q} B_{re,Q_1}^{min} + (1-\delta_{1,Q}) B_{av,Q_2} \\
&= \left(\delta_{1,Q}, 1-\delta_{1,Q}\right) \begin{pmatrix} B_{re,Q_1}^{min} \\ B_{av,Q_2} \end{pmatrix} \\
&= \left(\delta_{1,Q}, 1-\delta_{1,Q}\right) \begin{pmatrix} B_{av,Q_1} \\ B_{av,Q_2} \end{pmatrix} \\
&= \left(\delta_{1,Q_2}, 1-\delta_{1,Q_2}\right) \begin{pmatrix} B_{av,Q_1} \\ B_{av,Q_2} \end{pmatrix} \\
&= \boldsymbol{\delta}_{Q_2}{}^{T} \cdot \boldsymbol{B}_{av,Q_2}
\end{aligned}
\tag{VII.66}
$$

Therefore, it can be seen that for the case of $n = 2$, $B_{re,Q_2}^{min} = \boldsymbol{\delta}_{Q_2}{}^{T} \cdot \boldsymbol{B}_{av,Q_2}$ holds.

For the case of $n > 2$, mathematical induction is used. Firstly, let us prove the case of $n = 3$. According to **Equation VII.24.2**, there is

$$
\begin{aligned}
B_{re,Q_3}^{min} &= \delta_{2,Q} B_{re,Q_2}^{min} + (1-\delta_{2,Q}) B_{av,Q_3} \\
&= \delta_{2,Q}\left(\delta_{1,Q} B_{re,Q_1}^{min} + (1-\delta_{1,Q}) B_{av,Q_2}\right) + (1-\delta_{2,Q}) B_{av,Q_3} \\
&= \delta_{1,Q}\delta_{2,Q} B_{re,Q_1}^{min} + (1-\delta_{1,Q})\delta_{2,Q} B_{av,Q_2} + (1-\delta_{2,Q}) B_{av,Q_3} \\
&= \delta_{1,Q}\delta_{2,Q} B_{av,Q_1} + (1-\delta_{1,Q})\delta_{2,Q} B_{av,Q_2} + (1-\delta_{2,Q}) B_{av,Q_3} \\
&= \left(\delta_{1,Q_3}\delta_{2,Q_3}, (1-\delta_{1,Q_3})\delta_{2,Q_3}, 1-\delta_{2,Q_3}\right) \begin{pmatrix} B_{av,Q_1} \\ B_{av,Q_2} \\ B_{av,Q_3} \end{pmatrix} \\
&= \boldsymbol{\delta}_{Q_3}{}^{T} \cdot \boldsymbol{B}_{av,Q_3}
\end{aligned}
\tag{VII.67}
$$

Therefore, it can be seen that for the case of $n = 3$, $B_{re,Q_3}^{min} = \boldsymbol{\delta}_{Q_3}{}^{T} \cdot \boldsymbol{B}_{av,Q_3}$ holds.

Assuming that when $n = k\ (3 \le k < N)$, $B_{re,Q_k}^{min} = \boldsymbol{\delta}_{Q_k}{}^{T} \cdot \boldsymbol{B}_{av,Q_k}$ holds. Let us consider the case of $n = k + 1$. According to **Equation VII.24.2**, we have

$$
\begin{aligned}
B_{re,Q_{k+1}}^{min} &= \delta_{k,Q} B_{re,Q_k}^{min} + \left(1-\delta_{k,Q}\right) B_{av,Q_{k+1}} \\
&= \delta_{k,Q} \cdot \boldsymbol{\delta}_{Q_k}{}^{T} \cdot \boldsymbol{B}_{av,Q_k} + \left(1-\delta_{k,Q}\right) B_{av,Q_{k+1}} \\
&= \delta_{k,Q_{k+1}} \cdot \begin{pmatrix} \prod_{j=1}^{k-1} \delta_{j,Q_k}, \left(1-\delta_{1,Q_k}\right) \prod_{j=2}^{k-1} \delta_{j,Q_k}, \\ \dots, \left(1-\delta_{i-1,Q_k}\right) \prod_{j=i}^{k-1} \delta_{j,Q_k}, \dots, \\ (1-\delta_{k-2,Q_k}) \prod_{j=k-1}^{k-1} \delta_{j,Q_k}, 1-\delta_{k-1,Q_k} \end{pmatrix}_{1\times k} \cdot \boldsymbol{B}_{av,Q_k} \\
&\quad + \left(1-\delta_{k,Q_{k+1}}\right) B_{av,Q_{k+1}}
\end{aligned}
\tag{VII.68.1}
$$

Further derivation yields

$$B_{re,Q_{k+1}}^{min} = \delta_{k,Q_{k+1}} \cdot \begin{pmatrix} \prod_{j=1}^{k-1} \delta_{j,Q_{k+1}}, (1-\delta_{1,Q_{k+1}}) \prod_{j=2}^{k-1} \delta_{j,Q_{k+1}}, \\ \dots, (1-\delta_{i-1,Q_{k+1}}) \prod_{j=i}^{k-1} \delta_{j,Q_{k+1}}, \dots, \\ (1-\delta_{k-2,Q_{k+1}}) \prod_{j=k-1}^{k-1} \delta_{j,Q_{k+1}}, 1-\delta_{k-1,Q_{k+1}} \end{pmatrix}_{1\times k} \cdot \boldsymbol{B}_{av,Q_k}$$
$$+(1-\delta_{k,Q_{k+1}})B_{av,Q_{k+1}}$$

(VII.68.2)

And finally, we have

$$\begin{aligned} B_{re,Q_{k+1}}^{min} &= \begin{pmatrix} \prod_{j=1}^{k} \delta_{j,Q_{k+1}}, (1-\delta_{1,Q_{k+1}}) \prod_{j=2}^{k} \delta_{j,Q_{k+1}}, \\ \dots, (1-\delta_{i-1,Q_{k+1}}) \prod_{j=i}^{k} \delta_{j,Q_{k+1}}, \dots, \\ (1-\delta_{k-2,Q_{k+1}}) \prod_{j=k-1}^{k} \delta_{j,Q_{k+1}}, (1-\delta_{k-1,Q_{k+1}})\delta_{k,Q_{k+1}}, \\ 1-\delta_{k,Q_{k+1}} \end{pmatrix}_{1\times(k+1)} \cdot \boldsymbol{B}_{av,Q_{k+1}} \\ &= \boldsymbol{\delta}_{Q_{k+1}}{}^{T} \cdot \boldsymbol{B}_{av,Q_{k+1}} \end{aligned}$$

(VII.68.3)

Therefore, it can be seen that for the case of $n = k + 1$, $B_{re,Q_{k+1}}^{min} = \boldsymbol{\delta}_{Q_{k+1}}{}^{T} \cdot \boldsymbol{B}_{av,Q_{k+1}}$ also holds.

Based on the above mathematical induction, it can be concluded that $B_{re,Q_n}^{min} = \boldsymbol{\delta}_{Q_n}{}^{T} \cdot \boldsymbol{B}_{av,Q_n}$ holds for the case of $n > 2$. ■

**Note:**

1) In the above derivation, the relation $\delta_{i,Q_{i+1}} = \delta_{i,Q_{i+2}} = \cdots = \delta_{i,Q_{N-1}} = \delta_{i,Q_N} = \delta_{i,Q}$ is adopted for several times, where $Q_{i+1} \subseteq Q_{i+2} \subseteq \cdots \subseteq Q_{N-1} \subseteq Q_N = Q$ (see **Definition IV.19** and its note section);
2) The above **Corollary VII.1.11** reveals that the key factors of determining the minimum reachable access bandwidth of an information particle group are its vector of average access bandwidth requirements and its vector of switching functions. It is believed that this provides a useful guidance for further in-depth research on computations such as superpositions and splittings of information particle groups in the future.

**Corollary VII.1.12: The vector of switching functions of an information particle subgroup has and only has one component with its value being equal to one.**

**Given an information particle group $\boldsymbol{Q}$, which contains $\boldsymbol{N}\,(\geq \boldsymbol{2})$ information particles and has been sorted by performing an incremental sorting operation $\boldsymbol{Op_{Inc}(Q)}$. The vector of switching functions (Definition IV.20) of an information particle subgroup $\boldsymbol{Q_n}\ (\boldsymbol{Q_n} \subseteq \boldsymbol{Q}, \boldsymbol{2} \leq \boldsymbol{n} \leq \boldsymbol{N})$ is $\boldsymbol{\delta_{Q_n}}$. Then, it can be concluded that among $\boldsymbol{n}$ components of $\boldsymbol{\delta_{Q_n}}$, there is one and only one component with its value being equal to one.**

**Proof:**

As for the case of $n = 2$, according to **Definition IV.20**, we have

$$\boldsymbol{\delta}_{Q_2}{}^T = \left(\delta_{1,Q_2}, 1 - \delta_{1,Q_2}\right) \tag{VII.69}$$

According to **Definition IV.19**, the switching function $\delta_{1,Q_2}$ can only be taken as 0 or 1. It can be obtained that

$$\boldsymbol{\delta}_{Q_2}{}^T = \begin{cases} (1,0) & (\delta_{1,Q_2} = 1) \\ (0,1) & (\delta_{1,Q_2} = 0) \end{cases} \tag{VII.70}$$

Therefore, it can be seen that the proposition holds in the case of $n = 2$.

Next, let us consider the case of $n > 2$. In this case, according to **Definition IV.20**, the vector $\boldsymbol{\delta}_{Q_n}$ can be generally expressed as follows:

$$\boldsymbol{\delta}_{Q_n}{}^T = \left(\boldsymbol{\delta}_{Q_n}(1), \boldsymbol{\delta}_{Q_n}(2), \ldots, \boldsymbol{\delta}_{Q_n}(i), \ldots, \boldsymbol{\delta}_{Q_n}(n-1), \boldsymbol{\delta}_{Q_n}(n)\right)_{1\times n} \tag{VII.71.1}$$

where

$$\begin{cases} \boldsymbol{\delta}_{Q_n}(1) & \triangleq & \prod_{j=1}^{n-1} \delta_{j,Q_n} \\ \boldsymbol{\delta}_{Q_n}(i) & \triangleq & (1 - \delta_{i-1,Q_n}) \prod_{j=i}^{n-1} \delta_{j,Q_n}\ (2 \leq i \leq n-1) \\ \boldsymbol{\delta}_{Q_n}(n) & \triangleq & 1 - \delta_{n-1,Q_n} \end{cases} \tag{VII.71.2}$$

Firstly, let us prove that the $n$ components of $\boldsymbol{\delta}_{Q_n}$ cannot all be zero. Here, reduction to absurdity is adopted. Assuming that all the $n$ components of $\boldsymbol{\delta}_{Q_n}$ are 0, $\delta_{n-1,Q_n} = 1$ can be obtained from the premise that $\boldsymbol{\delta}_{Q_n}(n) = 0$. Then, from $\delta_{n-1,Q_n} = 1$ and $\boldsymbol{\delta}_{Q_n}(n-1) = 0$, $\delta_{n-2,Q_n} = 1$ can be derived. By repeating this reasoning process, we can obtain that $\delta_{j,Q_n} = 1\ (1 \leq j \leq n-1)$, from which $\boldsymbol{\delta}_{Q_n}(1) \triangleq \prod_{j=1}^{n-1} \delta_{j,Q_n} = 1$ can be derived. This contradicts the assumption that all the $n$ components of $\boldsymbol{\delta}_{Q_n}$ are 0. Therefore, it is impossible for all the $n$ components of $\boldsymbol{\delta}_{Q_n}$ being equal to 0.

Next, we will prove that only one term among $n$ components of $\boldsymbol{\delta}_{Q_n}$ can be equal to 1. Considering the fact that the $n$ components of $\boldsymbol{\delta}_{Q_n}$ cannot all be 0, there must be at least one component that is equal to 1. Below, we will consider three different cases one by one.

Firstly, assuming that $\boldsymbol{\delta}_{Q_n}(1) = 1$, $\delta_{j,Q_n} = 1\ (1 \leq j \leq n-1)$ is obtained, and then $\boldsymbol{\delta}_{Q_n}(i) = 0\ (2 \leq i \leq n)$ can be easily obtained (i.e. all the other $n-1$ components are 0).

Secondly, assuming that $\boldsymbol{\delta}_{Q_n}(n) = 1$, then $\delta_{n-1,Q_n} = 0$ can be obtained, and $\boldsymbol{\delta}_{Q_n}(i) = 0\ (1 \leq i \leq n-1)$ must hold (i.e. all the other $n-1$ components are 0).

Finally, assuming that $\boldsymbol{\delta}_{Q_n}(i) = 1\ (2 \leq i \leq n-1)$, we have $\delta_{i-1,Q_n} = 0\ (2 \leq i \leq n-1)$, and it can be easily obtained that $\boldsymbol{\delta}_{Q_n}(j) = 0\ (1 \leq j \leq i-1)$. On the other hand, since $\boldsymbol{\delta}_{Q_n}(i) = 1$, we can also obtain that $\delta_{j,Q_n} = 1\ (i \leq j \leq n-1)$. And then, $\boldsymbol{\delta}_{Q_n}(j) = 0\ (i+1 \leq j \leq n)$ can be obtained. Therefore, in this case, it can be concluded that all the other $n-1$ components must be 0. ■

## VIII. The joint optimization with the mechanism of network wave

Aiming at the disorder problem of the utilization of network resources commonly existing in multi-hop transmission networks, in [7], we propose the idea and the corresponding supporting theory, i.e. theory of network wave, by constructing volatility information transmission mechanism between the sending nodes and their corresponding receiving nodes of a pair of paths (composed of two primary paths), so as to improve the orderliness of the utilization of network resources. However, in this work, the problems of dividing information blocks into smaller granularity, i.e. information packets, and how to guarantee the delay requirements of all the information packets in end-to-end transmissions are not considered.

On the other hand, in the above proposed theory of particle access, the problem of how to guarantee the end-to-end delay of an information particle group on a multi-hop path is not considered. In view of this, in this section, the joint optimization of the mechanism of network wave and the mechanism of particle access is studied. For a given information particle group transmitted over a given primary path, a method to obtain the minimum reachable transmission bandwidth, which can guarantee the end-to-end delays of all the information particles contained in the information particle group, is proposed (***Theorem VIII.1*** and ***Algorithm VIII.1***). In **Theorem VIII.2**, it is proved that by setting the transmission period of the primary path as its intrinsic period, the minimum reachable transmission bandwidth of the end-to-end multi-hop transmission system will be optimized (that is, among all the minimum reachable transmission bandwidths corresponding to various configurable transmission periods, the minimum reachable transmission bandwidth corresponding to the intrinsic period is the smallest one), which is the significance of the joint optimization of the mechanism of network wave and the mechanism of particle access lying in! **The insight drawn from this conclusion is that the organic combination of the network wave mechanism and the particle access mechanism for multi-hop transmission applications forms a so-called "perfect match"! Furthermore,** in ***Theorem VIII.3***, **we prove that, given a primary path and an information particle group it bears, to guarantee the end-to-end transmission performance of the information particle group over the primary path, a lower bound exists for the peak reachable transmission bandwidth of the primary path (that is, if the peak reachable transmission bandwidth of the primary path is smaller than this value, no effective end-to-end transmission strategy can be found to satisfy the end-to-end transmission** **performance requirements of the information particle group). Finally,** in ***Theorem VIII.4***, **it is verified that this lower bound is reachable by adopting our proposed end-to-end transmission scheme combining the network wave mechanism and the particle access mechanism. This strongly proves the optimality of the proposed end-to-end transmission scheme.**

It is shown that the combination of the mechanism of network wave and the mechanism of particle access is an effective way to solve the problem of end-to-end multi-hop transmission. **Therefore, the corresponding information processing mechanism integrating the network wave mechanism and particle access mechanism in information processing networks is named as the "Wave-Particle Duality mechanism in Networks" (i.e. Mechanism of WPDN). Correspondingly, the related theory investigating this mechanism is referred to as the "Theory of Wave-Particle Duality in Networks" (i.e. Theory of WPDN).**

### *A. Descriptions of the problems*

At the time instant of $t = 0$, consider a primary path $P$ composed of $N + 1$ nodes (see **Definition III.1** in [7]), starting from node 1 (the source node) in $P$, an information particle group $Q$ (see **Definition IV.1**) containing $M\ (\geq 1)$ information particles is relayed through node 2, node 3,..., node $i$,..., node $N\ (\geq 1)$, and finally arrives at node $N + 1$ (the destination node).

**Problem VIII.1:** As shown in the sub-figure A of **Figure VIII.1**, assuming that the starting time instant of the initial time beat is $t_{Bb}\ (\geq 0)$ (that is, the initial time instant when node 1 is activated). With $T_{S_P}\ (T_{S_P} \geq T^*_{S_P})$ ($T_{S_P}$ and $T^*_{S_P}$ are the reachable period and the intrinsic period of $P$, respectively. See **Definition VII.2** and **Definition VII.4** in [7]) as its period, the primary path $P$ executes the **Algorithm X.1** (i.e., the algorithm for the volatile transmission of information blocks based on the reachable period of a primary path) proposed in [7]. During the execution of **Algorithm X.1**, the duration of each time beat is set to be $T_B\ (> 0)$, and the transmission bandwidth of each relaying node is $B$ (bits/s) (i.e., in the considered multihop transmission system, the transmission bandwidth of each relaying node is set to be equal to each other). After performing the incremental sorting operation $Op_{Inc}(Q)$ on the information particle group $Q$ (see **Definition IV.2**), the bearing capacity of the information particle $i\ (1 \leq i \leq M)$ in $Q$ and the deadline for its departure from node $N$ $(\geq 1)$ are $\ell_i$ and $t_{e,i}^{(N_1)}$ $(\geq t_{Bb} + N \cdot T_B)$, respectively.

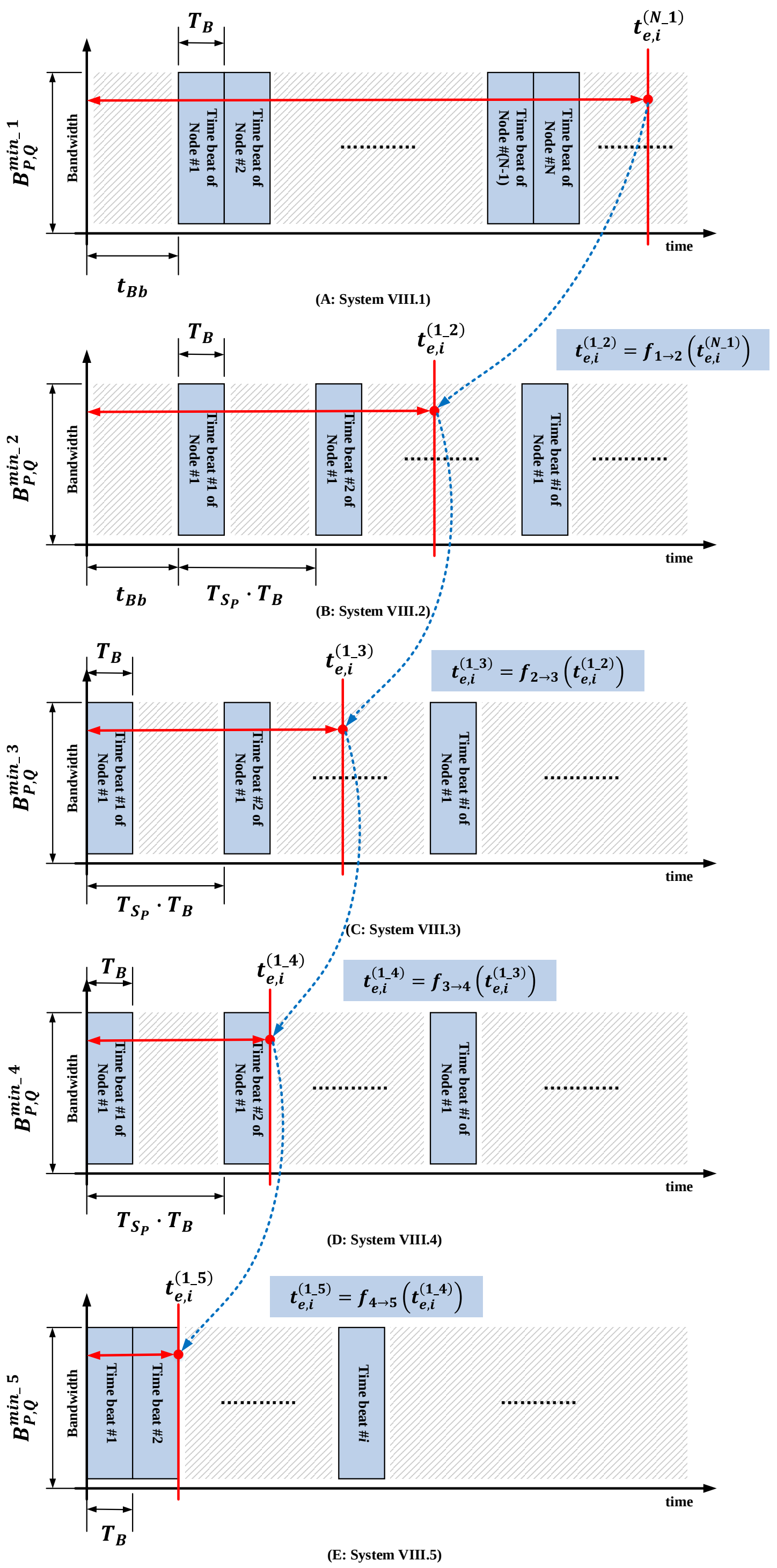

**Figure VIII.1** The mapping relationships of the deadlines of information particles between different systems

Then, what is the minimum reachable transmission bandwidth $B_{P,Q}^{min_1}$ (bits/s) of each relaying node that can ensure that all the $M$ information particles contained in $Q$ reach the destination node $N+1$ before their corresponding deadlines $t_{e,i}^{(N_1)}$ $(1 \leq i \leq M)$ (that is, the final time instant of each information particle leaving from the node $N$ is no later than its deadline)?

**Note:**

1) The first half of the superscript of $t_{e,i}^{(N_1)}$ represents the sequence number of the node which the information particle leaving from, and the second half represents the corresponding sequence number of the problem being considered;
2) the second half of the superscript of $B_{P,Q}^{min_1}$ indicates its corresponding sequence number of the problem being considered;
3) In order to make full use of bandwidth resources, the transmission of an information particle at each relaying node is allowed to span multiple time beats. Therefore, the moment when an information particle leaves a specific node is indeed the moment when the last information bit of the particle "leaves" the node;
4) For simplicity, it is assumed that the current active node completes the transmission of all the information particles (all or part of them) waiting to be transmitted at the end of each time beat when the node being considered is active;
5) During the execution of **Algorithm X.1**, the transmission bandwidth of each relay node is equal, and the duration of each time beat is identical. Furthermore, as for any sending node on the primary path, within every time beat when it is in the sending state, let it strictly replicate all transmission operations (including the transmission rate and transmission sequence for each information particles, etc.) performed by its predecessor node in the preceding time beat. In other words, the multi-hop transmission of information particle groups over the primary path implemented via **Algorithm X.1** can be analogous to object conveyance on a conveyor belt;
6) For all the information particles, from $t=0$ to the time when they finally reach the destination node of the primary path, they need to experience at least $t_{Bb} + N \cdot T_B$ time delays. Therefore, only the cases with deadlines $t_{e,i}^{(N_1)} \geq t_{Bb} + N \cdot T_B$ are considered here (that is, it is believed that the cases with deadlines $t_{e,i}^{(N_1)} < t_{Bb} + N \cdot T_B$ cannot be realized);

7) During the execution of **Algorithm X.1**, the source node can only transmit the pending information particles when the time beat assigned for its sending state arrives. Therefore, the source node must stay in the waiting-for-transmission state prior to the arrival of the corresponding sending time beat. The starting time instant of the initial time beat $t_{Bb}$ corresponds to the arrival instant of the first sending time beat of the source node. Furthermore, for scenarios in which **Algorithm X.1** has started execution at or before the current time instant, considering that the sending time beats of the source node arrive periodically with a period of $T_{S_P} \cdot T_B$, it is reasonable to assume that the relation of $0 \le t_{Bb} \le T_{S_P} \cdot T_B$ generally holds;
8) The system defined in **Problem VIII.1** is hereinafter referred to as "**System VIII.1**".

**Problem VIII.2:** As shown in the sub-figure B of **Figure VIII.1**, assuming that the starting time instant of the initial time beat is $t_{Bb}$ $(\ge 0)$. With $T_{S_P}$ $(T_{S_P} \ge T_{S_P}^*)$ as its period, the primary path $P$ executes the **Algorithm X.1** proposed in [7]. During the execution of **Algorithm X.1**, the duration of each time beat is set to be $T_B$ $(> 0)$, and the transmission bandwidth of each relaying node is $B$ (bits/s). Let us only consider the node 1 of $P$. After performing the incremental sorting operation $Op_{Inc}(Q)$ on the information particle group $Q$, the bearing capacity of the information particle $i$ $(1 \le i \le M)$ in $Q$ is $\ell_i$, and the deadline for its departure from node 1 is

$$t_{e,i}^{(1_2)} \triangleq t_{e,i}^{(N_1)} - (N-1) \cdot T_B \tag{VIII.1}$$

where $t_{e,i}^{(N_1)}$ is given in **Problem VIII.1**.

Then, what is the minimum reachable transmission bandwidth $B_{P,Q}^{min_2}$ (bits/s) of node 1 that can ensure that all the $M$ information particles contained in $Q$ leave from node 1 before their corresponding deadlines $t_{e,i}^{(1_2)}$ $(1 \le i \le M)$?

**Note:**

1) The system defined in **Problem VIII.2** is hereinafter referred to as "**System VIII.2**";
2) Through **Equation VIII.1**, **System VIII.1** is mapped to **System VIII.2**.

**Problem VIII.3:** As shown in the sub-figure C of **Figure VIII.1**, assuming that the starting time instant of the initial time beat is 0 (that is, the left shift operation is performed on the time axis relative to **System VIII.2**). With $T_{S_P}$ $(T_{S_P} \geq T_{S_P}^*)$ as its period, the primary path $P$ executes the **Algorithm X.1** proposed in [7]. During the execution of **Algorithm X.1**, the duration of each time beat is set to be $T_B$ $(> 0)$, and the transmission bandwidth of each relaying node is $B$ (bits/s). Let us only consider the node 1 of $P$. After performing the incremental sorting operation $Op_{Inc}(Q)$ on the information particle group $Q$, the bearing capacity of the information particle $i$ $(1 \leq i \leq M)$ in $Q$ is $\ell_i$, and the deadline for its departure from node 1 is

$$t_{e,i}^{(1_3)} \triangleq t_{e,i}^{(1_2)} - t_{Bb}$$

(VIII.2)

where both $t_{e,i}^{(1_2)}$ and $t_{Bb}$ are given in **Problem VIII.2**.

Then, what is the minimum reachable transmission bandwidth $B_{P,Q}^{min_3}$ (bits/s) of node 1 that can ensure that all the $M$ information particles contained in $Q$ leave from node 1 before their corresponding deadlines $t_{e,i}^{(1_3)}$ $(1 \leq i \leq M)$?

**Note:**

1) The system defined in **Problem VIII.3** is hereinafter referred to as "**System VIII.3**";
2) Through **Equation VIII.2**, **System VIII.2** is mapped to **System VIII.3**.

**Problem VIII.4:** As shown in the sub-figure D of **Figure VIII.1**, assuming that the starting time instant of the initial time beat is 0. With $T_{S_P}$ $(T_{S_P} \geq T_{S_P}^*)$ as its period, the primary path $P$ executes the **Algorithm X.1** proposed in [7]. During the execution of **Algorithm X.1**, the duration of each time beat is set to be $T_B$ $(> 0)$, and the transmission bandwidth of each relaying node is $B$ (bits/s). Let us only consider the node 1 of $P$. After performing the incremental sorting operation $Op_{Inc}(Q)$ on the information particle group $Q$, the bearing capacity of the information particle $i$ $(1 \leq i \leq M)$ in $Q$ is $\ell_i$, and the deadline for its departure from node 1 is

$$t_{e,i}^{(1_4)} \triangleq T_B \cdot \left(1 + T_{S_P} \cdot \left\lfloor \frac{t_{e,i}^{(1_3)} - T_B}{T_B \cdot T_{S_P}} \right\rfloor\right)$$

(VIII.3)

where $t_{e,i}^{(1_3)}$ is given in **Problem VIII.3**.

Then, what is the minimum reachable transmission bandwidth $B_{P,Q}^{min_4}$ (bits/s) of node 1 that can ensure that all the $M$ information particles contained in $Q$ leave from node 1 before their corresponding deadlines $t_{e,i}^{(1_4)}$ $(1 \leq i \leq M)$?

**Note:**

1) The system defined in **Problem VIII.4** is hereinafter referred to as "**System VIII.4**";
2) Through **Equation VIII.3**, **System VIII.3** is mapped to **System VIII.4**.

**Problem VIII.5:** As shown in the sub-figure E of **Figure VIII.1**, assuming that the starting time instant of the initial time beat is 0. Let us only consider the node 1 of $P$. Let node 1 remain active at all times (that is, it is always in the sending state), and it only outputs at the end of each time beat. After performing the incremental sorting operation $Op_{Inc}(Q)$ on the information particle group $Q$, the bearing capacity of the information particle $i$ $(1 \leq i \leq M)$ in $Q$ is $\ell_i$, and the deadline for its departure from node 1 is

$$t_{e,i}^{(1_5)} \triangleq t_{e,i}^{(1_4)} - T_B \cdot (T_{S_P} - 1) \cdot \left\lfloor \frac{t_{e,i}^{(1_4)} - T_B}{T_B \cdot T_{S_P}} \right\rfloor$$

(VIII.4)

where $t_{e,i}^{(1_4)}$ is given in **Problem VIII.4**.

Then, what is the minimum reachable transmission bandwidth $B_{P,Q}^{min_5}$ (bits/s) of node 1 that can ensure that all the $M$ information particles contained in $Q$ leave from node 1 before their corresponding deadlines $t_{e,i}^{(1_5)}$ $(1 \leq i \leq M)$?

**Note VIII.5:**

1) The system defined in **Problem VIII.5** is hereinafter referred to as "**System VIII.5**";
2) Through **Equation VIII.4**, **System VIII.4** is mapped to **System VIII.5**;
3) Through **Equation VIII.1**, **Equation VIII.2**, **Equation VIII.3** and **Equation VIII.4**, **System**

**VIII.1** is finally mapped to **System VIII.5**.

*B. Relationships between key parameters of each system*

In this subsection, we will prove that the minimum reachable transmission bandwidths corresponding to the five systems defined above are equal to each other. In this way, we can naturally transform the problem of solving **Problem VIII.1** into the problem of solving **Problem VIII.5** (and the method of solving **Problem VIII.5** has already been given in this paper (refer to **Theorem VII.1** and **Corollary VII.1.1**)). Furthermore, we will also prove that by setting the period of the primary path as its intrinsic period, it will be beneficial to minimize the end-to-end transmission bandwidth.

**Lemma VIII.1.1:** $\boldsymbol{B_{P,Q}^{min_1} = B_{P,Q}^{min_2}}$**.**

**Proof:**

First, let us prove that $B_{P,Q}^{min_1}$ of **System VIII.1** is also reachable for **System VIII.2**. Considering that $B_{P,Q}^{min_1}$ is reachable for **System VIII.1**, for all the information particles contained in information particle group $Q$, there is at least one set of corresponding time instants $t_i^{(N_1)}$ $(1 \le i \le M)$ of leaving node $N$ in **System VIII.1** satisfying the following relationships

$$t_i^{(N_1)} \le t_{e,i}^{(N_1)} \ (1 \le i \le M)$$

(VIII.5)

Let $t_i^{(1_1)}$ $(1 \le i \le M)$ indicate the time instant when an information particle $i$ in **System VIII.1** leaves node 1. We have

$$t_i^{(1_1)} = t_i^{(N_1)} - (N-1) \cdot T_B \ (1 \le i \le M)$$

(VIII.6)

Combing the above two equations, it can be obtained that

$$\begin{aligned} t_i^{(1_1)} &= t_i^{(N_1)} - (N-1) \cdot T_B \\ &\le t_{e,i}^{(N_1)} - (N-1) \cdot T_B \ (1 \le i \le M) \\ &= t_{e,i}^{(1_2)} \end{aligned}$$

(VIII.7)

Let $t_i^{(1_2)}$ $(1 \le i \le M)$ indicate the time instant when an information particle $i$ in **System VIII.2** leaves node 1. According to the relations of $t_i^{(1_2)} = t_i^{(1_1)}$ $(1 \le i \le M)$, time instants $t_i^{(1_1)}$ are mapped to **System VIII.2**. Based on the above equation, we can get

$$t_i^{(1_2)} \le t_{e,i}^{(1_2)} \ (1 \le i \le M) \tag{VIII.8}$$

This shows that when the transmission bandwidth of node 1 in **System VIII.2** is $B_{P,Q}^{min_1}$, for all the information particles contained in the information particle group $Q$, there is at least one set of time instants $t_i^{(1_2)}$ $(1 \le i \le M)$ of leaving node 1 satisfying the relationships of $t_i^{(1_2)} \le t_{e,i}^{(1_2)}$ $(1 \le i \le M)$. Therefore, the transmission bandwidth $B_{P,Q}^{min_1}$ is reachable for **System VIII.2**.

Next, let us prove that $B_{P,Q}^{min_2}$ of **System VIII.2** is also reachable for **System VIII.1**. Considering that $B_{P,Q}^{min_2}$ is reachable for **System VIII.2**, for all the information particles contained in information particle group $Q$, there is at least one set of corresponding time instants $t_i^{(1_2)}$ $(1 \le i \le M)$ of leaving node 1 in **System VIII.2** satisfying the following relationships

$$t_i^{(1_2)} \le t_{e,i}^{(1_2)} \ (1 \le i \le M) \tag{VIII.9}$$

According to the relations of $t_i^{(1_1)} = t_i^{(1_2)}$ $(1 \le i \le M)$, time instants $t_i^{(1_2)}$ $(1 \le i \le M)$ are mapped to **System VIII.1**. Based on the above equation, we have

$$\begin{aligned} t_i^{(N_1)} &= t_i^{(1_1)} + (N-1) \cdot T_B \\ &\le t_{e,i}^{(1_2)} + (N-1) \cdot T_B \ (1 \le i \le M) \\ &= t_{e,i}^{(N_1)} \end{aligned} \tag{VIII.10}$$

This shows that when the transmission bandwidths of all the relaying nodes in **System VIII.1** are set to be $B_{P,Q}^{min_2}$, for all the information particles contained in the information particle group $Q$, there is at least one set of time instants $t_i^{(N_1)}$ $(1 \le i \le M)$ of leaving node $N$ satisfying the relationships of $t_i^{(N_1)} \le t_{e,i}^{(N_1)}$ $(1 \le i \le M)$. Therefore, the transmission bandwidth $B_{P,Q}^{min_2}$ is reachable for **System VIII.1**.

To sum up, considering that $B_{P,Q}^{min_1}$ and $B_{P,Q}^{min_2}$ are the minimum reachable transmission bandwidths of **System VIII.1** and **System VIII.2**, respectively, it can be obtained that $B_{P,Q}^{min_1} = B_{P,Q}^{min_2}$ must hold. ■

**Lemma VIII.1.2: $B_{P,Q}^{min_2} = B_{P,Q}^{min_3}$.**

**Proof:**

First, let us prove that $B_{P,Q}^{min_2}$ of **System VIII.2** is also reachable for **System VIII.3**. Considering that $B_{P,Q}^{min_2}$ is reachable for **System VIII.2**, for all the information particles contained in information particle group $Q$, there is at least one set of corresponding time instants $t_i^{(1_2)}$ $(1 \le i \le M)$ of leaving node 1 in **System VIII.2** satisfying the following relationships

$$t_i^{(1_2)} \le t_{e,i}^{(1_2)} \ (1 \le i \le M)$$

(VIII.11)

According to the relations of $t_i^{(1_3)} = t_i^{(1_2)} - t_{Bb}$ $(1 \le i \le M)$, time instants $t_i^{(1_2)}$ $(1 \le i \le M)$ are mapped to **System VIII.3**. Based on the above equation, we have

$$\begin{aligned} t_i^{(1_3)} &= t_i^{(1_2)} - t_{Bb} \\ &\le t_{e,i}^{(1_2)} - t_{Bb} \ (1 \le i \le M) \\ &= t_{e,i}^{(1_3)} \end{aligned}$$

(VIII.12)

This shows that when the transmission bandwidth of node 1 in **System VIII.3** is set to be $B_{P,Q}^{min_2}$, for all the information particles contained in the information particle group $Q$, there is at least one set of time instants $t_i^{(1_3)}$ $(1 \le i \le M)$ of leaving node 1 satisfying the relationships of $t_i^{(1_3)} \le t_{e,i}^{(1_3)}$ $(1 \le i \le M)$. Therefore, the transmission bandwidth $B_{P,Q}^{min_2}$ is reachable for **System VIII.3**.

Next, let us prove that $B_{P,Q}^{min_3}$ of **System VIII.3** is also reachable for **System VIII.2**. Considering that $B_{P,Q}^{min_3}$ is reachable for **System VIII.3**, for all the information particles contained in information particle group $Q$, there is at least one set of corresponding time instants $t_i^{(1_3)}$ $(1 \le i \le M)$ of leaving node 1 in **System VIII.3** satisfying the following relationships

$$t_i^{(1_3)} \leq t_{e,i}^{(1_3)}\ (1 \leq i \leq M)$$

(VIII.13)

According to the relations of $t_i^{(1_2)} = t_i^{(1_3)} + t_{Bb}\ (1 \leq i \leq M)$, time instants $t_i^{(1_3)}$ are mapped to **System VIII.2**. Based on the above equation, we have

$$\begin{aligned} t_i^{(1_2)} &= t_i^{(1_3)} + t_{Bb} \\ &\leq t_{e,i}^{(1_3)} + t_{Bb}\ (1 \leq i \leq M) \\ &= t_{e,i}^{(1_2)} \end{aligned}$$

(VIII.14)

This shows that when the transmission bandwidth of node 1 in **System VIII.2** is set to be $B_{P,Q}^{min_3}$, for all the information particles contained in the information particle group $Q$, there is at least one set of time instants $t_i^{(1_2)}\ (1 \leq i \leq M)$ of leaving node 1 satisfying the relationships of $t_i^{(1_2)} \leq t_{e,i}^{(1_2)}\ (1 \leq i \leq M)$. Therefore, the transmission bandwidth $B_{P,Q}^{min_3}$ is reachable for **System VIII.2**.

To sum up, considering that $B_{P,Q}^{min_2}$ and $B_{P,Q}^{min_3}$ are the minimum reachable transmission bandwidths of **System VIII.2** and **System VIII.3**, respectively, it can be obtained that $B_{P,Q}^{min_2} = B_{P,Q}^{min_3}$ must hold. ■

**Lemma VIII.1.3: $B_{P,Q}^{min_3} = B_{P,Q}^{min_4}$.**

**Proof:**

First, let us prove that $B_{P,Q}^{min_3}$ of **System VIII.3** is also reachable for **System VIII.4**. Considering that $B_{P,Q}^{min_3}$ is reachable for **System VIII.3**, for all the information particles contained in information particle group $Q$, there is at least one set of corresponding time instants $t_i^{(1_3)}\ (1 \leq i \leq M)$ of leaving node 1 in **System VIII.3** satisfying the relationships of $t_i^{(1_3)} \leq t_{e,i}^{(1_3)}\ (1 \leq i \leq M)$. It can be seen from the fourth point in the note part of **Problem VIII.1** that for **System VIII.3**, the time instants when information particles leave node 1 can only be $T_B \cdot (1 + k \cdot T_{S_P})\ (k \geq 0, k \in \mathbb{Z})$. Therefore, the relationships of $t_i^{(1_3)} \leq t_{e,i}^{(1_3)}\ (1 \leq i \leq M)$ imply that

$$
\begin{aligned}
t_i^{(1_3)} &\leq T_B \cdot \left(1 + T_{S_P} \cdot \left\lceil \frac{t_{e,i}^{(1_3)} - T_B}{T_B \cdot T_{S_P}} \right\rceil \right) (1 \leq i \leq M) \\
&= t_{e,i}^{(1_4)}
\end{aligned}
$$

(VIII.15)

Furthermore, according to the relations of $t_i^{(1_4)} = t_i^{(1_3)}$ $(1 \leq i \leq M)$, time instants $t_i^{(1_3)}$ $(1 \leq i \leq M)$ are mapped to **System VIII.4**. We have

$$
t_i^{(1_4)} \leq t_{e,i}^{(1_4)} \ (1 \leq i \leq M)
$$

(VIII.16)

This shows that when the transmission bandwidth of node 1 in **System VIII.4** is set to be $B_{P,Q}^{min_3}$, for all the information particles contained in the information particle group $Q$, there is at least one set of time instants $t_i^{(1_4)}$ $(1 \leq i \leq M)$ of leaving node 1 satisfying the relationships of $t_i^{(1_4)} \leq t_{e,i}^{(1_4)}$ $(1 \leq i \leq M)$. Therefore, the transmission bandwidth $B_{P,Q}^{min_3}$ is reachable for **System VIII.4**.

Next, let us prove that $B_{P,Q}^{min_4}$ of **System VIII.4** is also reachable for **System VIII.3**. Considering that $B_{P,Q}^{min_4}$ is reachable for **System VIII.4**, for all the information particles contained in information particle group $Q$, there is at least one set of corresponding time instants $t_i^{(1_4)}$ $(1 \leq i \leq M)$ of leaving node 1 in **System VIII.4** satisfying the relationships of $t_i^{(1_4)} \leq t_{e,i}^{(1_4)}$ $(1 \leq i \leq M)$. Since

$$
\begin{aligned}
t_{e,i}^{(1_4)} &= T_B \cdot \left(1 + T_{S_P} \cdot \left\lceil \frac{t_{e,i}^{(1_3)} - T_B}{T_B \cdot T_{S_P}} \right\rceil \right) \\
&\leq T_B \cdot \left(1 + T_{S_P} \cdot \frac{t_{e,i}^{(1_3)} - T_B}{T_B \cdot T_{S_P}} \right) \quad (1 \leq i \leq M) \\
&= t_{e,i}^{(1_3)}
\end{aligned}
$$

(VIII.17)

we have

$$
t_i^{(1_4)} \leq t_{e,i}^{(1_4)} \leq t_{e,i}^{(1_3)} \ (1 \leq i \leq M)
$$

(VIII.18)

According to the relations of $t_i^{(1_3)} = t_i^{(1_4)}$ $(1 \leq i \leq M)$, time instants $t_i^{(1_4)}$ are mapped to

**System VIII.3**. We have

$$t_i^{(1_3)} \le t_{e,i}^{(1_3)} \; (1 \le i \le M)$$

(VIII.19)

This shows that when the transmission bandwidth of node 1 in **System VIII.3** is set to be $B_{P,Q}^{min_4}$, for all the information particles contained in the information particle group $Q$, there is at least one set of time instants $t_i^{(1_3)}$ $(1 \le i \le M)$ of leaving node 1 satisfying the relationships of $t_i^{(1_3)} \le t_{e,i}^{(1_3)}$ $(1 \le i \le M)$. Therefore, the transmission bandwidth $B_{P,Q}^{min_4}$ is reachable for **System VIII.3**.

To sum up, considering that $B_{P,Q}^{min_3}$ and $B_{P,Q}^{min_4}$ are the minimum reachable transmission bandwidths of **System VIII.3** and **System VIII.4**, respectively, it can be obtained that $B_{P,Q}^{min_3} = B_{P,Q}^{min_4}$ must hold. ■

**Lemma VIII.1.4: $B_{P,Q}^{min_4} = B_{P,Q}^{min_5}$.**

**Proof:**

First, let us prove that $B_{P,Q}^{min_4}$ of **System VIII.4** is also reachable for **System VIII.5**. Considering that $B_{P,Q}^{min_4}$ is reachable for **System VIII.4**, for all the information particles contained in information particle group $Q$, there is at least one set of corresponding time instants $t_i^{(1_4)}$ $(1 \le i \le M)$ of leaving node 1 in **System VIII.4** satisfying the relationships: $t_i^{(1_4)} \le t_{e,i}^{(1_4)}$ $(1 \le i \le M)$. Since both $\frac{t_i^{(1_4)} - T_B}{T_B \cdot T_{S_P}}$ and $\frac{t_{e,i}^{(1_4)} - T_B}{T_B \cdot T_{S_P}}$ are integers, we have

$$\left\lfloor \frac{t_i^{(1_4)} - T_B}{T_B \cdot T_{S_P}} \right\rfloor = \frac{t_i^{(1_4)} - T_B}{T_B \cdot T_{S_P}} \; (1 \le i \le M)$$

(VIII.20)

and

$$\left\lfloor \frac{t_{e,i}^{(1_4)} - T_B}{T_B \cdot T_{S_P}} \right\rfloor = \frac{t_{e,i}^{(1_4)} - T_B}{T_B \cdot T_{S_P}} \; (1 \le i \le M)$$

(VIII.21)

Mapping $t_i^{(1_4)}$ $(1 \le i \le M)$ and $t_{e,i}^{(1_4)}$ $(1 \le i \le M)$ to **System VIII.5**, it can be obtained that

$$\begin{aligned} t_i^{(1_5)} &= t_i^{(1_4)} - T_B \cdot (T_{S_P} - 1) \cdot \left\lfloor \frac{t_i^{(1_4)} - T_B}{T_B \cdot T_{S_P}} \right\rfloor \\ &= t_i^{(1_4)} - T_B \cdot (T_{S_P} - 1) \cdot \left( \frac{t_i^{(1_4)} - T_B}{T_B \cdot T_{S_P}} \right) (1 \le i \le M) \\ &= \frac{t_i^{(1_4)}}{T_{S_P}} + \frac{T_B \cdot (T_{S_P} - 1)}{T_{S_P}} \end{aligned} \tag{VIII.22}$$

and

$$\begin{aligned} t_{e,i}^{(1_5)} &= t_{e,i}^{(1_4)} - T_B \cdot (T_{S_P} - 1) \cdot \left\lfloor \frac{t_{e,i}^{(1_4)} - T_B}{T_B \cdot T_{S_P}} \right\rfloor \\ &= t_{e,i}^{(1_4)} - T_B \cdot (T_{S_P} - 1) \cdot \left( \frac{t_{e,i}^{(1_4)} - T_B}{T_B \cdot T_{S_P}} \right) (1 \le i \le M) \\ &= \frac{t_{e,i}^{(1_4)}}{T_{S_P}} + \frac{T_B \cdot (T_{S_P} - 1)}{T_{S_P}} \end{aligned} \tag{VIII.23}$$

Since $t_i^{(1_4)} \le t_{e,i}^{(1_4)}$ $(1 \le i \le M)$, and by further combing the above two equations, we have

$$t_i^{(1_5)} \le t_{e,i}^{(1_5)} \ (1 \le i \le M) \tag{VIII.24}$$

This shows that when the transmission bandwidth of node 1 in **System VIII.5** is set to be $B_{P,Q}^{min_4}$, for all the information particles contained in the information particle group $Q$, there is at least one set of time instants $t_i^{(1_5)}$ $(1 \le i \le M)$ of leaving node 1 satisfying the relationships of $t_i^{(1_5)} \le t_{e,i}^{(1_5)}$ $(1 \le i \le M)$. Therefore, the transmission bandwidth $B_{P,Q}^{min_4}$ is reachable for **System VIII.5**.

Next, let us prove that $B_{P,Q}^{min_5}$ of **System VIII.5** is also reachable for **System VIII.4**. Considering that $B_{P,Q}^{min_5}$ is reachable for **System VIII.5**, for all the information particles contained in information particle group $Q$, there is at least one set of corresponding time instants $t_i^{(1_5)}$ $(1 \le i \le M)$ of leaving node 1 in **System VIII.5** satisfying the relationships: $t_i^{(1_5)} \le t_{e,i}^{(1_5)}$ $(1 \le i \le M)$. Mapping $t_i^{(1_5)}$ $(1 \le i \le M)$ and $t_{e,i}^{(1_5)}$ $(1 \le i \le M)$ to **System VIII.4**, we have

$$t_i^{(1_4)} = T_{S_P} \cdot \left( t_i^{(1_5)} - \frac{T_B \cdot (T_{S_P} - 1)}{T_{S_P}} \right) (1 \le i \le M)$$

(VIII.25)

and

$$t_{e,i}^{(1_4)} = T_{S_P} \cdot \left( t_{e,i}^{(1_5)} - \frac{T_B \cdot (T_{S_P} - 1)}{T_{S_P}} \right) (1 \le i \le M)$$

(VIII.26)

Since $t_i^{(1_5)} \le t_{e,i}^{(1_5)}$ $(1 \le i \le M)$, by further combining the above two equations, it can be obtained that

$$t_i^{(1_4)} \le t_{e,i}^{(1_4)} \ (1 \le i \le M)$$

(VIII.27)

This shows that when the transmission bandwidth of node 1 in **System VIII.4** is set to be $B_{P,Q}^{min_5}$, for all the information particles contained in the information particle group $Q$, there is at least one set of time instants $t_i^{(1_4)}$ $(1 \le i \le M)$ of leaving node 1 satisfying the relationships of $t_i^{(1_4)} \le t_{e,i}^{(1_4)}$ $(1 \le i \le M)$. Therefore, the transmission bandwidth $B_{P,Q}^{min_5}$ is reachable for **System VIII.4**.

To sum up, considering that $B_{P,Q}^{min_4}$ and $B_{P,Q}^{min_5}$ are the minimum reachable transmission bandwidths of **System VIII.4** and **System VIII.5**, respectively, it can be obtained that $B_{P,Q}^{min_4} = B_{P,Q}^{min_5}$ must hold. ■

**Theorem VIII.1: Equivalent transformation theorem of minimum reachable transmission bandwidth.**

$$\boldsymbol{B_{P,Q}^{min_1} = B_{P,Q}^{min_2} = B_{P,Q}^{min_3} = B_{P,Q}^{min_4} = B_{P,Q}^{min_5}}$$

**Proof:**

According to **Lemma VIII.1.1**, **Lemma VIII.1.2**, **Lemma VIII.1.3** and **Lemma VIII.1.4**, the proposition can be instantly proved. ■

**Note:**

1) For the information particle group corresponding to **System VIII.5**, the minimum reachable

access bandwidth $B$ of this information particle group can be directly calculated by adopting the methods proposed in **Theorem VII.1** or **Corollary VII.1.1**. It can be easily known from the foregoing description of **System VIII.5** that: $B_{P,Q}^{min_5} = B$. Furthermore, the relations of $B_{P,Q}^{min_1} = B_{P,Q}^{min_2} = B_{P,Q}^{min_3} = B_{P,Q}^{min_4} = B_{P,Q}^{min_15} = B$ can be derived based on **Theorem VIII.1**. This constitutes the method for solving the minimum reachable transmission bandwidth $B_{P,Q}^{min_1}$ in **System VIII.1**.

**Theorem VIII.2: Positive coupling relationship theorem between the reachable period and the minimum reachable transmission bandwidth of a primary path.**

**In System VIII.1, given $N$ (the number of relaying nodes in a primary path $P$), $t_{Bb}$ (the starting time instant of the initial time beat) and $T_B$ (the duration of each time beat). Let the information particle group transmitted over the primary path $P$ be $Q$, and after the incremental sorting operation $Op_{Inc}(Q)$ is performed on $Q$, the deadline for the information particle $i$ $(1 \le i \le M)$ to leave the node $N$ $(\ge 1)$ is $t_{e,i}^{(N_1)}$ $(1 \le i \le M)$. With $T_{S_P}$ $(T_{S_P} \ge T_{S_P}^*)$ ($T_{S_P}$ and $T_{S_P}^*$ are the reachable period and the intrinsic period of $P$, respectively.) as its period, the primary path $P$ executes the Algorithm X.1 proposed in [7]. Let $B_{P,Q}^{min_1}(T_{S_P})$ denote the minimum reachable transmission bandwidth corresponding to the adopted period of $T_{S_P}$. The relation of $B_{P,Q}^{min_1}(T_{S_P}) \ge B_{P,Q}^{min_1}(T_{S_P}^*)$ holds.**

**Proof:**

Based on **Equation VIII.1** and **Equation VIII.2**, we have

$$t_{e,i}^{(1_3)} = t_{e,i}^{(N_1)} - (N-1) \cdot T_B - t_{Bb} \; (1 \le i \le M)$$

(VIII.28)

By combing with **Equation VIII.3**, it can be obtained that

$$t_{e,i}^{(1_4)} \triangleq T_B \cdot \left(1 + T_{S_P} \cdot \left\lfloor \frac{t_{e,i}^{(N_1)} - N \cdot T_B - t_{Bb}}{T_B \cdot T_{S_P}} \right\rfloor\right) \; (1 \le i \le M)$$

(VIII.29)

Furthermore, combing **Equation VIII.23** and **Equation VIII.29**, we can get

$$
\begin{aligned}
t_{e,i}^{(1_5)}(T_{S_P}) &= \frac{t_{e,i}^{(1_4)}}{T_{S_P}} + \frac{T_B \cdot (T_{S_P} - 1)}{T_{S_P}} \\
&= \frac{t_{e,i}^{(1_4)} - T_B}{T_{S_P}} + T_B \qquad\qquad (1 \le i \le M) \\
&= T_B \cdot \left( \left\lceil \frac{t_{e,i}^{(N_1)} - N \cdot T_B - t_{Bb}}{T_B \cdot T_{S_P}} \right\rceil + 1 \right)
\end{aligned}
$$

(VIII.30)

and since $T_{S_P} \ge T_{S_P}^*$, we have

$$
\begin{aligned}
t_{e,i}^{(1_5)}(T_{S_P}) &= T_B \cdot \left( \left\lceil \frac{t_{e,i}^{(N_1)} - N \cdot T_B - t_{Bb}}{T_B \cdot T_{S_P}} \right\rceil + 1 \right) \\
&\le T_B \cdot \left( \left\lceil \frac{t_{e,i}^{(N_1)} - N \cdot T_B - t_{Bb}}{T_B \cdot T_{S_P}^*} \right\rceil + 1 \right) (1 \le i \le M) \\
&= t_{e,i}^{(1_5)}(T_{S_P}^*)
\end{aligned}
$$

(VIII.31)

where $t_{e,i}^{(1_5)}(T_{S_P})$ and $t_{e,i}^{(1_5)}(T_{S_P}^*)$ represent the deadlines for the information particle $i$ $(1 \le i \le M)$ leaving the node 1 with $T_{S_P}$and $T_{S_P}^*$ being the corresponding transmission period in **System VIII.5**, respectively. It can be obtained from **Equation VIII.31** and **Corollary VI.1.1** that

$$
\begin{aligned}
B_{P,Q}^{min_5}(T_{S_P}) &= B_{re,Q}^{min}(T_{S_P}) \\
&\ge B_{re,Q}^{min}(T_{S_P}^*) \\
&= B_{P,Q}^{min_5}(T_{S_P}^*)
\end{aligned}
$$

(VIII.32)

where $B_{re,Q}^{min}(T_{S_P})$ and $B_{re,Q}^{min}(T_{S_P}^*)$ represent the minimum reachable access bandwidths in **System VIII.5** with the transmission period being $T_{S_P}$ and $T_{S_P}^*$, respectively. Furthermore, according to **Theorem VIII.1**, we have

$$
\begin{aligned}
B_{P,Q}^{min_1}(T_{S_P}) &= B_{P,Q}^{min_5}(T_{S_P}) \\
&\ge B_{P,Q}^{min_5}(T_{S_P}^*) \\
&= B_{P,Q}^{min_1}(T_{S_P}^*)
\end{aligned}
$$

(VIII.33)

Hence, it can be concluded that the relation of $B_{P,Q}^{min_1}(T_{S_P}) \ge B_{P,Q}^{min_1}(T_{S_P}^*)$ holds. ■

**Note:**

1) This theorem shows that by setting the transmission period of the primary path as its intrinsic

period $T_{S_P}^{*}$, the minimum reachable transmission bandwidth of **System VIII.1** will be optimized (that is, among all the minimum reachable transmission bandwidths corresponding to various configurable transmission periods, the minimum reachable transmission bandwidth corresponding to the transmission period of $T_{S_P}^{*}$ is the smallest one), which is the significance of joint optimization of the mechanism of network wave and the mechanism of particle access lying in! The insight drawn from this conclusion is that the organic combination of the network wave mechanism and the particle access mechanism for multi-hop transmission applications forms a so-called "perfect match"!

***C. The joint optimization algorithm of the mechanism of network wave and the mechanism of particle access***

In this subsection, we first describe the algorithm steps (that is, the following **Algorithm VIII.1**) for calculating the minimum reachable transmission bandwidth of **System VIII.1**. Next, we describe the algorithm steps (that is, the following **Algorithm VIII.2**) of allocating each information particle of an information particle group into each successive active time beats of node 1.

**Algorithm VIII.1: algorithm steps for calculating the minimum reachable transmission bandwidth of System VIII.1.**

1: BEGIN

2: Map the deadline $t_{e,i}^{(N_1)}$ of the information particle $i\ (1 \le i \le M)$ in **System VIII.1** to the deadline $t_{e,i}^{(1_2)} = t_{e,i}^{(N_1)} - (N-1) \cdot T_B$ of the information particle $i$ in **System VIII.2**;

3: Map the deadline $t_{e,i}^{(1_2)}$ of the information particle $i\ (1 \le i \le M)$ in **System VIII.2** to the deadline $t_{e,i}^{(1_3)} = t_{e,i}^{(1_2)} - t_{Bb}$ of the information particle $i$ in **System VIII.3**;

//According to **Theorem VIII.2**, in order to optimize the transmission bandwidth, the //transmission period of the primary path is set as its intrinsic period $T_{S_P}^*$

4: Map the deadline $t_{e,i}^{(1_3)}$ of the information particle $i\ (1 \le i \le M)$ in **System VIII.3** to the deadline $t_{e,i}^{(1_4)} = T_B \cdot \left(1 + T_{S_P}^* \cdot \left\lfloor \frac{t_{e,i}^{(1_3)} - T_B}{T_B \cdot T_{S_P}^*} \right\rfloor\right)$ of the information particle $i$ in **System VIII.4**;

5: Map the deadline $t_{e,i}^{(1_4)}$ of the information particle $i\ (1 \le i \le M)$ in **System VIII.4** to the deadline $t_{e,i}^{(1_5)} = t_{e,i}^{(1_4)} - T_B \cdot (T_{S_P}^* - 1) \cdot \left\lfloor \frac{t_{e,i}^{(1_4)} - T_B}{T_B \cdot T_{S_P}^*} \right\rfloor$ of the information particle $i$ in **System VIII.5**;

6: For the information particle group in **System VIII.5**, apply **Theorem VII.1** or **Corollary VII.1.1** to calculate the minimum reachable access bandwidth $B$ of the information particle group;

7: Get the minimum reachable transmission bandwidth as $B_{P,Q}^{min_5} = B$;

8: According to **Theorem VIII.1**, the minimum reachable transmission bandwidth of **System VIII.1** is $B_{P,Q}^{min_1} = B_{P,Q}^{min_5}$, that is, when executing **Algorithm X.1** proposed in [7], the transmission bandwidths from node 1 to node $N$ are configured as $B_{P,Q}^{min_1}$.

9: END

**Note:**

1) It is required that the information particle group must be incrementally sorted by following the operation of $Op_{Inc}(Q)$ before the algorithm is executed;
2) In each mapping step, the bearing capacity $\ell_i \, (1 \le i \le M)$ of each information particle remains unchanged.

**Algorithm VIII.2: the algorithm steps to calculate the time beats for the primary path (that is, for node 1) transmitting the information particle $\boldsymbol{i} \, (\mathbf{1} \le \boldsymbol{i} \le \boldsymbol{M})$ and the corresponding allocations of transmission resources.**

1: BEGIN

2: Initialize the time pointer as $t_p = 0$;

3: Initialize the sequence number of the information particle being processed as $i = 1$;

4: WHILE ($i \le M$) DO

//The processing procedure follows the increasing of the sequence number of
//the information particle being processed until all the information particles in the
//information particle group are processed

5: The sequence number of the starting time beat of the transmission of information particle $i$ at node 1 is $n_\alpha = \left\lceil \frac{t_p}{T_B} \right\rceil$;

6: The sequence number of the ending time beat of the transmission of information particle $i$ at node 1 is $n_\beta = \left\lceil \frac{t_p + \ell_i / B_{P,Q}^{min_1}}{T_B} \right\rceil$;

7: IF ($n_\alpha = n_\beta$) THEN

// In this case, the information particle $i$ will be transmitted out within one time
//beat of node 1

The amount of information transmitted by the information particle $i$ in the time beat $n_\alpha$ is $\ell_i$ (bits);

ELSE

// In this case (that is, $n_\alpha < n_\beta$), the information particle $i$ will be transmitted out
//within multiple time beats of node 1

As for the information particle $i$, the amount of information transmitted within the time beat $n_\alpha$ is $B_{P,Q}^{min_1} \cdot \left(\left\lceil \frac{t_p}{T_B} \right\rceil \cdot T_B - t_p\right)$ (bits);

As for the information particle $i$, the amount of information transmitted within the time beat $n_\beta$ is $B_{P,Q}^{min_1} \cdot \left[t_p + \frac{\ell_i}{B_{P,Q}^{min_1}} - (n_\beta - 1) \cdot T_B\right]$ (bits);

IF ($n_\alpha < n_\beta - 1$) THEN

// In this case, the transmission of information particle $i$ at node 1 will span
//more than two time beats

As for the information particle $i$, the amount of information transmitted between the time beat $n_\alpha$ and $n_\beta$ (excluding the time beat $n_\alpha$ and $n_\beta$) is $B_{P,Q}^{in_1} \cdot T_B \cdot (n_\beta - n_\alpha - 1)$ (bits);

END IF

END IF

Refresh the time pointer as $t_p \Leftarrow t_p + \frac{\ell_i}{B_{P,Q}^{min_1}}$;

Refresh the sequence number of the information particle being processed as $i \Leftarrow i + 1$;

END WHILE

END

**Note:**

1) For the better understanding of the above algorithm, please refer to **Figure VIII.2**;

2) The information particle group has been incrementally sorted by following the operation of

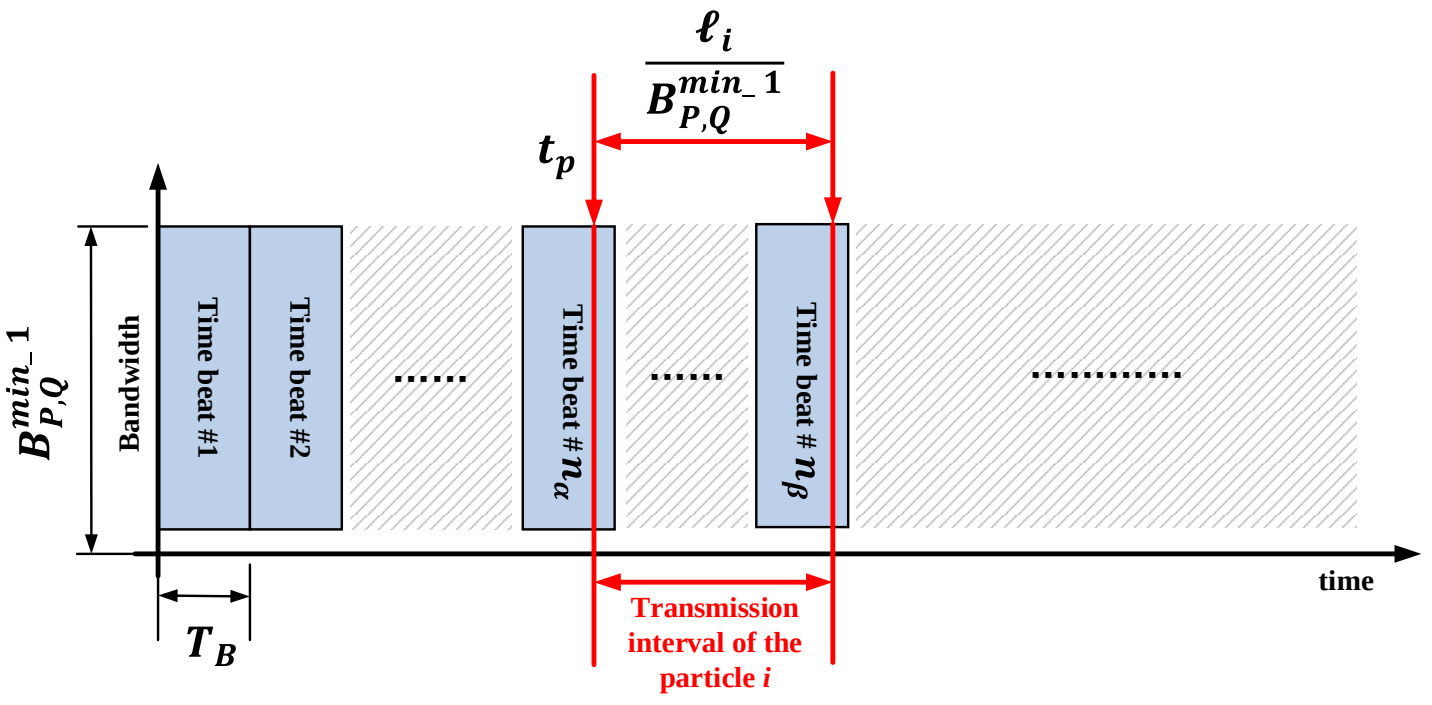

**Figure VIII.2** the relationships between some key time points in **Algorithm VIII.2**
(The sequence number of a time beat shown in the figure indicates
the counter of the time beats when node 1 is active)

$Op_{Inc}(Q)$ before the algorithm is executed;

3) The "sequence number of a time beat" in the algorithm is serve as the counter of the time beats when node 1 is active. For example, if the sequence number is $n$, it corresponds to the time beat when node 1 is active for the $n$th time;

4) The description of the algorithm only focuses on how the amount of information carried by the information particles is distributed among the time beats corresponding to node 1. As for the transmission of these information particles in other relaying nodes, the **Algorithm X.1** in [7] is strictly followed and executed accordingly;

5) The above algorithm does not specify how the information bits arranged in a specific time beat should be transmitted in the corresponding time beat. In fact, any transmission method that can ensure that the information bits arranged in a specific time beat are transmitted before the end of the time beat is considered feasible.

### *D. Existence and reachability of the lower bound on the peak reachable transmission bandwidth of a primary path*

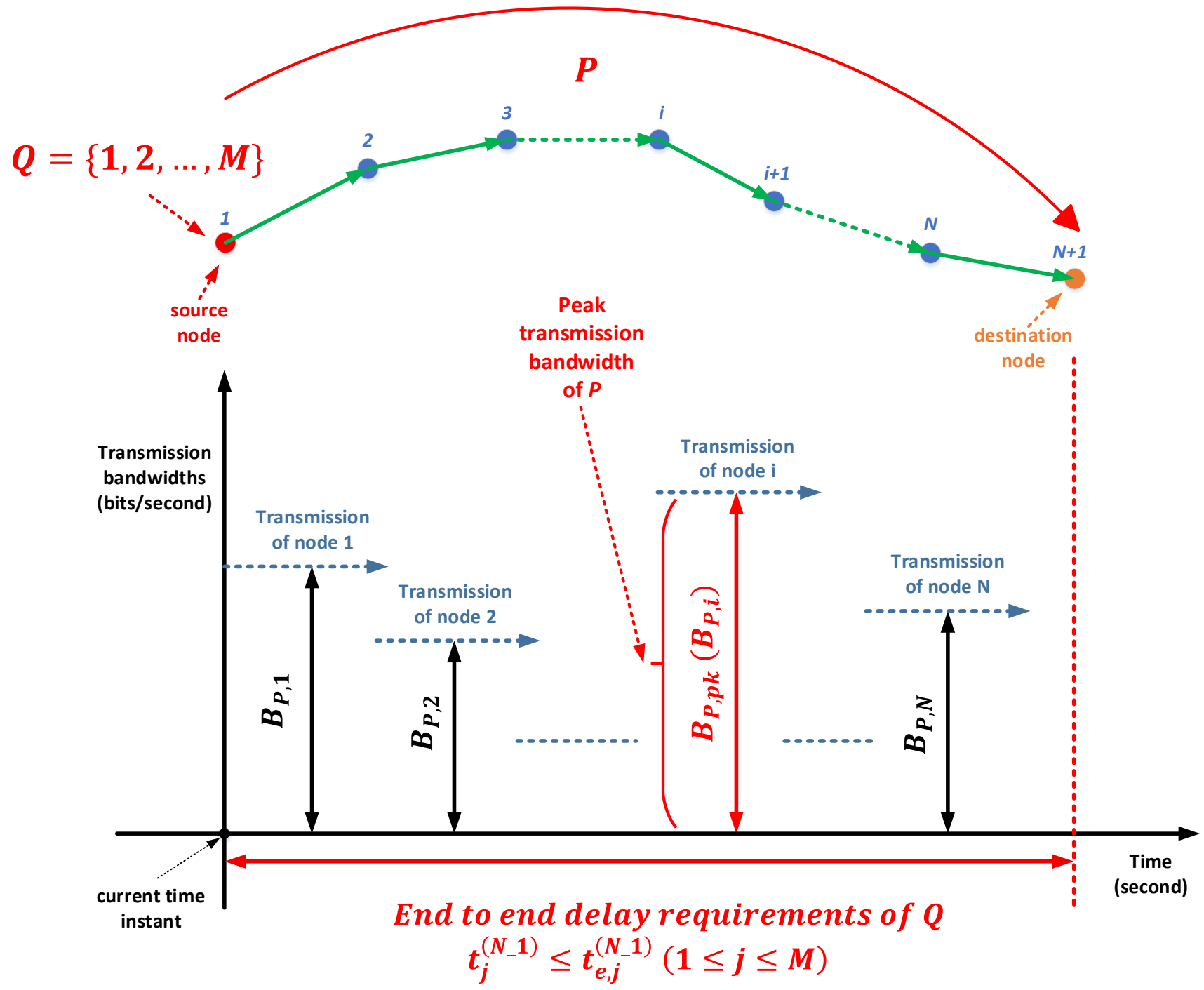

**Figure VIII.3** End-to-end transmission of an information particle group $Q$ over a primary path $P$ with reachable transmission bandwidth configured for each sending node

In this subsection, we further conduct an in-depth study of the system described in **Problem VIII.1**, namely **System VIII.1**. We still start from the current time instant $t = 0$, as illustrated in **Figure VIII.3**, consider a primary path $P$ composed of $N + 1$ nodes (see **Definition III.1** in [7]), starting from node 1 (the source node) in $P$, an information particle group $Q$ (see **Definition IV.1**) containing $M\ (\geq 1)$ information particles is relayed through node 2, node 3,..., node $i$,..., node $N\ (\geq 1)$, and finally arrives at node $N + 1$ (the destination node). After performing the incremental sorting operation $Op_{Inc}(Q)$ on the information particle group $Q$ (see **Definition IV.2**), the bearing capacity of the information particle $i\ (1 \leq i \leq M)$ in $Q$ and the deadline for its departure from node $N\ (\geq 1)$ are $\ell_j$ and $t_{e,i}^{(N_1)}$ $(\geq t_{Bb} + N \cdot T_B)$, respectively. To ensure that all information particles contained in $Q$ are transmitted from the source node to the destination node prior to their respective deadlines (i.e., satisfying $t_j^{(N_1)} \leq t_{e,j}^{(N_1)}\ (1 \leq j \leq M)$), appropriate reachable transmission bandwidth is allocated to each sending node on the primary path $P$. Specifically, it is assumed that the reachable transmission bandwidth allocated to the sending node $n_i$ $(n_i \in P, 1 \leq i \leq N)$ is $B_{P,i}$ $(1 \leq i \leq N)$ (unit: bits/second). Among all sending nodes on the primary path $P$, if the value $B_{P,j}$ of the reachable transmission bandwidth configured for node $n_j$ $(1 \leq j \leq N)$ is the maximum among the reachable transmission bandwidths configured for all sending nodes (i.e., $B_{P,j} \geq B_{P,i}$ $(1 \leq i, j \leq N, i \neq j)$), then $B_{P,j}$ is referred to as the peak reachable transmission bandwidth of the primary path $P$, which is denoted as $B_{P,pk}$.

In this subsection, we focus on investigating the following two key problems:

1) **Key Problem 1:** Given a primary path $P$ and an information particle group $Q$, to guarantee the end-to-end transmission performance of $Q$ over $P$ (i.e., all information particles contained in $Q$ are transmitted from the source node to the destination node before their respective deadlines), does there exist a lower bound for the peak reachable transmission bandwidth $B_{P,pk}$? In other words, is there a value such that if $B_{P,pk}$ is smaller than it, no effective end-to-end transmission strategy can be found to satisfy the performance requirements for the end-to-end transmission of $Q$;
2) **Key Problem 2:** If such a lower bound exists, is this bound achievable?

**Note:**

1) The difference between the system considered in this subsection and **System VIII.1** described in **Problem VIII.1** is that the reachable transmission bandwidth allocated to each sending node here is not required to be equal to one another;
2) It does not rule out the scenario where more than one sending node on the primary path has a reachable transmission bandwidth equal to the peak reachable transmission bandwidth of the primary path. An example is the case in which the reachable transmission bandwidth allocated to every sending node on the primary path is identical.

To facilitate the subsequent discussion, several necessary definitions are presented in the following.

**Definition VIII.1: An effective transmission strategy for an information particle group $Q$ over a primary path $P$**

Over the effective survival time interval $D_Q^{int}$ (**Definition IV.4**) of an information particle group $Q$, a time-varying function matrix $N \times M$ satisfying all the following conditions is defined as an effective transmission strategy for the information particle group $Q$ over a primary path $P$, denoted as $\boldsymbol{V}_{P,Q}^{ES}(t)$ $(t \in D_Q^{int})$.

$$\boldsymbol{V}_{P,Q}^{ES}(t) \triangleq \begin{pmatrix} v_{1,1}^{ES}(t) & v_{1,2}^{ES}(t) & \dots\dots & v_{1,M}^{ES}(t) \\ \dots\dots & \dots\dots\dots & \dots\dots & \dots\dots\dots \\ v_{i,1}^{ES}(t) & v_{i,2}^{ES}(t) & \dots\dots & v_{i,M}^{ES}(t) \\ \dots\dots & \dots\dots\dots & \dots\dots & \dots\dots\dots \\ v_{N,1}^{ES}(t) & v_{N,2}^{ES}(t) & \dots\dots & v_{N,M}^{ES}(t) \end{pmatrix}_{N \times M}$$

(VIII.34)

1) $v_{i,j}^{ES}(t) \in S_{v_j^{ES}}^{U}$ $(1 \le i \le N, 1 \le j \le M)$, where $S_{v_j^{ES}}^{U}$ denotes the universal set of effective processing strategies for information particle $j$ $(j \in Q)$ (**Definition III.8** and **Definition III.9**);
2) $\sum_{j \in Q} v_{i,j}^{ES}(t) \le B_{P,i}$ $\left(1 \le i \le N, t \in D_Q^{int}\right)$;
3) In the primary path $P$, if an interference relationship exists between sending node $n_i$ and sending node $n_k$ $(1 \le i, k \le N, i \ne k)$ (see **Definition IV.3** in [7]), then $\left(\sum_{j \in Q} v_{i,j}^{ES}(t)\right) \cdot \left(\sum_{j \in Q} v_{k,j}^{ES}(t)\right) = 0$ $\left(t \in D_Q^{int}\right)$.

**Note:**

1) **Condition 1** requires that $v_{i,j}^{ES}(t)$ $(1 \le i \le N, 1 \le j \le M)$ must be an effective processing strategy for information particle $j$ $(j \in Q)$. Therefore, the time-varying function $v_{i,j}^{ES}(t)$ is an effective transmission strategy performed by sending node $n_i$ $(n_i \in P, 1 \le i \le N)$ for information particle $j$ $(j \in Q)$. Furthermore, the value $v_{i,j}^{ES}(t)$ of this function at time instant $t$ $\left(t \in D_Q^{int}\right)$ can also be regarded as the instantaneous transmission rate (or instantaneous processing rate) of node $n_i$ for information particle $j$ at time instant $t$;
2) It can also be observed from the requirement of **Condition 1** that each sending node $n_i$ $(n_i \in P, 1 \le i \le N)$ on the primary path $P$ completes the transmission of all the information bits with a total volume of $\ell_j$ carried by information particle $j$ $(j \in Q)$ before its deadline $t_{e,j}^{(N_1)}$ (a more concise notation $t_{e,j}$ is adopted when no confusion arises), i.e., $\int_{t \in D_Q^{int}} v_{i,j}^{ES}(t)\, dt = \ell_j$ $(1 \le i \le N, 1 \le j \le M)$ holds;

3) **Condition 2** requires that the total instantaneous transmission rate (i.e., $\sum_{j\in Q} v_{i,j}^{ES}(t)$) of each sending node $n_i$ $(n_i \in P, 1 \le i \le N)$ on the primary path $P$ at any time instant within the effective survival time interval $D_Q^{int}$ of the information particle group $Q$ shall not exceed the reachable transmission bandwidth $B_{P,i}$ allocated to this node;

4) **Condition 3** requires that any two mutually interfering sending nodes on the primary path $P$ shall not be in the sending state simultaneously (i.e., shall not be active at the same time);

5) If the reachable transmission bandwidth configuration for each sending node is inappropriate, an effective transmission strategy satisfying the above conditions may not exist (this is exactly the question addressed in **Theorem VIII.3** in the subsequent discussion).

**Definition VIII.2: Global active time interval of a sending node in a primary path $P$**

Given an effective transmission strategy $V_{P,Q}^{ES}(t)$ for an information particle group $Q$ over a primary path $P$ (assuming such a strategy exists). Suppose that sending node $n_i$ $(n_i \in P, 1 \le i \le N)$ remains active over the closed time interval $[t_a, t_b]$ $([t_a, t_b] \subseteq D_Q^{int})$ (i.e., $\sum_{j\in Q} v_{i,j}^{ES}(t) > 0$ $(t \in [t_a, t_b])$, and this closed time interval $[t_a, t_b]$ is referred to as a local active time interval of sending node $n_i$ $(n_i \in P, 1 \le i \le N)$ over the time interval $D_Q^{int}$. On this basis, the union of all local active time intervals of sending node $n_i$ over the time interval $D_Q^{int}$ is defined as a global active time interval of sending node $n_i$ over the time interval $D_Q^{int}$ under the given information particle group $Q$ and the effective transmission strategy $V_{P,Q}^{ES}(t)$ for $Q$ on the primary path $P$, which is denoted as $D_{P,i}^{int}$ (it follows from the definition that $D_{P,i}^{int} \subseteq D_Q^{int}$ holds).

**Lemma VIII.3.1: Given a primary path $P$ consisting of $N+1$ nodes. Let the information particle group carried on $P$ be $Q$, and let an effective transmission strategy for $Q$ over $P$ be given as $V_{P,Q}^{ES}(t)$ $\left(t \in D_Q^{int}\right)$. If an interference relationship exists between sending node $n_i$ and sending node $n_k$ $(1 \le i, k \le N, i \ne k)$ on the primary path $P$, and the global active time intervals corresponding to $n_i$ and $n_k$ are $D_{P,i}^{int}$ $(D_{P,i}^{int} \subseteq D_Q^{int})$ and $D_{P,k}^{int}$ $(D_{P,k}^{int} \subseteq D_Q^{int})$, respectively, then**

$$D_{P,i}^{int} \cap D_{P,k}^{int} = \emptyset \qquad \textbf{(VIII.35)}$$

**Proof:**

Proof by contradiction is adopted. Assume that the proposition does not hold. Then it follows from **Definition VIII.2** that there exists at least one time instant $t$ $((t \in D_Q^{int}) \wedge (t \in D_{P,i}^{int}) \wedge (t \in D_{P,k}^{int}))$ at which both node $n_i$ and node $n_k$ are active, i.e., the following formula holds

$$\left(\sum_{j\in Q} v_{i,j}^{ES}(t)\right)\cdot\left(\sum_{j\in Q} v_{k,j}^{ES}(t)\right) > 0$$

(VIII.36)

It follows from **Definition VIII.1** that this contradicts the known premise that $\boldsymbol{V}_{P,Q}^{ES}(t)$ $\left(t \in D_Q^{int}\right)$ is an effective transmission strategy for $Q$ over $P$. ■

**Theorem VIII.3: Existence theorem for the lower bound of peak reachable transmission bandwidth of a primary path.**

**Given a primary path $P$ consisting of $N+1$ nodes, and its intrinsic interference intensity is $I_{S_P}^*$ (see Definition V.14 in [7]). The peak reachable transmission bandwidth of $P$ is $B_{P,pk}$. An information particle group $Q$ carried on $P$ contains $M$ information particles, and its minimum reachable access bandwidth is $B_{re,Q}^{min}$. If**

$$B_{P,pk} < B_{re,Q}^{min} \cdot I_{S_P}^*$$

**, then no effective transmission strategy for $Q$ over $P$ exists.**

**Proof:**

We proceed by contradiction. Suppose the proposition is false, i.e., suppose that when $B_{P,pk} < B_{re,Q}^{min} \cdot I_{S_P}^*$, there exists at least one effective transmission strategy, i.e. $\boldsymbol{V}_{P,Q}^{ES}(t)$ $\left(t \in D_Q^{int}\right)$, for $Q$ over $P$.

First, perform the incremental sorting operation $Op_{Inc}(Q)$ on the information particle group $Q$. Let $t_{e,j}$ $(1 \le j \le M)$ and $\ell_j$ $(1 \le j \le M)$ denote the deadline and the bearing capacity of the information particle $j$ $(1 \le j \le M)$ in $Q$, respectively. Next, construct an auxiliary information particle group $Q^V$, which also contains $M$ information particles. Let $t_{e,j}^V$ $(1 \le j \le M)$ and $\ell_j^V$ $(1 \le j \le M)$ denote the deadline and the bearing capacity of the information particle $j$ $(1 \le j \le M)$ in $Q^V$, respectively. Furthermore, let the following two relations hold

$$t_{e,j}^V \triangleq t_{e,j} \ (1 \le j \le M)$$

(VIII.37)

and

$$\ell_j^V \triangleq I_{S_P}^* \cdot \ell_j \ (1 \le j \le M)$$

(VIII.38)

Let $B_{re,Q^V}^{min}$ denote the minimum reachable access bandwidth of the information particle group $Q^V$. Combining the above two equations, it follows from **Theorem VI.6** that

$$B_{re,Q^V}^{min} = B_{re,Q}^{min} \cdot I_{S_P}^*$$

(VIII.39)

We shall next prove that the peak reachable transmission bandwidth $B_{P,pk}$ of $P$ is a reachable access bandwidth of the information particle group $Q^V$ (**Definition IV.11**).

To this end, starting from the aforementioned effective transmission strategy $\boldsymbol{V}_{P,Q}^{ES}(t)$ $\left(t \in D_Q^{int}\right)$ whose existence was assumed earlier, we construct an effective processing strategy for the information particle group $Q^V$ (**Definition IV.6** and **Definition IV.7**). Arbitrarily select $I_{S_P}^*$ sending nodes from the set of sending nodes of $P$ that can form a maximum second-order interference subset within $P$ (see **Definition V.12** in [7]). Without loss of generality, suppose that the selected $I_{S_P}^*$ sending nodes forming such a maximum second-order interference subset in $P$ are denoted as $n_{i_k}$ $(n_{i_k} \in P, 1 \le i_k \le N, 1 \le k \le I_{S_P}^*)$. Given the effective transmission strategy $\boldsymbol{V}_{P,Q}^{ES}(t)$ $\left(t \in D_Q^{int}\right)$ for $Q$ over $P$, let the global active time interval of sending node $n_{i_k}$ $(1 \le k \le I_{S_P}^*)$ corresponding to the effective transmission strategy $\boldsymbol{V}_{P,Q}^{ES}(t)$ $\left(t \in D_Q^{int}\right)$ be $D_{P,i_k}^{int}$ $(D_{P,i_k}^{int} \subseteq D_Q^{int}, 1 \le k \le I_{S_P}^*)$. According to **Lemma VIII.3.1**, we have

$$D_{P,i_{k_1}}^{int} \cap D_{P,i_{k_2}}^{int} = \emptyset$$

(VIII.40)

where $D_{P,i_{k_1}}^{int}, D_{P,i_{k_2}}^{int} \subseteq D_Q^{int}, 1 \le k_1, k_2 \le I_{S_P}^*, k_1 \ne k_2$ . Construct $M$ time-varying functions $\mu_j(t)$ $(1 \le j \le M, t \in D_Q^{int})$ over the effective survival time interval $D_Q^{int}$ of the information particle group $Q$, i.e.,

$$\mu_j(t) \triangleq \begin{cases} 0 & \left(t \in D_Q^{int} \backslash \left(\bigcup_{k=1}^{I_{S_P}^*} D_{P,i_k}^{int}\right)\right) \\ v_{i_k,j}^{ES}(t) & \left(t \in D_{P,i_k}^{int}, 1 \le k \le I_{S_P}^*\right) \end{cases}$$

(VIII.41)

Next, we verify that the time-varying functions $\mu_j(t)$ $(1 \le j \le M, t \in D_Q^{int})$ constructed above constitute an effective processing strategy for the corresponding information particle $j$ $(j \in$ $Q^V$) contained in the information particle group $Q^V$ in accordance with the four conditions specified in **Definition III.8**:

1) **Verification of the first condition:** Let $D_{Q^V,j}^{int}$ $(1 \le j \le M)$ denote the effective survival time interval of the information particle $j$ $(j \in Q^V)$ contained in $Q^V$. It can be known from the construction procedure of $Q^V$ that $D_{Q^V,j}^{int} \subseteq D_{Q^V}^{int} = D_Q^{int}$ $(1 \le j \le M)$. It can be observed from **Equation VIII.41** that $\mu_j(t)$ $(1 \le j \le M, t \in D_Q^{int})$ is well-defined in $D_Q^{int}$. Accordingly, $\mu_j(t)$ $(1 \le j \le M, t \in D_Q^{int})$ must also be well-defined in $D_{Q^V,j}^{int}$. That is, for any time instant $t$ $\left(t \in D_{Q^V,j}^{int}\right)$, there exists a real number $\mu_j(t)$ (derived from **Equation VIII.41**) corresponding thereto. Hence, this condition is satisfied;
2) **Verification of the second condition:** It can be seen from **Equation VIII.41** that $\mu_j(t) \ge 0$ $(t \in D_Q^{int})$. Therefore, this condition is satisfied;
3) **Verification of the third condition:** It follows from the foregoing assumption that the transmission strategy $\boldsymbol{V}_{P,Q}^{ES}(t)$ $\left(t \in D_Q^{int}\right)$ is effective. Therefore, by condition 2 of **Definition VIII.1**, it follows that the transmission strategy $\boldsymbol{V}_{P,Q}^{ES}(t)$ should satisfy $\sum_{j \in Q} v_{i,j}^{ES}(t) \le B_{P,i}$ $\left(1 \le i \le N, t \in D_Q^{int}\right)$. Furthermore, we have $v_{i,j}^{ES}(t) \le B_{P,pk}$ $\left(1 \le i \le N, 1 \le j \le M, t \in D_Q^{int}\right)$. Combining with **Equation VIII.41**, it readily follows that $\mu_j(t) \le B_{P,pk}$ $(t \in D_Q^{int}, 1 \le j \le M)$, hence the function $\mu_j(t)$ $(t \in D_Q^{int}, 1 \le j \le M)$ is bounded. Therefore, this condition is satisfied.
4) **Verification of the fourth condition:** Owing to $v_{i,j}^{ES}(t) \in S_{v_j^{ES}}^U$ $(1 \le i \le N, 1 \le j \le M)$, we can derive the following relations

$$\begin{aligned} \ell_j &= \int_{t \in D_Q^{int}} v_{i,j}^{ES}(t)\, dt \\ &= \int_{t \in D_{P,i}^{int}} v_{i,j}^{ES}(t)\, dt \end{aligned} \qquad (1 \le i \le N, 1 \le j \le M)$$

(VIII.42)

Integrating the time-varying function $\mu_j(t)$ $(1 \le j \le M, t \in D_Q^{int})$ over a specified time interval yields

$$
\begin{aligned}
\int_{t\in D_{Q^V,j}^{int}} \mu_j(t)\,dt &= \int_{t\in D_Q^{int}} \mu_j(t)\,dt \\
&= \int_{t\in D_Q^{int}\setminus\left(\cup_{k=1}^{I_{S_P}^*} D_{P,i_k}^{int}\right)} \mu_j(t)\,dt \\
&\quad +\sum_{k=1}^{I_{S_P}^*}\left(\int_{t\in D_{P,i_k}^{int}} \mu_j(t)\,dt\right) \qquad (1\le j\le M) \\
&= 0+\sum_{k=1}^{I_{S_P}^*}\left(\int_{t\in D_{P,i_k}^{int}} v_{i_k,j}^{ES}(t)\,dt\right) \\
&= I_{S_P}^*\cdot \ell_j \\
&= \ell_j^V
\end{aligned}
$$

(VIII.43)

In the above derivation, **Equation VIII.42** (namely, $\int_{t\in D_{P,i_k}^{int}} v_{i_k,j}^{ES}(t)\,dt=\ell_j\ (1\le j\le M)$) and **Equation VIII.38** (namely, $\ell_j^V \triangleq I_{S_P}^*\cdot \ell_j\ (1\le j\le M)$ ) are utilized. Consequently, this condition holds.

Combining the verifications of the four foregoing conditions, we have proven that $\mu_j(t)$ $(1\le j\le M, t\in D_Q^{int})$ is an effective processing strategy for the corresponding information particle $j$ $(j\in Q^V)$. In accordance with **Definition IV.6** and **Definition IV.7**, it can be concluded that $\left(\mu_1(t),\dots,\mu_j(t),\dots,\mu_M(t)\right)$ $(1<j<M, t\in D_Q^{int})$ constitutes an effective processing strategy for the information particle group $Q^V$. Furthermore, considering that the transmission strategy $\boldsymbol{V}_{P,Q}^{ES}(t)$ $\left(t\in D_Q^{int}\right)$ is effective, the following relation holds with reference to **Equation VIII.41**

$$
\sum_{j\in Q^V}\mu_j(t)=\begin{cases}0\ (\le B_{P,pk}) & \left(t\in D_Q^{int}\setminus\left(\bigcup_{k=1}^{I_{S_P}^*} D_{P,i_k}^{int}\right)\right)\\ \sum_{j=1}^{M} v_{i_k,j}^{ES}(t)\ (\le B_{P,i_k}\le B_{P,pk}) & \left(t\in D_{P,i_k}^{int}, 1\le k\le I_{S_P}^*\right)\end{cases}
$$

(VIII.44)

which gives

$$
\sum_{j\in Q^V}\mu_j(t)\le B_{P,pk}\ (t\in D_Q^{int})
$$

(VIII.45)

It can be known from the definition of a reachable access bandwidth for an information particle group (**Definition IV.11**) that $B_{P,pk}$ is one reachable access bandwidth of the information particle group $Q^V$. On the other hand, combining the definition of the minimum reachable access bandwidth of an information particle group (**Definition IV.12**) with **Equation VIII.39**, we obtain

$$
\begin{aligned}
B_{P,pk} &\geq B_{re,Q^V}^{min} \\
&= B_{re,Q}^{min} \cdot I_{S_P}^*
\end{aligned}
$$

(VIII.46)

This contradicts the foregoing assumption that $B_{P,pk} < B_{re,Q}^{min} \cdot I_{S_P}^*$. ■

**Theorem VIII.4: Reachability theorem for the lower bound of peak reachable transmission bandwidth of a primary path.**

**In System VIII.1, given a primary path $P$ composed of $N+1$ nodes, its intrinsic interference intensity and intrinsic period are denoted by $I_{S_P}^*$ and $T_{S_P}^*$, respectively (with $I_{S_P}^* = T_{S_P}^*$ satisfied). An information particle group $Q$ carried on $P$ contains $M$ information particles, and its minimum reachable access bandwidth is $B_{re,Q}^{min}$. The system executes Algorithm X.1 proposed in [7] periodically with period $T_{S_P}^*$. During algorithm execution, each time beat has a fixed duration of $T_B$, and the starting time instant of the initial time beat is $t_{Bb}$ (with $0 \leq t_{Bb} \leq T_{S_P}^* \cdot T_B$ satisfied). Let $B_{P,Q}^{min_1}$ be the minimum reachable transmission bandwidth when $P$ bears $Q$. Then the following relation holds**

$$\lim_{T_B \to 0} B_{P,Q}^{min_1} = B_{re,Q}^{min} \cdot I_{S_P}^*$$

**(VIII.47)**

**Proof:**

It follows from **Theorem VIII.1** that the minimum reachable transmission bandwidth $B_{P,Q}^{min_1}$ attainable by **System VIII.1** is equal to the minimum reachable transmission bandwidth $B_{P,Q}^{min_5}$ achievable by **System VIII.5** mapped from **System VIII.1**. Accordingly, we hereinafter convert the problem of proving $\lim_{T_B \to 0} B_{P,Q}^{min_1}$ into the problem of proving $\lim_{T_B \to 0} B_{P,Q}^{min_5}$; that is, we first verify the validity of $\lim_{T_B \to 0} B_{P,Q}^{min_5} = B_{re,Q}^{min} \cdot I_{S_P}^*$, and thereby establish $\lim_{T_B \to 0} B_{P,Q}^{min_1} = B_{re,Q}^{min} \cdot I_{S_P}^*$.

Perform the incremental sorting operation $Op_{Inc}(Q)$ on the information particle group $Q$ borne by the primary path $P$. Let $t_{e,j}^{(N_1)}$ $(1 \leq j \leq M)$ and $\ell_j$ $(1 \leq j \leq M)$ denote the deadline and bearing capacity of the information particle $j$ $(1 \leq j \leq M)$ in $Q$, respectively. When mapping the deadline $t_{e,j}^{(N_1)}$ $(1 \leq j \leq M)$ of each information particle belonging to the information particle group $Q$ carried on the primary path $P$ in **System VIII.1** to the deadline $t_{e,j}^{(1_5)}$ $(1 \le j \le M)$ of the corresponding information particle in the information particle group $Q^{(1_5)}$ of **System VIII.5**, the mapping relation specified in **Equation VIII.30** can be adopted. For the convenience of elaboration, this relational expression is rearranged and rewritten as follows

$$t_{e,j}^{(1_5)} = T_B \cdot \left( \left\lfloor \frac{t_{e,j}^{(N_1)} - N \cdot T_B - t_{Bb}}{T_B \cdot T_{S_P}^{*}} \right\rfloor + 1 \right) \; (1 \le j \le M)$$

(VIII.48)

The bearing capacity $\ell_j$ $(1 \le j \le M)$ of each information particle in the information particle group $Q$ carried by the primary path $P$ in **System VIII.1** remains unchanged when mapped to the bearing capacity of the corresponding information particle in the information particle group $Q^{(1_5)}$ in **System VIII.5**.

We next derive the limit of the deadline $t_{e,j}^{(1_5)}$ $(1 \le j \le M)$ corresponding to each individual information particle in **System VIII.5** under the condition $T_B \to 0$. Two inequalities can be sequentially obtained from **Equation VIII.48** as follows

$$\begin{aligned} t_{e,j}^{(1_5)} &> T_B \cdot \left( \frac{t_{e,j}^{(N_1)} - N \cdot T_B - t_{Bb}}{T_B \cdot T_{S_P}^{*}} - 1 + 1 \right) \\ &\ge T_B \cdot \left( \frac{t_{e,j}^{(N_1)} - N \cdot T_B - T_{S_P}^{*} \cdot T_B}{T_B \cdot T_{S_P}^{*}} \right) \qquad (1 \le j \le M) \\ &= \frac{t_{e,j}^{(N_1)} - T_B \cdot (N + T_{S_P}^{*})}{T_{S_P}^{*}} \\ &\triangleq \alpha_j \end{aligned}$$

(VIII.49.1)

and

$$\begin{aligned} t_{e,j}^{(1_5)} &\le T_B \cdot \left( \frac{t_{e,j}^{(N_1)} - N \cdot T_B - t_{Bb}}{T_B \cdot T_{S_P}^{*}} + 1 \right) \\ &\le T_B \cdot \left( \frac{t_{e,j}^{(N_1)} - N \cdot T_B}{T_B \cdot T_{S_P}^{*}} + 1 \right) \qquad (1 \le j \le M) \\ &= \frac{t_{e,j}^{(N_1)} - T_B \cdot (N - T_{S_P}^{*})}{T_{S_P}^{*}} \\ &\triangleq \beta_j \end{aligned}$$

(VIII.49.2)

Combining **Equation VIII.49.1** and **Equation VIII.49.2**, we obtain the following relation

$$\alpha_j < t_{e,j}^{(1_5)} \leq \beta_j \ (1 \leq j \leq M)$$

(VIII.49.3)

It readily follows from **Equation VIII.49.1** and **Equation VIII.49.2** that

$$\begin{aligned} \lim_{T_B \to 0} \alpha_j &= \lim_{T_B \to 0} \beta_j \\ &= \frac{t_{e,j}^{(N_1)}}{T_{S_P}^*} \ (1 \leq j \leq M) \\ &= \frac{t_{e,j}^{(N_1)}}{I_{S_P}^*} \end{aligned}$$

(VIII.50)

Combining **Equation VIII.49.3** and **Equation VIII.50**, we obtain

$$\lim_{T_B \to 0} t_{e,j}^{(1_5)} = \frac{t_{e,j}^{(N_1)}}{I_{S_P}^*} \ (1 \leq j \leq M)$$

(VIII.51)

Subsequently, we derive the minimum reachable transmission bandwidth $B_{P,Q}^{min_5}$ achievable by **System VIII.5** under condition $T_B \to 0$, and ultimately acquire the limit of the minimum reachable transmission bandwidth $B_{P,Q}^{min_1}$ attainable in **System VIII.1** (since $B_{P,Q}^{min_1} = B_{P,Q}^{min_5}$ holds, it follows that $\lim_{T_B \to 0} B_{P,Q}^{min_1} = \lim_{T_B \to 0} B_{P,Q}^{min_5}$). Let $B_{re,Q^{(1_5)}}^{min}$ denote the minimum reachable access bandwidth of the information particle group $Q^{(1_5)}$ in **System VIII.5**. It can be deduced from **Corollary VII.1.1** that

$$B_{re,Q^{(1_5)}}^{min} = \max_{j=1,\dots,M} \left\{ B_{av,Q_j^{(1_5)}} \right\}$$

(VIII.52)

Herein, $B_{av,Q_j^{(1_5)}}$ denotes the average access bandwidth requirement of the subgroup of information particles $Q_j^{(1_5)}$ ($Q_j^{(1_5)} \subseteq Q^{(1_5)}$) consisting of all information particles with sequence numbers no greater than $j$ $(1 \leq j \leq M)$ within the information particle group $Q^{(1_5)}$. Combining the definition of the average access bandwidth requirement of an information particle group (**Definition IV.8**), we obtain

$$\begin{aligned} B_{re,Q^{(1_5)}}^{min} &= \max_{j=1,\dots,M}\left\{\frac{\ell_{Q_j^{(1_5)}}}{L_{D_{Q_j^{(1_5)}}^{int}}}\right\} \\ &= \max_{j=1,\dots,M}\left\{\frac{\sum_{k=1}^{j}\ell_k}{L_{D_{Q_j^{(1_5)}}^{int}}}\right\} \\ &= \max_{j=1,\dots,M}\left\{\frac{\sum_{k=1}^{j}\ell_k}{t_{e,j}^{(1_5)}}\right\} \end{aligned}$$

(VIII.53)

Furthermore, combining **Equation VIII.51**, we obtain

$$\begin{aligned} \lim_{T_B\to 0} B_{re,Q^{(1_5)}}^{min} &= \lim_{T_B\to 0}\left(\max_{j=1,\dots,M}\left\{\frac{\sum_{k=1}^{j}\ell_k}{t_{e,j}^{(1_5)}}\right\}\right) \\ &= \max_{j=1,\dots,M}\left\{\frac{\sum_{k=1}^{j}\ell_k}{\lim_{T_B\to 0}\left(t_{e,j}^{(1_5)}\right)}\right\} \\ &= \max_{j=1,\dots,M}\left\{\frac{\sum_{k=1}^{j}\ell_k}{t_{e,j}^{(N_1)}}\right\}\cdot I_{S_P}^{*} \\ &= B_{re,Q}^{min}\cdot I_{S_P}^{*} \end{aligned}$$

(VIII.54)

Based on the foregoing formula and relation $B_{P,Q}^{min_5} = B_{re,Q^{(1_5)}}^{min}$ (see the note section of **Theorem VIII.1**), we obtain

$$\begin{aligned} \lim_{T_B\to 0} B_{P,Q}^{min_5} &= \lim_{T_B\to 0} B_{re,Q^{(1_5)}}^{min} \\ &= B_{re,Q}^{min}\cdot I_{S_P}^{*} \end{aligned}$$

(VIII.55)

Since $\lim_{T_B\to 0} B_{P,Q}^{min_1} = \lim_{T_B\to 0} B_{P,Q}^{min_5}$, we finally derive the result from **Equation VIII.55**

$$\lim_{T_B\to 0} B_{P,Q}^{min_1} = B_{re,Q}^{min}\cdot I_{S_P}^{*}$$

(VIII.56)

■

**Note:**

1) The assumption that $0 \le t_{Bb} \le T_{S_P}^{*}\cdot T_B$ holds in this proposition is made because it is assumed that **System VIII.1** has been running **Algorithm X.1** proposed in [7] at or prior to the current time instant $(t = 0)$ (see **Clause 7** in the note section of **Problem VIII.1** for further details);

2) It can be observed from **Equation VIII.49.1** and **Equation VIII.49.2** that the following condition holds for all $j\ (1 \le j \le M)$

$$T_B \ll \frac{t_{e,j}^{(N_1)}}{(N + T_{S_P}^*)}\ (1 \le j \le M)$$

(VIII.57)

then $t_{e,j}^{(1_5)} \approx \frac{t_{e,j}^{(N_1)}}{I_{S_P}^*}\ (1 \le j \le M)$ holds. Furthermore, an approximate relation can be ultimately obtained

$$B_{P,Q}^{min_1} \approx B_{re,Q}^{min} \cdot I_{S_P}^*$$

(VIII.58)

Therefore, in **System VIII.1**, provided that a sufficiently small duration $T_B$ of a time beat is selected in accordance with **Equation VIII.57**, the peak reachable transmission bandwidth of the primary path can sufficiently approximate the lower bound, i.e. $B_{re,Q}^{min} \cdot I_{S_P}^*$, by executing **Algorithm X.1** proposed in [7].

## IX. Conclusions

Aiming at some problems existing in the current QoS guarantee mechanism in large-scale networks (i.e. poor scalability, coarse granularity for provided service levels, poor fairness between different service levels, and improving delay performance at the expense of sacrificing some resource utilization), this paper puts forward the basic idea and theory of particle access. In the proposed particle access mechanism, the network allocates access resources to the information particle group which is composed of all the information particles to be transmitted, so as to ensure that the occupied bandwidth resources are minimized on the premise of meeting the delay requirements of each information particle. Specifically, in the paper, the concepts of both information particle and information particle group are defined; The relationships between the average access bandwidth requirement, the peak average access bandwidth requirement, and the minimum reachable access bandwidth of an information particle group are analyzed; The influences of time attribute and attribute of bearing capacity of an information particle group on the minimum reachable access bandwidth are analyzed. The equal scaling theorem for information particle groups is proved; Effective methods for the calculation of the minimum reachable access bandwidth of an information particle group are given, and a particle access algorithm based on dynamically adjusting the minimum reachable access bandwidth is proposed; Furthermore, several properties of the minimum reachable access bandwidth(s) of information particle group(s) are proved; Finally, the joint optimization of the mechanism of network wave and the mechanism of particle access is studied. For a given information particle group transmitted over a given primary path, a method to obtain the minimum reachable transmission bandwidth, which can guarantee the end-to-end delays of all the information particles contained in the information particle group, is proposed; It is proven that, given a primary path and an information particle group it bears, to guarantee the end-to-end transmission performance of the information particle group over the primary path, there exists a lower bound for the peak reachable transmission bandwidth of the primary path, and this lower bound is achievable.

The research results of this paper provide a new idea for further improving the QoS mechanism of large-scale networks, and lay the corresponding theoretical foundation.

In the following, some important conclusions are summarized:

1) In **Theorem V.1**, it is pointed out that the minimum reachable access bandwidth of an

information particle group is equal to its peak average access bandwidth requirement;

2) In **Theorem VI.3** and **Theorem VI.5**, it is proved that for a given information particle group which meets a certain bandwidth constraint, it can still meet the bandwidth constraint by adjusting the deadline attribute and/or the bearing capacity attribute of a newly arrived information particle;
3) **Theorem VI.6** presents the equal scaling property of an information particle group. That is, proportionally scaling the deadline attribute and/or bearing capacity attribute of each information particle contained in the information particle group will induce the corresponding equal scaling of the minimum reachable access bandwidth of the information particle group;
4) In **Theorem VII.1** and **Corollary VII.1.1**, we respectively provide methods for calculating the minimum reachable access bandwidth of an information particle group;
5) In **Corollary VII.1.6**, the fact that the minimum reachable access bandwidth decreases at the inflection point of the access bandwidth requirement of an information particle group is proved;
6) In **Algorithm VII.1**, a particle access algorithm based on dynamically adjusting the minimum reachable access bandwidth is proposed;
7) The upper bound (**Corollary VII.1.7**) and the lower bound (**Corollary VII.1.8**) of the minimum reachable access bandwidth of an information particle group are given, respectively;
8) The property of linear superpositions of multiple information particle groups under specific constraints is proved (**Corollary VII.1.9**). The property of linear splitting of an information particle group is also proved (**Corollary VII.1.10**);
9) A closed form expression of representing the minimum reachable access bandwidth of an information particle subgroup as the product of its vector of switching functions and its vector of average access bandwidth requirements is given (**Corollary VII.1.11**). And it is proved that the vector of switching functions has and only has one component with its value being equal to one (**Corollary VII.1.12**);
10) In **Theorem VIII.1**, **Algorithm VIII.1** and **Algorithm VIII.2**, for a given information particle group transmitted over a given primary path, a method to obtain the minimum reachable transmission bandwidth, which can guarantee the end-to-end delays of all the information particles contained in the information particle group, is proposed;
11) In **Theorem VIII.3** and **Theorem VIII.4**, the existence and achievability of the lower bound

of the peak reachable transmission bandwidth of a primary path are proven. This strongly demonstrates that the end-to-end transmission scheme proposed based on the theory of wave-particle duality in networks is optimal.

**References**

[1] A. S. Tanenbaum, D. J. Wetherall, “Computer networks (5th ed.)”, Prentice Hall, 2011.

[2] R. Braden, L. Zhang, S. Berson, S. Herzog, S. Jamin, “Resource ReSerVation Protocol (RSVP)”, IETF RFC 2205, 1997.

[3] K. Nichols, S. Blake, F. Baker, D. Black, “Definition of the Differentiated Services Field (DS Field) in the IPv4 and IPv6 Headers”, IETF RFC 2474, 1998.

[4] B. Davie, A. Charny, J.C.R. Bennett, K. Benson, J.Y. Le Boudec, W. Courtney, S. Davari, V. Firoiu, D. Stiliadis, “An Expedited Forwarding PHB (Per-Hop Behavior)”, IETF RFC 3246, 2001.

[5] J. Heinanen, F. Baker, W. Weiss, J. Wroclawski, “Assured Forwarding PHB Group”, IETF RFC 2597, 1999.

[6] Bo LI, G. A. Abdulwahab Mohammed, Mao YANG, Zhongjiang YAN, “The Design Methodology for MAC Strategies and Protocols Supporting Ultra-low Delay Services in Next Generation IEEE 802.11 WLAN”, International Conference on Internet of Things as a Service 2020.

[7] Bo LI, Mao YANG, and Zhongjiang YAN, “Theory of Network Wave”, arXiv preprint, ID: 2203.05241 (v.9), 2026.

**Appendix A：Symbol List**

| | |
|---|---|
| $\ell_i$ | The bearing capacity of an information particle (refer to Definition III.3) |
| $t_{b,i}$ | The initial time instant of an information particle (refer to Definition III.4) |
| $t_{e,i}$ | The deadline of an information particle (refer to Definition III.5) |
| $D_i^{int}$ | The effective survival time interval of an information particle (refer to Definition III.6) |
| $L_{D_i^{int}}$ | The effective survival time span of an information particle (refer to Definition III.7) |
| $v_i^{ES}(t)$ | An effective processing strategy performed by the system for an information particle $i$ (refer to Definition III.8) |
| $S_{v_i^{ES}}^{U}$ | The universal set of effective processing strategies for an information particle $i$ (refer to Definition III.9) |
| $B_{av,i}$ | The average access bandwidth requirement of an information particle (refer to Definition III.10) |
| $Q$ | An information particle group (refer to Definition IV.1) |
| $Op_{Inc}(Q)$ | The incremental sorting operation performed on an information particle group $Q$ (refer to Definition IV.2) |
| $\ell_Q$ | The bearing capacity of an information particle group (refer to Definition IV.3) |
| $D_Q^{int}$ | The effective survival time interval of an information particle group (refer to Definition IV.4) |
| $L_{D_Q^{int}}$ | The effective survival time span of an information particle group (refer to Definition IV.5) |
| $S_{Q^{ES}}^{U}$ | The universal set of effective processing strategies performed on an information particle group $Q$ (refer to Definition IV.6) |
| $B_{av,Q}$ | The average access bandwidth requirement of an information particle group $Q$ (refer to Definition IV.8) |
| $B_{av,Q}^{pk}$ | The peak average access bandwidth requirement of an information |
| | particle group $Q$ (refer to Definition IV.9) |
| $B_{re,Q}$ | The reachable access bandwidth of an information particle group $Q$ (refer to Definition IV.11) |
| $S^{U}_{B_{re,Q}}$ | The universal set of reachable access bandwidths for an information particle group $Q$ (refer to Definition IV.12) |
| $B^{min}_{re,Q}$ | The minimum reachable access bandwidth of an information particle group $Q$ (refer to Definition IV.12) |
| $\eta_{Q\|B_{re,Q}}$ | The access efficiency of an information particle group $Q$ under a given reachable access bandwidth and within its effective survival time interval (refer to Definition IV.14) |
| $N_{cr,Q}$ | The inflection point of the required access bandwidth of an information particle group $Q$ (refer to Definition IV.15) |
| $Q^{*}_{cr}$ | The largest principal subgroup of an information particle group $Q$ (refer to Definition IV.16) |
| $Q_i$ | After an incremental sorting operation $Op_{Inc}(Q)$ is performed, $Q_i$ $(Q_i \subseteq Q)$ is the information particle subgroup composed of all the information particles in $Q$ whose sorting sequence numbers are not greater than $i$ (refer to Definition IV.17) |
| $Q^i$ | An information particle group with an index number $i$ (refer to Definition IV.18) |
| $\delta_{i,Q}$ | Switching function of an information particle group $Q$ with sequence number $i$ (refer to Definition IV.19) |
| $\boldsymbol{\delta}_{Q_n}$ | Vector of switching functions of an information particle subgroup $Q_n$ $(Q_n \subseteq Q)$ (refer to Definition IV.20) |
| $\boldsymbol{B}_{av,Q_n}$ | Vector of average access bandwidth requirements of an information particle subgroup $Q_n$ $(Q_n \subseteq Q)$ (refer to Definition IV.21) |
| $t_{Bb}$ | The starting time instant of the initial time beat (refer to Problem VIII.1) |
| $T_B$ | The duration of a time beat (refer to Problem VIII.1) |
| $t^{(N_1)}_{e,i}$ | The deadline of an information particle leaving from node $N$ in System |
| | VIII.1 (refer to Problem VIII.1) |
| $B_{P,Q}^{min_1}$ | The minimum reachable transmission bandwidth of each relaying node in System VIII.1 (refer to Problem VIII.1) |
| $t_{e,i}^{(1_2)}$ | The deadline of an information particle $i$ leaving from node 1 in System VIII.2 (refer to Problem VIII.2) |
| $B_{P,Q}^{min_2}$ | The minimum reachable transmission bandwidth of node 1 in System VIII.2 (refer to Problem VIII.2) |
| $t_{e,i}^{(1_3)}$ | The deadline of an information particle $i$ leaving from node 1 in System VIII.3 (refer to Problem VIII.3) |
| $B_{P,Q}^{min_3}$ | The minimum reachable transmission bandwidth of node 1 in System VIII.3 (refer to Problem VIII.3) |
| $t_{e,i}^{(1_4)}$ | The deadline of an information particle $i$ leaving from node 1 in System VIII.4 (refer to Problem VIII.4) |
| $B_{P,Q}^{min_4}$ | The minimum reachable transmission bandwidth of node 1 in System VIII.4 (refer to Problem VIII.4) |
| $t_{e,i}^{(1_5)}$ | The deadline of an information particle $i$ leaving from node 1 in System VIII.5 (refer to Problem VIII.5) |
| $B_{P,Q}^{min_5}$ | The minimum reachable transmission bandwidth of node 1 in System VIII.5 (refer to Problem VIII.5) |
| $\boldsymbol{V}_{P,Q}^{ES}(t)$ | An effective transmission strategy for an information particle group $Q$ over a primary path $P$ (refer to Definition VIII.1) |
| $D_{P,i}^{int}$ | Global active time interval of a sending node $n_i$ in a primary path $P$ (refer to Definition VIII.2) |
| | |
| | |